\documentclass{aa}
 
\usepackage{graphicx}

\usepackage{txfonts}

\begin{document}

\title{The elliptical galaxies NGC\,1052 and NGC\,7796:}

\subtitle{stellar populations and abundance $\alpha$/Fe ratio
\thanks{Based on observations made in
{\it Observat\'orio do Pico dos Dias} (OPD), which is operated by LNA
({\it Laborat\'orio Nacional de Astrof\'\i sica}), Brazil.}
}

\author{
A. de C. Milone\inst{1}, 
M. G. Rickes\inst{2}
and M. G. Pastoriza\inst{2}}

\institute{
Instituto Nacional de Pesquisas Espaciais (INPE),
Coordena\c c\~ao de C\^encias Espaciais e Atmosf\'ericas,
Divis\~ao de Astrof\'\i sica,
Av. dos Astronautas 1758, S\~ao Jos\'e dos Campos, SP,
12227-010, Brazil\\
\email{acmilone@das.inpe.br}
\and
Universidade Federal do Rio Grande do Sul (UFRGS),
Instituto de F\'\i sica,
Departamento de Astronomia,
Av. Bento Gon\c calves 9500, Porto Alegre, RS, 
91501-970, Brazil\\
}

\date{Received 30 November 2005 / Accepted ...}

\offprints{A. de C. Milone}

%
%
\abstract
{Understanding how each early-type galaxy forms and evolves
is one of the objectives of the
extragalactic astrophysics and cosmology.
The spatial distribution of the stellar populations inside a spheroidal system
and their kinematical properties
supply important informations about the formation process.
Specifically, the reconstruction of the star formation history
is crucial in this context.}
{We have performed a detailed stellar population analysis
using long slit spectroscopic observations up to almost one effective radius
of two different early-type galaxies of low density regions of the local Universe:
NGC\,1052, a E4 Liner prototype of a loose group that has a stellar rotating disc,
and NGC\,7796, a E1 of the field which shows a kinematically distinct core.
The mean luminosity-weighted stellar age, metallicity, and $\alpha$/Fe ratio
along both photometric axes of them
have been obtained in order to reconstruct the star formation history
in their kinematically distinct subsystems.}
{We have measured Lick indices and computed their radial gradients.
They were compared with the predicted ones of simple stellar population models.
We have also applied a stellar population synthesis.}
{The star characteristics are associated with their kinematics:
they are older and $\alpha$-enhanced in the bulge of NGC\,1052
and core of NGC\,7796, while
they show a strong spread of $\alpha$/Fe and age along the disc of NGC\,1052
and an outwards radial decreasing of them outside the core of NGC\,7796.
The age variation is possibly connected to the $\alpha$/Fe one.}
{Both galaxies were formed by processes
in which the star formation occurred firstly at the bulge (NGC\,1052)
and nucleus (NGC\,7796) 12-15 Gyr ago on short timescales (0.1-1 Gyr)
providing an efficient chemical enrichment by SN-II.
In the disc of NGC\,1052,
there is some spread of age and formation timescales around its stars.
In NGC\,7796, the star formation timescale
had some outwards radial increasing along both axes.}

\maketitle

%
%
\section{Introduction}

A great effort has been spent to understand
how the stellar populations are formed and evolve in the early-type galaxies.
The early-type galaxies (or their nuclei) are not composed
by a single-aged stellar population
but they seem as a mix of simple stellar populations.
Recent discoveries show that ellipticals are not merely a one-parameter family
as a function of the global initial mass only
(Yoshii \& Arimoto 1987).
It is known they are a two-parameter family, given by the fundamental plane.
Many ellipticals have signatures of interaction with the environment.
These galaxies might have had different star formation histories,
with stellar populations differing in metallicity and/or age
(Worthey, Faber \& Gonzalez 1992).
Crucial information on the above issues have been compiled
from the radial gradients of the metal line-strength indices
for field ellipticals
(Davies, Sadler \& Peletier 1993 and
Kobayashi \& Arimoto 1999)
and for cluster ones
(Jorgensen 1997 and
Mehlert et al. 2003).
The metal line-strength gradients in early-type galaxies, for example,
can vary considerably, ranging from essentially featureless to
structured profiles showing e.g. changes of slope possibly associated
with kinematically decoupled cores
(Bender \& Surma 1992 and
Morelli et al. 2004),
or anomalies in the stellar populations
(Carollo \& Danziger 1994).

More recent studies about the stellar content of early-type galaxies
have analyzed two dimensional spectroscopic data
of their central parts in terms of kinematics and population parameters 
(e.g. Davies et al. 2001 and
McDermid et al. 2006).
Davies et al. (2001) have found two independent kinematic spatial subsystems
in the E3 galaxy NGC\,4365 which have the same luminosity-weighted age
(nearly 14 Gyr) and the same overabundance for the Magnesium-to-Iron ratio
indicating a common star formation history for both kinematically distinct
components.
McDermid et al. (2006) showed a summary of the results of
two integral-field spectroscopic surveys using two instruments
(SAURON and OASIS).
These results let to analyze in details the relationship between 
kinematically distinct components
and their host galaxies basically in terms of the stellar age distribution
(presence or absent of young populations).

The metallicity radial gradients in bulges or early-type galaxies
are related to the formation process of these spheroidal systems.
A monolithic dissipative collapse of gas clouds
associated with intense star formation
can form a strong stellar metallicity gradient
(Carlberg 1984,
Chiosi \& Carraro 2002 and
Kawata 1999).
On the other hand, it can also be induced by a hierarchical merging
(Mihos \& Hernsquist 1994 and
Bekki \& Shioya 1999).
Kobayashi (2004),
in a general point of view, concludes that
the metallicity and age gradients are dependent on the galaxy merging history.
Forbes, S\'anchez-Bl\'azquez \& Proctor (2005) 
found there is a direct correlation between the strength of the metallicity gradient
and the mass of the early-types.

Many galaxies show in the central region
the Mg/Fe stellar abundance ratio larger than the solar one
(i.e. an overabundance of the alpha-elements relative to the iron peak elements)
which is interpreted as a consequence of the chemical enrichment
given by type II supernovae relative to type Ia ones
(Idiart, Michard \& de Freitas Pacheco 2003).
However,
Calcium is found underabundant relative to Iron in ellipticals,
through the analysis of different absorption lines of it,
despite beeing an alpha-element like Magnesium:
the Ca II triplet at 8600 {\AA}
(Saglia et al.2002),
and
the blue Lick index Ca4227
(Thomas, Maraston \& Bender, 2003b).

Moreover, the radial gradients of Mg line-strength indices of Lick System
can have equal or different behavior of those of Fe indices
due to the galaxy formation process
and the time scales of the star formation events.

Other studies have shown that the radial dependency of stellar $\alpha$/Fe,
or specifically the Mg/Fe abundance ratio,
can solve the star formation history inside a galaxy.
Pipino, Matteucci \& Chiappini (2006)
have found that
NGC\,4697 was formed by an outside-in process
because the Mg/Fe ratio is increasing outwards in that galaxy
based on a chemical evolution analysis.
However, the majority of the stellar populations studies
in ellipticals have shown that the oversolar Mg/Fe abundance ratio
is a global characteristic of an E galaxy
(e.g. Mehlert et al 2003).
It is still not obvious that the Mg/Fe abundance ratio
is correlated to the global mass
and/or the central velocity dispersion of the early-type galaxies
(e.g. Henry \& Worthey 1999 and
Thomas et al. 2005).

At least, three kind of chemical evolution models have been proposed for ellipticals
based on an occurrence of a internal galactic wind
which rules the end of the star formation
(Matteucci et al. 1998).
During a long or short star burst,
some amount of kinetical and thermal energy is transferred to the interstellar medium
by the supernova explosions.
When this energy is greater than the gravitational energy of the system,
the interstellar material can be driven out of the galaxy
by a global wind, which is produced by the supernovae.
Thus the wind interrupts the star formation
and the stellar populations evolve passively.
One of them is called as the classic wind model
(Larson 1974a,
Larson 1974b,
Arimoto \& Yoshii 1987 and
Matteucci \& Tornamb\`e 1987), which
says that the star formation stops later in the more massive galaxies
than less massive ones because their deeper gravitational potential wells.
The consequence is to provide higher Mg/Fe abundance ratio
in less massive galaxies (disagreeing to the observed).
The other is named as the inverse wind model
(Matteucci 1994),
which establishes that the galactic wind occurs earlier
in the more massive galaxies providing an efficient star formation
and a higher Mg/Fe ratio.
It is according to the relation Mg$_2$-$\sigma_{v}^{0}$
(or log$<$Fe$>$-$\sigma_{v}^{0}$ that has a weaker positive correlation).
These chemical evolution models are based on
the monolithic collapse of a gas cloud.
The last one was also proposed by
Matteucci (1994).
It suggests that more massive ellipticals are formed by merging of proto-clouds
providing higher relative velocities to the cloud-cloud collisions.
The result is a more star formation effectiveness for the more massive galaxies 
so that galactic wind can happen earlier providing more Mg enrichment 
relative to Fe.

We have selected for the analysis of stellar populations
two elliptical galaxies
with intermediated masses ($\sim$ 10$^{11}$\,$M_{\odot}$)
of low density regions of the local Universe:
NGC\,1052, which belongs to a loose group,
and NGC\,7796 of the field.
They are different in terms of stellar kinematics:
NGC\,1052 has a stellar rotating disc
and NGC\,7796 a kinematically distinct core.
Some parameters of them are given in Table 1.

The E4 galaxy {\bf NGC\,1052} is well studied for several aspects:
it is classified as a LINER prototype, and
it has a nuclear jet at radio, optical and X-ray frequencies
(Kadler et al. 2004).
This elliptical has neutral hydrogen too.
It is third brightest member of a group with 11 galaxies identified by
Giuricin et al. (2000).
At V, R and I bands the isophotes, that follow de Vaucouleurs profile, 
change the ellipticity outwards from 0.10 up to 0.35 
and they have an small isophotal twist too
(Bender et al. 1988).
Radial gradients of the $(U-R)$ and $(B-R)$ colours
were observed in this galaxy
(Peletier et al. 1990).
The presence of nuclear gas was detected through HST Nicmos observation by
Ravindranath et al. (2001).
The NGC\,1052 internal kinematics was analyzed by several authors.  
Binney et al. (1990),
using dynamical models and photometric/spectroscopic data,
have concluded that this elliptical
has a rotating disc with inclination angle $i$ = 90$^{o}$
(a pure axisymmetric component)
and it shows $M/L$ = 4.5$h_{50}$ in R band and 8.0$h_{50}$ in B band.
Fried \& Illingworth (1994)
have calculated its rotation parameter
$(V_{rot}^{max}/\sigma_{v}^{0})^{\star}$ = 0.83 $\pm$ 0.05
and the logarithmic velocity dispersion radial gradient
${\Delta log(\sigma_{v}) \over \Delta log(r)}$ = -0.010 $\pm$ 0.020.
The Lick index Mg$_2$ radial gradients along both photometric axes
were firstly determined by
Couture \& Hardy (1988).
Carollo, Danziger \& Buson (1993) have measured the gradients
of some Lick indices along the E-W direction only.
Central values of several Lick indices of NGC\,1052
were obtained by some works (e.g.
Trager et al. 1998,
Beuing et al. 2002 and
Thomas et al. 2005)
but the majority have only published the central Mg$_2$ (e.g.
Terlevich et al. 1981, who were the first ones).
The stellar population analysis of NGC\,1052 made by 
Raimann et al. (2001), using long slit spectroscopic data over
one direction only, indicates a larger spatial spread in age
than in metallicity so that the old metal-rich populations
dominate the nucleus but the young population of 1 Gyr
becomes important outside of the nuclear region.
Thomas et al. (2005) have derived stellar age, [Z/H], [$\alpha$/Fe]
for an extended sample of early-type galaxy nuclei.
For NGC\,1052, these parameters are:
21.7 Gyr, [Z/H] = +0.222 and [$\alpha$/Fe] = +0.390 dex.
On the other hand, the recent study of
Pierce et al. (2005)
about the stellar ages, metallicities and abundances ratios
of the region inside 0.3 $r_{e}$ of NGC\,1052 and
a sample of 16 globular clusters of it have found that
its nucleus has a luminosity-weighted age $\sim$ 2 Gyr and
[Fe/H] $\sim$ +0.6 dex.
They have not found any strong radial gradients
in either age and metallicity, but there is a strong gradient in
$\alpha$-element abundance whose central value is very high.
Moreover, for all observed globular clusters of NGC\,1052,
they have obtained an age of $\sim$ 13 Gyr.
These results indicate NGC\,1052 is a result of a recent merger
(with starburst!)
which did not induce the formation of a young population of globular cluster.
Apparently, it is a puzzle for the evolutionary history of
this LINER elliptical galaxy.

{\bf NGC\,7796} is an isolated E+ galaxy,
or E1 using the ellipticity at 25 mag.arcmin$^{-2}$.
According to JHK$_{s}$ surface photometry of
Rembold et al. (2002),
this galaxy shows boxy isophotes with constant ellipticity and position angle.
Its surface brightness profile follows the r$^{1/4}$ law
with a decreasing value of $r_{e}$
from 16.3 arcsec (J band) to 10.7 arcsec (K$_{s}$ band).
According to
Ferrari et al. (2002),
NGC\,7796 presents an warm
and cold dust components with uniform spherical distributions 
inside the central region.
A counter rotating core for the stellar component was detected by
Bertin et al. (1994).
In addition, the stellar velocity dispersion along the major photometric axis
changes from 230 km.sec$^{-1}$ at the r = 20 arcsec
up to 270 km.sec$^{-1}$ at the center.
The radial profile of Mg$_2$ was also observed by
Bertin et al. (1994) and
central values of several Lick indices were only obtained by
Beuing et al. (2002) and
Thomas et al. (2005).
The stellar population for the central region was estimated having
an age of 11.8 Gyr, [Z/H] = +0.248 dex and [$\alpha$/Fe] = +0.344 dex
(Thomas et al. 2005).

The goal of this paper is to study in detail the radial distribution of
the stellar populations along the main photometric directions
in two different ellipticals,
a LINER of a group NGC\,1052 and the field galaxy NGC\,7796,
in order to understand the formation/evolution process of these galaxies,
specially the star formation histories.
The paper is organized as follows:
Section 2 presents the observations, data reduction
and kinematical measurements;
Section 3 deals the Lick index measurements;
Section 4 shows the radial gradients of some Lick indices;
Section 5 presents the comparisons with the simple stellar population models;
Section 6 describes the stellar population synthesis approach and
Section 7 appoints the star formation histories of these galaxies.
Finally, in Section 8 a general discussion of the results
and conclusions are drawn.

%
%
\section{Long slit spectroscopic observations, reductions and
kinematical measurements}

Long slit spectroscopic observations
of NGC\,1052 and NGC\,7796 along their major and minor photometric axes
were carried out on 1999 (August 12$^{th}$ and October 11$^{th}$ nights)
at the Cassegrain focus with a Boller \& Chivens spectrograph
of the 1.60m telescope of the {\it Observat\'orio do Pico dos Dias} (OPD)
operated by the {\it Laborat\'orio Nacional de Astrof\'\i sica} (LNA).
The photometric parameters of NGC\,1052 and NGC\,7796 
taken from the RC3 Catalogue and
the extra-galactic data base NED are listed in Table 1;
$H_{0}$=75 km.s$^{-1}$.Mpc$^{-1}$ has been adopted.

The slit width was 2.08 arcsec and its length was 230 arcsec.
The spatial angular scale was 1.092 arcsec.pixel$^{-1}$.
Adopting $h_{0}$ = 0.75,
this corresponds to the linear scale of 111 pc.pixel$^{-1}$ for NGC\,1052
($R$ distance of 1723 km.s$^{-1}$ from
Faber et al. 1989)
and 213 pc.pixel$^{-1}$ for NGC\,7796
($cz_{helio}$ as distance indicator, Tab. 1).
The average seeing was $FWHM_{seeing}$ = 2.0 arcsec.

The spectral range at the wavelength of rest
is $\lambda\lambda$4320-6360\,{\AA}
and the sampling is 2.01\,\AA.pixel$^{-1}$
using a grating of 600 lines.mm$^{-1}$.
The mean instrumental spectral resolution at $\lambda\lambda$4400-6200\,{\AA}
has $\sigma_{inst}$ = 71 km.s$^{-1}$.

The total galaxy spectral exposures were divided
in three equal of 30 minutes each one for the major axis observations
and 2 $\times$ 30 minutes for the minor axis ones
in order to perform the ``cosmic rays'' elimination and
to obtain better signal-to-noise ratio for their aperture spectra.

Spectra of one G and seven K giants of the Lick sample
(Worthey 1994 and
http://astro.wsu.edu/worthey/)
were collected in order to calibrate the absorption line-strengths
to the Lick System.
These stellar spectra were also used to perform
the cross correlations with the galaxy spectra following the
Tonry \& Davis (1979) method.
Spectrophotometric standard stars were observed as well.

%
\renewcommand{\tabcolsep}{2mm}
\begin{table}[htbp]
\footnotesize
\begin{center}
\begin{tabular} {l r r}
\hline 
\hline
Parameter                      & NGC 1052        & NGC 7796      \\
\hline
Classifications                & E4, LINER, Sy2  & E+            \\
$PA$                           & 120$^{\circ}$  & 168$^{\circ}$\\
$\epsilon$                     & 0.3082          & 0.1290        \\
$M_{B}$                        & -20.50          & -20.79        \\
$B_{T}$                        & 11.41           & 12.46         \\
$r_{e}$ (arcsec)               & 33.7            & 21.2          \\
$\mu_{e}$ (mag.arcmin$^{-2}$)  & 12.15           & 12.20         \\
$r_{25}$ (arcsec)              & 90.6            & 65.6          \\
$E(B-V)$                       & 0.027           & 0.01          \\
$(B-V)_{e}$                    & 1.01            & 1.00          \\
$cz_{helio}$ (km.s$^{-1}$)     & 1510$\pm$6      & 3290$\pm$24   \\
Redshift                       & 0.00504         & 0.01097       \\
$<\sigma_{v}^{0}>$ (km.s$^{-1}$)& 208$\pm$33    & 259$\pm$11    \\
$<$Mg$_{2}$$>$ (mag)           & 0.299$\pm$0.015 & 0.237$\pm$0.003 \\
Size$_{25}$ (arcmin$\times$arcmin) & 3.63\,$\times$\,2.51 
& 2.34\,$\times$\,2.04 \\
Environment                    & Group           & Field         \\
\hline
\end{tabular}
\end{center}
\caption{Data obtained in
the RC3 Catalogue,
the NASA/IPAC Extragalactic Database (NED)
which is operated by the Jet Propulsion Laboratory,
California Institute of Technology,
under contract with the National Aeronautics and Space Administration
and the extended version of the
Lyon-Meudon Extragalactic DAtabase (the HyperLEDA) operated by
the {\it Centre de Recherche Astronomique de Lyon}
(for the mean values of the central velocity dispersion and Mg$_{2}$ only).}
\label{data}
\end{table}
%

%
%
\subsection {Data reduction, spectrum extraction and sky subtraction}

The digital images were processed and reduced using
The NOAO Optical Astronomy Packages of IRAF.
Firstly, the images were properly bias subtracted and flat field corrected.
The cosmic ray hits above specific flux ratio thresholds
were also removed in all of them.

The one-dimensional stellar spectra were extracted
using apertures characterized by the FWHM of their profiles.
The sky subtractions of them were made using
two regions distant 55 arcsec from the profile centers
and a second order chebyshev function
to represent the sky level across the slit.
These spectra were wavelength calibrated ($rms$ = 0.2-0.4 {\AA})
using the line identifications in the Helium-Argon frames
which were obtained immediately before or after each exposure.

The aperture spectra along the major axis of NGC\,1052 ($PA$ = 120$^{o}$)
were extracted for the radial distances
of 0.00, 1.10, 3.56, 6.84, 11.80, 20.21
and 37.03 arcsec each side of the center.
They are adjacent each other
and each aperture is always greater than the previous one
in order to improve the spectral quality.
The symmetric spectral extractions along its minor axis ($PA$ = 210$^{o}$)
were for the radial distances of 0.00, 2.18, 4.91, 9.42 and 18.0 arcsec.
The obtained spectral signal-to-noise ratio per Angstrom, $S/N$({\AA}$^{-1}$)
has been from 9 (minor axis) and 36 (major axis)
for the most external spectra
up to 34 (minor axis) and 83 (major axis) for the central ones.
The S/N({\AA}$^{-1}$) was measured in the spectral range
$\lambda\lambda$5800-5850\,{\AA} (wavelength of rest) according to the simple 
expression derived for the photon statistic by
Cardiel et al. (1998).

The aperture spectra along the major axis of NGC\,7796 ($PA$ = 168$^{o}$)
were obtained for the radial distances
of 0.00, 1.10, 3.55, 7.10 and 13.21 arcsec
and along the minor axis
for the distances of 0.00, 1.09, 3.55, 6.93 and 12.5 arcsec.
The spectral (S/N)(\AA$^{-1}$) has been from
17 (minor axis) or 26 (major axis) for the most external spectra
up to 47 (minor axis) or 61 (major axis) for the central ones.

The extractions of the long slit aperture spectra of both galaxies
adopted second order chebyshev functions
to fit the background level across the slit at each wavelength.
For NGC\,1052,
the sky windows were localized from 81 arcsec of the galactic center.
For NGC\,7796,
the sky windows stayed at r $\geq$ 67.5 arcsec for the major axis extractions
and at r$\geq$ 55.7 arcsec (0.91 r$_{25}^{cor}$) for the minor axis ones.
There were some flux residuals
due to the poor subtraction of the telluric lines
in the outer regions of the galaxies.
However, these residuals are not in the spectrum windows
of the Lick indices (except for Fe5406; see details in Sect. 3).

In the sky background windows, the contribution of the galaxy spectrum was
smaller than $\approx$7.6\% relative to the sky level in the OPD/LNA
where $<$V$_{sky}>$ = 21.2 mag.arcsec$^{-2}$ in the dark nights
of August/October on 1999.
For the central aperture of an unique exposure of NGC\,1052 (major axis),
the relative error of the sky subtraction increases from $\sim$1.4\% in the red
up to $\sim$2.2\% in the blue region.
For a respective exposure of NGC\,7796, we have obtained
$\sim$1.9\% in the red and $\sim$3.1\% in the blue.
For the most distant apertures of an exposure for NGC\,1052 (major axis),
this error is $\sim$6.4\% in the red and $\sim$9.5\% in the blue;
and for an analog exposure of NGC\,7796,
it is $\sim$5.7\% in the red and $\sim$8.9\% in the blue.

The relative difference between the sensitivity functions of
distinct flux standard stars was $\leq$ 2.0\%
and $\leq$ 0.8\% in the runs of August and October respectively.
The influence of the flux calibration on the Lick index measurements
is discussed in Section 3.

In summary, all aperture spectra of both galaxies were wavelength calibrated
($rms$ $\leq$ 0.30 {\AA})
using He-Ar frames (obtained immediately before or after of each exposure),
accordingly summed,
flux calibrated,
deredden using the respective line-of-sight Galactic extinction,
put at the rest wavelength using the observed recession velocities
(from the cross correlation measurements)
and flux normalized at $\lambda$5870 {\AA}.
The parameters and kinematical measurements of all long slit
spectroscopic extractions
are given in Tables 2 and 3 for NGC\,1052
and Tables 4 and 5 for NGC\,7796.

%
\renewcommand{\tabcolsep}{1.7mm}	
\begin{table}[htbp]
\caption{Parameters and stellar kinematical measurements
of the long slit spectroscopic extraction regions
along the major axis of NGC\,1052.}
\begin{tabular} {r r r r r r}
\hline
\hline
Radius    & Area  & S/N &  $V_{rot}$  &$\sigma_{v}$& R$_{cc}$ \\
(arcsec)&(arcsec$^{2}$)&({\AA}$^{-1}$)&(km.s$^{-1}$)&(km.s$^{-1}$)&   \\
\hline
0.00      & 4.53  & 83  &    0$\pm$18 & 223$\pm$10 & 18.5 \\
\hline
1.10\,SE  & 4.53  & 78  &  -19$\pm$18 & 221$\pm$09 & 18.5 \\
1.08\,NW  & 4.53  & 78  &  +16$\pm$18 & 225$\pm$09 & 19.3 \\
\hline
3.56\,SE  & 5.68  & 62  &  -36$\pm$16 & 210$\pm$11 & 22.2 \\
3.54\,NW  & 5.68  & 59  &  +44$\pm$16 & 220$\pm$09 & 22.2 \\
\hline 
6.84\,SE  & 7.95  & 48  &  -40$\pm$17 & 211$\pm$14 & 20.5  \\
6.81\,NW  & 7.95  & 45  &  +75$\pm$17 & 216$\pm$14 & 21.7  \\
\hline
11.80\,SE & 12.73 & 42  &  -60$\pm$18 & 213$\pm$17 & 18.7  \\
11.78\,NW & 12.73 & 40  &  +79$\pm$17 & 212$\pm$16 & 20.6  \\
\hline 
20.21\,SE & 22.26 & 36  &  -83$\pm$21 & 212$\pm$18 & 13.8  \\
20.19\,NW & 22.26 & 36  & +108$\pm$18 & 182$\pm$16 & 15.6  \\
\hline 
37.03\,SE & 47.69 & 30  & -105$\pm$30 & 180$\pm$18 &  7.2  \\
37.01\,NW & 47.69 & 30  & +149$\pm$30 & 169$\pm$17 &  7.2  \\
\hline
\end{tabular}
\begin{list} {Table Notes.}
\item Column 1: Distance of the extraction region to the galaxy center.
SE and NW correspond to Southeast and Northwest respectively.
\item Column 6: cross correlation factor R$_{cc}$
of the best stellar template (HR\,8924).
\end{list}
\label{extracoes1}
\end{table}
%

%
\renewcommand{\tabcolsep}{1.7mm}	
\begin{table}[htbp]
\caption{Parameters and stellar kinematical measurements
of the long slit spectroscopic extraction regions
along the minor axis of NGC\,1052.}
\begin{tabular} {r r r r r r}
\hline
\hline
Radius    & Area   & S/N & $V_{rot}$    & $\sigma_{v}$& R$_{cc}$ \\
(arcsec)&(arcsec$^{2}$)&({\AA}$^{-1}$)&(km.s$^{-1}$)&(km.s$^{-1}$)&   \\
\hline
0.00      &  4.53  & 34  &    0$\pm$18  & 236$\pm$24  & 17.1 \\
\hline
2.17\,SW  &  4.53  & 25  &  +11$\pm$18  & 234$\pm$32  & 18.0 \\
2.19\,NE  &  4.53  & 24  &  -21$\pm$20  & 233$\pm$32  & 15.0 \\
\hline
4.90\,SW  &  6.82  & 17  &   +8$\pm$18  & 214$\pm$33  & 17.7 \\
4.92\,NE  &  6.82  & 16  &   +1$\pm$19  & 220$\pm$37  & 15.6 \\
\hline 
9.41\,SW  & 11.92  & 12  &   -5$\pm$21  & 233$\pm$47  & 13.7 \\
9.43\,NE  & 11.92  & 12  &   +5$\pm$19  & 177$\pm$25  & 14.0 \\
\hline
18.00\,SW & 23.80  &  9  &   -6$\pm$29  & 184$\pm$28  &  7.5 \\
18.01\,NE & 23.80  &  9  &  -10$\pm$29  & 144$\pm$18  &  6.6 \\
\hline
\end{tabular}
\begin{list} {}
\item Column 1: SW and NE correspond to Southwest and Northeast respectively.
\item Column 6: as in Table 2.
\end{list}
\label{extracoes2}
\end{table}
%

%
\renewcommand{\tabcolsep}{1.7mm}	
\begin{table}[htbp]
\caption{Parameters and stellar kinematical measurements
of the long slit spectroscopic extraction regions
along the major axis of NGC\,7796.}
\begin{tabular} {r r r r r r}
\hline
\hline
Radius   & Area  & S/N & $V_{rot}$   &$\sigma_{v}$& R$_{cc}$ \\
(arcsec)&(arcsec$^{2}$)&({\AA}$^{-1}$)&(km.s$^{-1}$)&(km.s$^{-1}$)&   \\
\hline
0.00     & 4.53  & 61  &    0$\pm$14 & 261$\pm$10 & 26.6  \\
\hline
1.09\,S  & 4.53  & 56  &   -3$\pm$13 & 257$\pm$11 & 28.1  \\
1.09\,N  & 4.53  & 56  &   +8$\pm$15 & 274$\pm$14 & 24.0  \\
\hline
3.55\,S  & 5.68  & 41  &   +4$\pm$16 & 255$\pm$20 & 21.9  \\
3.55\,N  & 5.68  & 40  &   +1$\pm$14 & 271$\pm$20 & 25.4  \\
\hline 
7.10\,S  & 9.09  & 32  &  +20$\pm$17 & 237$\pm$26 & 17.2  \\
7.10\,N  & 9.09  & 31  &   -8$\pm$18 & 260$\pm$30 & 16.4  \\
\hline
13.21\,S & 16.35 & 26  &  +50$\pm$24 & 240$\pm$33 & 10.6  \\
13.21\,N & 16.35 & 26  &  +25$\pm$24 & 243$\pm$34 & 10.9  \\
\hline
\end{tabular}
\begin{list} {Table Notes.}
\item Column 1: S and N correspond to South and North respectively.
\item Column 6: as in Table 2.
\end{list}
\label{extracoes3}
\end{table}
%

%
\renewcommand{\tabcolsep}{1.7mm}	
\begin{table}[htbp]
\caption{Parameters and stellar kinematical measurements
of the long slit spectroscopic extraction regions
along the minor axis for NGC\,7796.}
\begin{tabular} {r r r r r r}
\hline
\hline
Radius   & Area  & S/N & $V_{rot}$   &$\sigma_{v}$& R$_{cc}$ \\
(arcsec)&(arcsec$^{2}$)&({\AA}$^{-1}$)&(km.s$^{-1}$)&(km.s$^{-1}$)&   \\
\hline
0.00     & 4.53  & 47  &    0$\pm$14 & 261$\pm$16 & 24.7 \\
\hline
1.09\,E  & 4.53  & 44  &   -2$\pm$15 & 267$\pm$18 & 23.9 \\
1.09\,W  & 4.53  & 44  &   -1$\pm$15 & 260$\pm$18 & 24.4 \\
\hline
3.55\,E  & 5.68  & 31  &  +12$\pm$16 & 265$\pm$26 & 19.5 \\
3.55\,W  & 5.68  & 32  &   -8$\pm$16 & 253$\pm$28 & 20.3 \\
\hline 
6.93\,E  & 8.40  & 22  &  +18$\pm$21 & 253$\pm$41 & 13.1 \\
6.93\,W  & 8.40  & 22  &  -35$\pm$21 & 224$\pm$32 & 12.3 \\
\hline
12.50\,E & 14.77 & 17  &  +28$\pm$34 & 264$\pm$40 &  7.3 \\
12.50\,W & 14.77 & 17  &  -23$\pm$23 & 215$\pm$33 & 10.6 \\
\hline
\hline
\end{tabular}
\begin{list} {}
\item Column 1: E and W correspond to East and West respectively.
\item Column 6: as in Table 2.
\end{list}
\label{extracoes4}
\end{table}
%

%
%
\subsection{Stellar kinematical measurements with external comparisons}

The line-of-sight rotation curve and
the line-of-sight velocity dispersion radial profile
of the stellar component
along both photometric axes were measured for both galaxies
using our kinematical data:
$cz_{helio}$ and $\sigma_{v}$ of the aperture spectra.
We have also calculated the rotational parameter
(V$_{rot}^{max}$/$\sigma_{v}^{0}$)$^{\star}$,
the kinematical misalignment $\Psi_{k}$
and the logarithmic radial gradient of the velocity dispersion
${\Delta \log \sigma_{v} \over \Delta \log r}$
for both galaxies.
Our data have been compared with the literature data.

We have been adopted for this purpose the cross correlation method
(Tonry \& Davies 1979)
using the RVSAO
(Radial Velocity Package of the Smithsonian Astrophysical Observatory)
on IRAF enviroment
(Kurtz \& Mink 1998).
The uncertainty of the geocentric radial velocity of each aperture spectrum
is computed by $\frac{3}{8}\frac{FWHM_{peak}}{(1+r)}$, where
FWHM$_{peak}$ is the FWHM of the peak of the cross correlation function, 
$r$ is the ratio between the height of this peak and
the amplitude of a sinusoidal noise for the function
(Kurtz \& Mink 1998).
The heliocentric radial velocities of the stellar templates (K giants)
came from the
Duflot et al. (1995) Catalogue.

We had to apply an empirical calibration
in order to transform the FWHM$_{peak}$
into the galaxy velocity dispersion as made for example by
de la Rosa, de Carvalho \& Zepf (2001).
This takes into account the instrumental resolution
and closely obeys the relation
FWHM$_{peak}$ =
2 $\sqrt{\ln 4}\times\sqrt{\sigma_{v}^2 + 2 \times \sigma_{inst}}$,
where $\sigma_{inst}$ corresponds to 
the mean $\sigma_{peak}$ of the stellar auto-correlations
divided by $\sqrt{2}$.
A parabolic function fitting was adopted
to fit the half of the height of the correlation peak.
A third order polynomial function FWHM$_{peak}$($\sigma_v$)
was obtained adopting correlations
between artificially broadened spectra of six observed stars
(HR\,6136, HR\,6159, HR\,6299, HR\,8841 and HR\,8924)
and the non-broadened stellar ones.
The mean instrumental resolution
$\sigma_{inst}$ = 71 $\pm$ 10 km.s$^{-1}$
($\lambda\lambda$4400-6200\,{\AA})
was estimated by the template-template correlations.
Gaussian convolutions were made to represent $\sigma_{v}$ 
of 50 up to 400 km.s$^{-1}$ with steps of 50 km.s$^{-1}$.

The velocity dispersion errors were estimated
as a function of the signal-to-noise ratio and $\sigma_{v}$ itself.
Six levels of artificial Poisson noises
were applied to all artificially broadened spectra of HR\,8924
in order to represent the S/N variation of 14 up to \,85\,\AA$^{-1}$.
For each noisy broadened spectrum,
the velocity dispersion was measured and compared with the actual value.
After that, 
a second order polynomial function of the S/N(\AA$^{-1}$)
was constructed for each one of the $\sigma_{v}$ values
(of 50 up to 400 km.s$^{-1}$) 
to estimate the relative error of the velocity dispersion.
For example, the typical error of $\sigma_{v}$ = 200 km.s$^{-1}$
varies from 4.\% up to 16.\%
when the S/N(\AA$^{-1}$) changes from 85 to 14 \AA$^{-1}$.

The stellar kinematical results of all observed galaxy regions
are shown in Tables 2 and 3 (NGC\,1052) and Tabs. 4 and 5 (NGC\,7796),
including the R$_{cc}$ factor of the respective best cross correlation.
Our measurements of $cz_{helio}$ of the central aperture
are in agreement with the catalog values, i.e. 
the relative differences are smaller than
9\% for NGC\,1052 and 3.5\% for NGC\,7796.

Figures 1a and 1b show respectively the rotation curve
and the velocity dispersion radial profile of NGC\,1052
along its major axis.
Figures 2a and 2b show the same its the minor axis.
Our results are also plotted together with those of 
Binney et al. (1990) for both photometric directions and
Fried \& Illingworth (1994) for the major axis only.
Figures 3a and 3b show respectively the rotation curve
and the velocity dispersion radial profile of NGC\,7796
along its major photometric axis.
Figures 4a and 4b show the same for its minor axis.
Our data are plotted together with those of
Bertin et al. (1994)  
whose observations were made along the major axis only (Figs. 3a-b).

%
\begin{figure}[htbp]
\resizebox{\hsize}{!}{\includegraphics{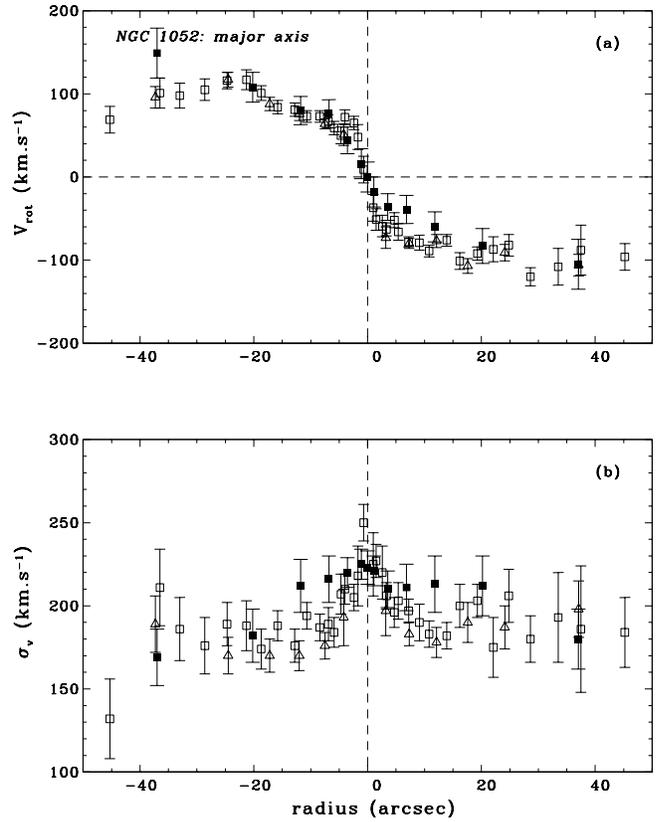}}
\caption{The stellar kinematical measurements
along the major axis of NGC\,1052.
{\bf a)} The line-of-sight velocity rotation curve;
{\bf b)} the line-of-sight velocity dispersion radial profile. 
Our data are plotted using solid squares.
The comparisons with the data of
Binney et al. (1990), open squares, and
Fried \& Illingworth (1994), open triangles, are presented as well.}
\label{sgvrot1}
\end{figure}  
%

%
\begin{figure}[htbp]
\resizebox{\hsize}{!}{\includegraphics{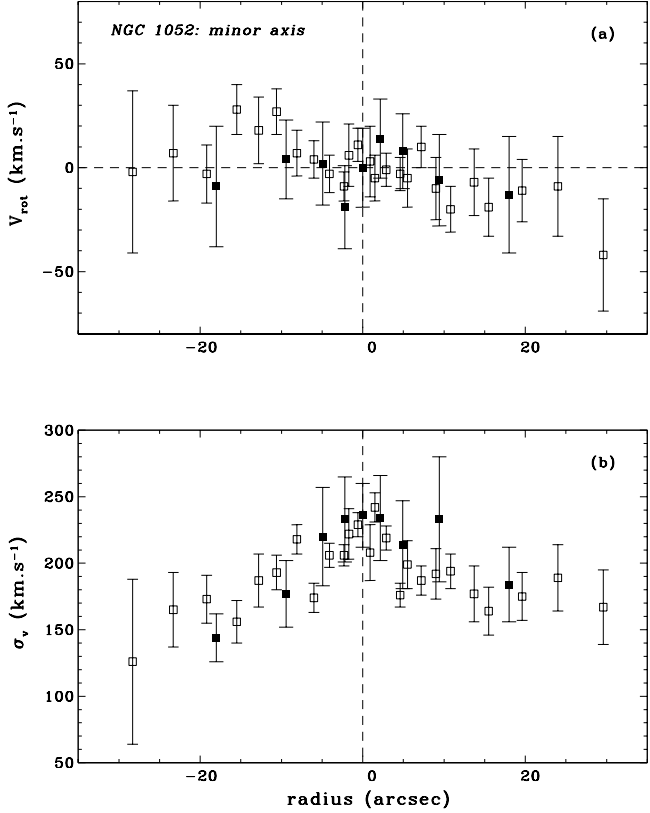}}
\caption{The stellar kinematical measurements
along the minor axis of NGC\,1052.
{\bf a)} The line-of-sight velocity rotation curve;
{\bf b)} the line-of-sight velocity dispersion radial profile. 
Our data are plotted using solid squares.
The data of
Binney et al. (1990), open squares, is presented as well.}
\label{sgvrot2}
\end{figure}  

In order to compare our results about velocity dispersion with the literature,
the observed $\sigma_{v}^{obs}$ of the central spectra
must be corrected of the aperture effect 
considering a typical value for the logarithmic radial gradient
of $\sigma_{v}$
(Jorgensen, Franx \& Kjaergaard 1995).

Taking into account the observations over both axes
and adopting $\frac{\Delta\log \sigma_{v}}{\Delta \log r}$ = -0.040 from
Jorgensen, Franx \& Kjaergaard (1995),
the mean values of $\sigma_{v}^{0}$ of 
NGC\,1052 and NGC\,7796 are
213$\pm$9 km.s$^{-1}$
and 250$\pm$8 km.s$^{-1}$ respectively.
They are in excellent agreement to the published ones.
For NGC\,1052, LEDA presents $\sigma_{v}^{0}$ = 208$\pm$33 km.s$^{-1}$
from 24 measurements including
Binney et al. (1990) and
Fried \& Illingworth (1994)
who measured 240$\pm$10 km.s$^{-1}$ and 195$\pm$15 km.s$^{-1}$
respectively.
For NGC\,7796, LEDA shows $\sigma_{v}^{0}$ = 259$\pm$11 km.s$^{-1}$
from two studies only;
one of them is Bertin et al. (1994) who measured 265$\pm$9 km.s$^{-1}$.
The individual $\sigma_{v}^{0}$ for NGC\,1052 are:
207$\pm$10 km.s$^{-1}$ (major axis)
and 220$\pm$24 km.s$^{-1}$ (minor axis).
For NGC\,7796 they are:
250$\pm$10 km.s$^{-1}$ (major axis)
and 251$\pm$16 km.s$^{-1}$ (minor axis).

For {\bf NGC\,1052}, the rotation curve along the major axis
is similar to those acquired by
Binney et al. (1990) and
Fried \& Illingworth (1994); see Fig. 1a.
The rotation curve over the minor axis is in agreement to that of
Binney et al. (1990); see Fig. 2a.
Along the major axis, despite the present worse spatial resolution,
our result shows nuclear stellar rotation like a rigid body
(up to r = 4-7 arcsec with a projected angular velocity
of 10 km.s$^{-1}$.arcsec$^{-1}$),
a discontinuity region at r = 7-12 arcsec like a plateau,
and a similar rigid body rotation outwards after this rotation discontinuity
with an smaller projected angular velocity
($\sim$ 2 km.s$^{-1}$.arcsec$^{-1}$).
However, our result points out an increasing of V$_{rot}$
outside of the observed region,
differently from
Binney et al. (1990) and
Fried \& Illingworth (1994)
data that have both shown a constant value.
Along the minor axis,
no residual rotation is detected for the stellar component. 
The present line-of-sight velocity dispersion radial profile
along the major axis is similar
to those ones of
Binney et al. (1990) and
Fried \& Illingworth (1994);
see Fig. 1b.
The $\sigma_v$ profile along the minor axis
is in agreement to that of
Binney et al. (1990): Fig. 2b.
For the major axis profile,
the main difference is that the data of
Binney et al. (1990) and
Fried \& Illingworth (1994)
have shown more negative $\sigma_v$ gradient (up to r = 10 arcsec)
than our data.
The calculated rotational parameter
(V$_{rot}^{max}$/$\sigma_{v}^{0}$)$^{\star}$
assumes the nuclear velocity dispersion $\sigma_{v}^{0}$ 
corrected by the aperture effect
using the own logarithmic $\sigma_v$ radial gradient measured
along the major axis direction in the present work
(${\Delta \log \sigma_{v} \over \Delta \log r}$ = -0.046 $\pm$ 0.015).
Our maximum line-of-sight rotational velocity V$_{rot}^{max}$
is the average of the ones measured
in each side of the V$_{rot}$ radial profile
which were calculated using a third polynomial fitting for the velocity curve
that takes into account the errors in V$_{rot}$ only.
NGC\,1052 is a spheroidal system flattened by the rotational residual motion 
around the minor axis.
Its rotational parameter, considering all observed region
($\leq$ 3.8 kpc $\sim$ 0.9 r$_{e}$), 
is (V$_{rot}^{max}$/$\sigma_{v}^{0}$)$^{\star}$ = 0.89 ($\pm$ 0.14)
which is in close agreement to the one found by
Fried \& Illingworth (1994).
Its maximum rotational velocity
is 122 $\pm$ 13 km.s$^{-1}$ in the observed region,
but it still appears to increase outwards.
Its nuclear velocity dispersion is 204 $\pm$ 10 km.s$^{-1}$
(Coma normalized using own $\sigma_{v}$ radial gradient).
Its line-of-sight velocity dispersion radial profile inside 0.9 r$_{e}$
is nearly symmetric.
We could classify NGC\,1052 as an oblate rotator,
but it is better to conclude that
NGC\,1052 has a fast rotating stellar component
because its rotational parameter does not reach unity.
The estimated kinematic misalignment $\Psi_{est}$ is about 3$^{o}$,
if the maximum projected rotational velocity along the minor axis is not zero.
This indicates that the stellar disc of NGC\,1052
has an inclination angle very near to 90$^{o}$ or, in other words,
the photometric axes of this elliptical are direct projections
of the kinematical axes.
This corroborates the result of
Binney et al. (1990).

For {\bf NGC\,7796},
the rotation curve and $\sigma_{v}$ radial profile,
both along the major axis,
are in agreement with those ones of
Bertin et al. (1994)
up to r = 14 arcsec; see Figures 3a and 3b.
NGC\,7796 has no significant stellar rotation
around both axes.
Its rotational parameter
(V$_{rot}^{max}$/$\sigma_{v}^{0}$)$^{\star}$ is 0.15 ($\pm$ 0.04).
The velocity dispersion profiles
inside the observed region ($\leq$ 4.3 kpc $\sim$ 0.6 r$_{e}$)
are nearly uniform; see Figures 3b and 4b.
The logarithmic $\sigma_v$ radial gradient measured
over the major axis in the present work is
${\Delta \log \sigma_{v} \over \Delta \log r}$ = -0.028 $\pm$ 0.030.
The nuclear velocity dispersion is 253 $\pm$ 10 km.s$^{-1}$
(Coma normalized using this $\sigma_v$ gradient).
Its $\Psi_{est}$ is very high, $\sim$ 62$^{o}$.
Bertin et al. (1994),
whose long slit observations have greater spatial resolution
(1.78 arcsec.pixel$^{-1}$),
have concluded that NGC\,7769 has a counter rotating core inside r = 4 arcsec
that our observations marginally corroborate.
But, if the major axis rotation curve is considered
up to r = 8 arcsec we would detect a kinematically
decoupled inner component (r $\leq$ 3.5 arcsec) that has no rotation
while ``the main body'' has a very small rotation (14 km.s$^{-1}$).
The minor axis rotation curve (Fig. 4a),
considered only up to r = 7 arcsec,
does not show any rotation inside r = 1.5 arcsec
and it presents a tiny rotation (26 km.s$^{-1}$) greater
than another one about the major axis.
Therefore, the core of NGC\,7796 shows complex kinematics.

%
\begin{figure}[htbp]
\resizebox{\hsize}{!}{\includegraphics{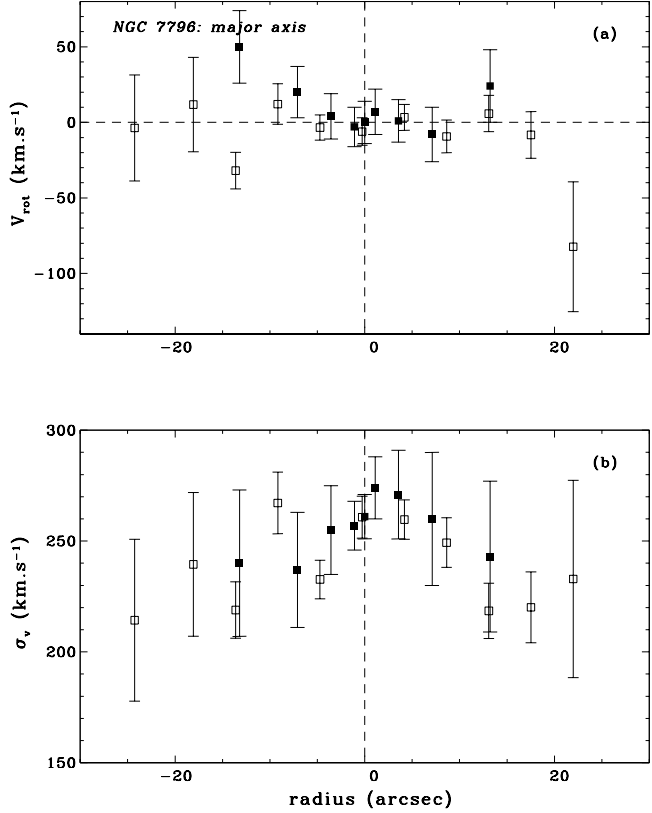}}
\caption{The stellar kinematical measurements
along the major axis of NGC\,7796.
{\bf a)} The line-of-sight velocity rotation curve;
{\bf b)} the line-of-sight velocity dispersion radial profile.
Our data are plotted using solid squares.
The data of
Bertin et al. (1994),
open squares, is presented as well.}
\label{sgvrot3}
\end{figure}  
%

%
\begin{figure}[htbp]
\resizebox{\hsize}{!}{\includegraphics{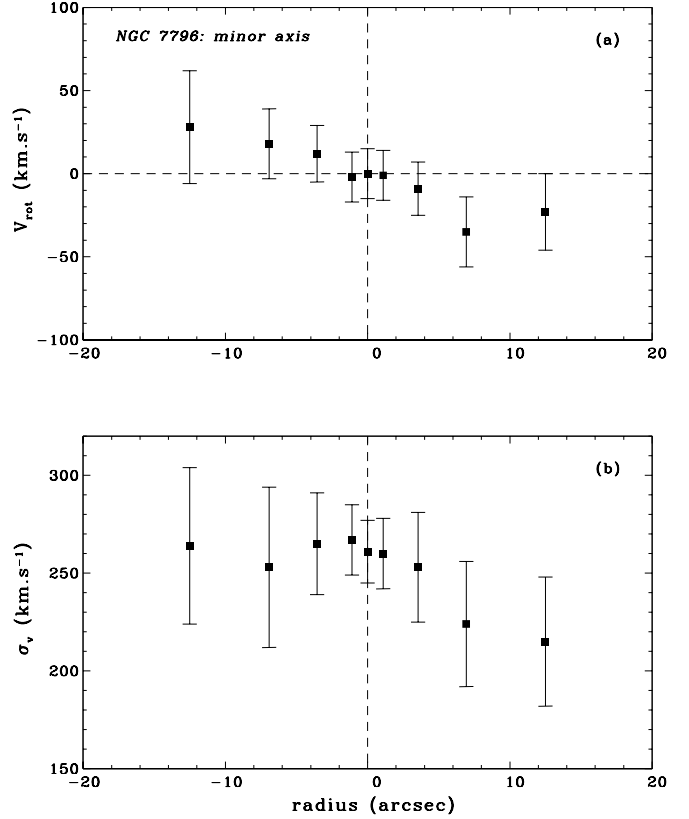}}
\caption{The stellar kinematical measurements
along the minor axis of NGC\,7796.
{\bf a)} The line-of-sight velocity rotation curve;
{\bf b)} the line-of-sight velocity dispersion radial profile.}
\label{sgvrot4}
\end{figure}  
%

%
%
\section {Measurements of the Lick indices} 

The Lick indices are measurements of optical absorption lines
of spectra of stars and stellar composite systems like globular clusters and galaxies.
They have been usually employed to quantity
the luminosity-weighted mean ages, metallicities and abundance ratios
(as $\alpha$-elements/Fe) of the composite stellar populations 
with ages $\geq$ 2 Gyr.
The Lick indices of atomic lines are measured as equivalent widths $EW$
({\AA} unity) considering a linear local pseudo-continuum,
which is defined by two wavelength windows
side by side of the central feature bandpass [$\lambda_1$, $\lambda_2$].
The Lick indices of molecular lines quantify
the absorbed flux in the index bandpass relative to that pseudo-continuum
and it is expressed in the magnitude scale (here denoted by $MAG$).
These line-strengths were originally measured in flux non-calibrated spectra
with low resolution using one dimensional image detector
(Image Digital Scanner, IDS, of the Lick Observatory).
The spectral resolution of the Lick System
is actually variable in the range $\lambda\lambda$4000-6000 {\AA}:
FWHM$_{Lick}$ = 8.4 - 11.5 {\AA}
with greater values in the red and blue edges
(Worthey \& Ottaviani 1997).
The Lick indices are defined to be
independent of the line-of-sight velocity dispersion of the stellar system.
Therefore, the spectral broadening effect due to the velocity dispersion
must be subtracted in quadrature.

The EW indices can be transformed into the MAG index scale and vice-verse:
$MAG = -2.5 \log [1 - \frac{EW}{(\lambda_2 - \lambda_1)}]$ and
$EW = (\lambda_2 - \lambda_1) (1 - 10^{-0.4\,MAG})$

%
%
\subsection {Lick calibration and index errors}

The current instrumental resolution $\sigma_{inst}$ is nearly 71 km.s$^{-1}$
considering the interval $\lambda\lambda$4400-6200 {\AA}
so that $FWHM$ $\sim$ 3 {\AA} at $\lambda$5300 {\AA}.
The respective one of the Lick System
is $\sigma_{Lick}$ = 220 km.s$^{-1}$,
or FWHM$_{Lick}$ $\approx$ 9.2\,{\AA} at approximately the same range
(Worthey \& Ottaviani 1997).
Note that $\sigma(km.s^{-1}) = \frac{c.FWHM}{2\sqrt{\ln 4}\lambda}$.
Therefore, we had to put all galaxy and IDS sample star spectra nearly to the
Lick spectral resolution in order
to measure suitably the absorption line indices.
To make this, all spectra were broadened using adequate Gaussians.
The appropriate spectral broadening $\sigma_{broad}$ was obtained
from the result of the cross correlation of the actual Lick/IDS spectra
(http://astro.wsu.edu/worthey/)
with the observed stellar sample,
subtracting the instrumental spectral resolution in quadrature,
i.e. $\sigma_{Lick}^2$ = $\sigma_{inst}^2$ + $\sigma_{broad}^2$.

Firstly, all galaxy and star spectra were accurately reduced to the wavelength of rest 
and broadened to the Lick resolution accordingly.
The indices of Fe4383 to Na D, including Mg$_1$ and Mg$_2$, were measured.
However, two steps are still necessary
in order to transform the index measurements to the Lick System:
(i) a linear correction, and
(ii) a correction for the line-of-sight velocity dispersion $\sigma_{v}$
(Worthey \& Ottaviani 1997).
Both steps must be carefully done and they must be also used
to estimate the final errors of the indices Lick.
It is still necessary to compare the indices to the published ones,
specifically those for the galaxy nuclei.
We have adopted eight Lick standard stars (see Sect. 2).

The first step is made through linear correlations between
the actual published stellar Lick indices and
the measured ones in our adequately broadened stellar spectra,
like $INDEX_{Lick}$ = A + B$\times INDEX_{our}$ (see Table 6).
The angular coefficient stayed in the range 0.63 $\leq$ B $\leq$ 1.27,
and fitting $rms$ was smaller than 0.37 {\AA} for the $EW$ indices
excluding the Fe5015 index ($rms$ = 0.63 {\AA})
and $\leq$ 0.0062 mag for the $MAG$ ones.
The range of the index values measured 
in the broadened spectra of the eight Lick standards
were similar to those ones observed in the galaxies, i.e.
Fe4383 in [4.50, 8.50 {\AA}],
Ca4455 in [0.50, 2.00 {\AA}],
Fe4531 in [3.00, 4.50 {\AA}],
Fe4668 in [5.00, 9.00 {\AA}],
H$\beta$ in [0.50, 2.10 {\AA}],
Fe5015 in [5.00, 7.00 {\AA}],
Mg b in [2.00, 5.00 {\AA}],
Fe5270 in [2.80, 4.20 {\AA}],
Fe5335 in [2.20, 4.20 {\AA}],
Fe5406 in [1.40, 3.00 {\AA}],
Fe5709 in [0.90, 1.60 {\AA}],
Fe5782 in [0.60, 1.50 {\AA}],
Na D in [1.80, 5.40 {\AA}],
Mg$_1$ in [0.010, 0.230 mag], and
Mg$_2$ in [0.100, 0.400 mag].

The second step is based on the galaxy velocity dispersion 
that broadens the absorption lines of each spectrum.
Except for the H$\beta$ Lick index,
the general consequence is a decreasing of the index value.
A simple way is to compute some artificial broadenings
of our Lick observed stellar spectra to different $\sigma_{broad}$
representative of the line-of-sight galaxy $\sigma_{v}$.
Again, we have assumed spectral broadening by Gaussian convolutions
for $\sigma_{v}$ going from 50 to 400 km.s$^{-1}$ (with 50 km.s$^{-1}$ steps).
The $F(\sigma_{v})$ correction factor is the ratio of the measured index in
the stellar observed spectrum to the measured index
in the spectrum of a given representative $\sigma_{v}$
denoted by $\frac{EW(index)_0}{EW(index)_{\sigma_{v}}}$
where $EW(index)_0$ is the equivalent width of the index
for an unbroadened spectrum (i.e. $\sigma_{v}$ = 0 km.s$^{-1}$).
We have computed a second order polynomial fit 
in order to determine the mean correction factor
as a function of the actual velocity dispersion for each Lick index,
$F(\sigma_{v})$ = a + b$\times\sigma_{v}$ + c$\times\sigma_{v}^{2}$);
see Tab. 6.
The respective fitting $rms$ was always smaller than 0.08 {\AA},
except for the Ca4455 index (0.14 {\AA}).

%
\renewcommand{\tabcolsep}{5mm}
\begin{table*}[htbp]
\caption{Coefficients of the calibration to the Lick System for:
the linear correction, $INDEX_{Lick}$ = A + B$\times INDEX_{our}$,
and the velocity dispersion correction,
$F(\sigma_{v})$ = a + b$\times\sigma_{v}$ + c$\times\sigma_{v}^{2}$,
with their respective errors.}
\begin{tabular} {l|r r r r r r r}
\hline
\hline
\multicolumn{8}{c}{{\it Fitting \,coefficients \,and \,errors}} \\
\hline
Index   & A      & B       & $rms$   & a & b & c & $rms$ \\
\hline
$EW$    & ({\AA})&         & ({\AA}) 
&       & (10$^{-4}$km$^{-1}$.s) & (10$^{-6}$km$^{-2}$.s$^2$) &   \\
Fe4383  & 0.6920 & 0.94864 & 0.2720  & 0.9990 & 0.0699 & 2.2919 & 0.0149 \\
Ca4455  & 1.3488 & 0.62615 & 0.2025  & 1.0085 &-5.0818 & 8.5130 & 0.1418 \\
Fe4531  &-0.3046 & 1.22950 & 0.3710  & 1.0083 &-4.6272 & 1.6660 & 0.0574 \\
Fe4668  & 0.1884 & 1.10060 & 0.3419  & 1.0005 &-0.6561 & 1.3835 & 0.0159 \\
H$\beta$&-0.3155 & 1.12340 & 0.2026  & 1.0045 &-5.3309 & 1.7006 & 0.0754 \\
Fe5015  &-0.5575 & 1.18350 & 0.6288  & 0.9913 & 3.5982 & 1.7919 & 0.0286 \\
Mg b    &-0.2062 & 1.10380 & 0.2202  & 1.0019 &-1.3020 & 2.6367 & 0.0261 \\
Fe5270  &-0.7634 & 1.27300 & 0.1708  & 0.9959 & 1.9521 & 1.7809 & 0.0215 \\
Fe5335  &-0.3890 & 1.19720 & 0.1422  & 1.0005 &-1.5167 & 6.1129 & 0.0186 \\
Fe5406  &-0.2449 & 1.22160 & 0.0725  & 1.0030 &-1.8058 & 5.2013 & 0.0298 \\
Fe5709  & 0.0660 & 1.16050 & 0.2494  & 1.0041 &-2.5330 & 3.0240 & 0.0370 \\
Fe5782  & 0.0465 & 1.23470 & 0.2014  & 0.9933 & 3.0617 & 4.7034 & 0.0420 \\
Na D    & 0.0183 & 0.97408 & 0.0962  & 1.0012 &-0.8740 & 1.3064 & 0.0116 \\
$MAG$   & ($mag$)&        & ($mag$)  &        &        &        &        \\
Mg$_1$  & 0.0305 & 0.98239 & 0.0057  & 0.9989 & 0.6306 & 0.3009 & 0.0258 \\
Mg$_2$  & 0.0153 & 1.09010 & 0.0062  & 0.9995 & 0.1950 & 0.1189 & 0.0074 \\
\hline
\end{tabular}
\label{lick1}
\end{table*}

The final errors of the Lick indices for each aperture spectrum
were computed considering a Poisson error of the respective measurement,
the error of the computed heliocentric velocity 
and the error propagations due to both steps of the Lick transformation. 
The Poisson uncertainty of each index,
$\delta(EW)$ or $\delta(MAG)$, was estimated
as a function of the S/N({\AA}$^{-1}$) of each extracted spectrum
following the simple expressions of
Cardiel et al. (1998).

In Table 7, the errors of the Lick indices and their respective propagations
due to the Lick calibration are shown, as example, for two extreme
spectral S/N({\AA}$^{-1}$) ratios (83 from the central spectrum of NGC\,1052
and 26 from one most external of NGC\,7796).
For higher S/N, the final errors of the Lick indices
are basically determinated by the Lick transformation
and, for lower S/N, the intrinsic errors of the measurements
become important to determine the final errors,
except for Ca4455 and Fe5335.

%
\renewcommand{\tabcolsep}{2mm}
\begin{table}[htbp]
\caption{The errors of the Lick indices for
two spectra with different S/N ratios:
the Poisson error plus the cz$_{helio}$ uncertainty
and the Lick System calibration errors.}
\begin{tabular} {l|r r r}
\hline
\hline
\multicolumn{4}{c}{{\it Lick \,index \,errors}} \\
\hline
Index   & Poisson+cz      & linear corr.  & $\sigma_v$ corr.  \\
\hline
S/N=83 {\AA}$^{-1}$  &   &               &         \\
$EW$    & ({\AA})         & ({\AA})       & ({\AA})  \\
Fe4383  & 0.128           & 0.170         & 0.173   \\
Ca4455  & 0.065           & 0.088         & 0.385   \\
Fe4531  & 0.091           & 0.292         & 0.165   \\
Fe4668  & 0.137           & 0.258         & 0.178   \\
H$\beta$& 0.060           & 0.151         & 0.187   \\
Fe5015  & 0.114           & 0.466         & 0.047   \\
Mg b    & 0.049           & 0.159         & 0.208   \\
Fe5270  & 0.054           & 0.145         & 0.088   \\
Fe5335  & 0.065           & 0.122         & 0.173   \\
Fe5406  & 0.048           & 0.070         & 0.125   \\
Fe5709  & 0.037           & 0.180         & 0.067   \\
Fe5782  & 0.036           & 0.153         & 0.115   \\
Na D    & 0.038           & 0.063         & 0.095   \\
$MAG$   & (mag)           & (mag)         & (mag)    \\
Mg$_1$  & 0.0011          & 0.0038        & 0.0047  \\
Mg$_2$  & 0.0013          & 0.0044        & 0.0024  \\
\hline
S/N=26 {\AA}$^{-1}$  &   &               &         \\
$EW$    & ({\AA})         & ({\AA})       & ({\AA})  \\
Fe4383  & 0.424           & 0.140         & 0.340   \\
Ca4455  & 0.196           & 0.010         & 0.728   \\
Fe4531  & 0.266           & 0.372         & 0.302   \\
Fe4668  & 0.409           & 0.313         & 0.314   \\
H$\beta$& 0.155           & 0.175         & 0.068   \\
Fe5015  & 0.306           & 0.537         & 0.512   \\
Mg b    & 0.157           & 0.181         & 0.320   \\
Fe5270  & 0.158           & 0.201         & 0.230   \\
Fe5335  & 0.194           & 0.172         & 0.535   \\
Fe5406  & 0.147           & 0.114         & 0.245   \\
Fe5709  & 0.120           & 0.207         & 0.092   \\
Fe5782  & 0.112           & 0.188         & 0.219   \\
Na D    & 0.134           & 0.058         & 0.132   \\
$MAG$   & (mag)           & (mag)         & (mag)    \\
Mg$_1$  & 0.0034          & 0.0037        & 0.0035  \\
Mg$_2$  & 0.0039          & 0.0049        & 0.0025  \\
\hline
\end{tabular}
\label{lick2}
\end{table}
%

%
%
\subsection {Influence of the sky subtraction and flux calibration}

The subtraction of the sky background was satisfactory
for all galaxy and stellar spectra.
However, the poor subtraction of the telluric emission line
HgI$\lambda$5460.74 {\AA} has only affected the Fe5406 index
in the outer regions of the galaxies.
The residual of this line subtraction stays in the Fe5406 red continuum in
the aperture spectra of NGC\,1052 for r $\geq$ 11.78 arcsec (major axis)
or for all apertures (minor axis).
The regions of all sky line residuals have been changed
in the minor axis spectra of NGC\,1052 by linear interpolations. 
For NGC\,7796, the red continuum and bandpass of Fe5406
are affected by that sky line beyond the radial distance of
7.10 arcsec (major axis) or 6.93 arcsec (minor axis).
The other sky line residuals do not affect the Lick index measurements
because they do not occur in the index bandpasses
as well as their continuum windows.
 
All Lick indices were also measured in the flux non-calibrated spectra ($fnc$)
of the galaxies according to the original procedure of the Lick/IDS System.
We have computed the index differences,
$\Delta INDEX$ = $INDEX_{fc}$ - $INDEX_{fnc}$,
where $fc$ denotes flux calibrated spectra.
There is no systematic disagreement for all indices
because these differences are comparable or smaller than their errors:
$\overline{\Delta EW}$ $\approx$ +0.02 {\AA} and \linebreak
$\overline{\Delta MAG}$ $\approx$ -0.009 mag.
As an example, for the central major axis spectrum of NGC\,1052
whose S/N ratio is 83 {\AA}$^{-1}$, these differences are:
$\Delta EW$ ({\AA}) =
+0.066 for Fe4383,
+0.010 for Ca4455,
+0.024 for Fe4531,
+0.015 for Fe4668,
-0.009 for H$\beta$,
-0.184 for Fe5015,
+0.006 for Mg b,
+0.003 for Fe5270,
-0.005 for Fe5335,
-0.003 for Fe5406,
-0.003 for Fe5709,
-0.005 for Fe5782 and 
-0.010 for Na D;
and $\Delta MAG$ (mag) =
-0.0103 for Mg$_1$, and
-0.0102 for Mg$_2$.
For the major axis spectrum of NGC\,1052 at r = 20.21 arcsec,
whose S/N ratio is 36 {\AA}$^{-1}$, they are:
$\Delta EW$ ({\AA}) =
+0.102 for Fe4383,
+0.015 for Ca4455,
-0.003 for Fe4531,
-0.041 for Fe4668,
-0.007 for H$\beta$,
-0.063 for Fe5015,
-0.004 for Mg b,
+0.003 for Fe5270,
-0.008 for Fe5335,
-0.003 for Fe5406,
-0.003 for Fe5709,
-0.002 for Fe5782 and 
-0.010 for Na D;
and $\Delta MAG$ (mag) =
-0.0118 for Mg$_1$, and
-0.0115 for Mg$_2$.
Therefore, if there is some influence of the flux normalization precision
over the Lick index measurements, this is negligible.

%
%
\subsection {Emission line corrections of some indices for NGC\,1052}

For the aperture spectra of NGC\,1052
up to the radial distance r = 3.56 arcsec (major axis)
and up to r = 4.92 arcsec (minor axis),
some Lick indices are affected by nebular emission lines of this LINER.
The H$\beta$ index, of course, is directly contaminated
by its respective gas emission.
The Mg$_1$, Mg$_2$ and Fe5015 features are severely affected by
the [O III]$\lambda$4958{\AA} emission line
present in their blue continuum windows.
The bandpass of Mg$_2$ is still contaminated 
by the [N I]$\lambda$5198-5200{\AA} emission doublet,
hereafter as [N I]$\lambda$5199{\AA}.
Additionally, Fe5015 is strongly disturbed by
the [O III]$\lambda$5007{\AA} in its red continuum as well.
The Mg b index has less contamination by the [N I]$\lambda$5199{\AA}
in its red continuum window.
These indices must be carefully corrected,
specifically the Mg b whose respective correction is more reliable.
The corrections of Mg b, Mg$_1$ and Mg$_2$ were computed
following the simple and accurate procedure described by
Goudfrooij \& Emsellem (1996).
The H$\beta$ correction should be only applied through
a suitable composite absorption-line templates
that were actually constructed using the stellar population synthesis approach
(Sect. 6).
However, it is redundant because the synthesis method supplies
directly the information which would be obtained
from this stellar age indicator Lick index.
On the other hand, the composite stellar population templates
of the synthesis technique had to be adopted for computing the
emission line corrections of the Magnesium indices.
These corrections are based on the equivalent widths of
the nebular lines that must be measured in the pure emission spectra
at the Lick/IDS resolution,
i.e. after the subtraction of an absorption line template
from the observed spectrum.
The corrections due to the presence of an emission line
inside an index continuum window or an index bandpass
are computed as a function of the emission line equivalent width
$EW_{em.line}$ measured in the continuum window
or the index bandpass respectively,
where $a(INDEX)$ and $b(INDEX)$ are constants of each index.
For Mg b, $a$ $\approx$ 1.127 and for Mg$_1$, $a$ $\approx$ 0.4475.
For Mg$_2$, $a$ $\approx$ 0.4158 and $b$ $\approx$ -1.086 mag.

$$ \Delta EW = a(INDEX) \times EW_{em.line} $$

$$ \Delta MAG = b(INDEX) \times \frac{EW_{em.line}}
{\Delta \lambda_i - EW_{INDEX}} $$

The Mg$_1$ and Mg$_2$ corrections due to the [O III]$\lambda$4958{\AA} line
(in their blue continuums) were computed in the $EW$ scale
and they were after transformed to the $MAG$ one.
Table 8 presents the equivalent widths of the
[O III]$\lambda$4958{\AA} and [N I]$\lambda$5199{\AA} lines,
measured exactly in the index continuum windows
(at $\lambda\lambda$4895.125-4957.625{\AA},
the blue one of Mg$_1$ and Mg$_2$,
and at $\lambda\lambda$5191.375-5206.375{\AA},
the red one of Mg b)
and the [N I]$\lambda$5199{\AA} line in the Mg$_2$ bandpass
($\lambda\lambda$5154.125-5196.625 {\AA}) as well as
for the inner regions of NGC\,1052.
The equivalent widths of the emission lines
were measured adopting a single Gaussian profile fit.

After taking into account the emission line corrections
for the measurements of Mg$_1$, Mg$_2$ (2 corrections) and Mg b,
finally these indices were calibrated to the Lick scale.
Beyond those radial distances for NGC\,1052,
H$\beta$, Mg b, Mg$_1$ and Mg$_2$ were directly measured
without corrections.
The Fe5015 index was not corrected neither adopted for both galaxies.

For NGC\,7796, emission lines were not detected in the aperture spectra
and the residual spectra as well (Sect 6).

%
\renewcommand{\tabcolsep}{4mm} 
\begin{table*}[htbp]
\caption{Equivalent widths of [O III]$\lambda$4958{\AA} line,
measured in the blue continuum windows of Mg$_1$ and Mg$_2$
(that are the same),
and [N I]$\lambda$5199{\AA} line,
measured either in the red continuum of Mg b and Mg$_2$ bandpass,
for the aperture spectra of the more central regions of NGC\,1052
along both axes.}
\begin{tabular} {l|c c c }
\hline
\hline
\multicolumn{4}{c}{$EW$ \,(\AA)} \\
\hline
R (arcsec)
& [O III]$\lambda$4958$_{\lambda\lambda 4895.125-4957.625}$
& [N I]$\lambda$5199$_{\lambda\lambda 5191.375-5206.375}$
& [N I]$\lambda$5199$_{\lambda\lambda 5154.125-5196.625}$ \\ 
\hline
major axis &            &              &               \\
\hline
0.00     & -5.0$\pm$0.3 & -1.4$\pm$0.1 & -1.2$\pm$0.1  \\
\hline
1.10\,SE & -5.5$\pm$0.5 & -1.0$\pm$0.1 & -0.9$\pm$0.1  \\
3.56\,SE & -4.1$\pm$0.6 & -0.8$\pm$0.1 & -0.7$\pm$0.1  \\
\hline
1.08\,NW & -4.5$\pm$0.2 & -0.9$\pm$0.1 & -0.8$\pm$0.1  \\
3.54\,NW & -2.1$\pm$0.7 & -0.4$\pm$0.1 & -0.4$\pm$0.1  \\
\hline
minor axis &            &              &               \\
\hline
0.00     & -4.1$\pm$0.2 & -0.70$\pm$0.05 & --  \\
\hline
2.17\,SW & -3.1$\pm$1.5 & -0.88$\pm$0.09 & --  \\
4.90\,SW & --          & -0.81$\pm$0.10 & --  \\
\hline
2.19\,NE & -3.2$\pm$0.6 & -0.82$\pm$0.05 & --  \\
4.92\,NE & --          & -0.15$\pm$0.10 & -3.1$\pm$0.2  \\
\hline
\end{tabular}
\label{emissions1}
\end{table*}
%

%
\renewcommand{\tabcolsep}{2mm} 
\begin{table*}[htbp]
\caption{Lick indices as a function of the distance to the galaxy center -
NGC\,1052, major axis.}
\begin{tabular} {l|c c c c c c c c }
\hline
\hline
\multicolumn{9}{c}{$EW$ \,(\AA) \,indices} \\
\hline
R (arcsec)
& Fe4383 & Ca4455
& Fe4531 & Fe4668
& Mg\, b & Fe5270 
& Fe5335 & Fe5406 \\
\hline
0.00
& 6.14 $\pm$ 0.30 & 2.35 $\pm$ 0.30
& 4.42 $\pm$ 0.32 & 8.81 $\pm$ 0.35
& 4.21 $\pm$ 0.23 & 2.46 $\pm$ 0.17 
& 2.61 $\pm$ 0.21 & 2.02 $\pm$ 0.15 \\
\hline
1.10\,SE 
& 6.52 $\pm$ 0.30 & 2.50 $\pm$ 0.30
& 3.60 $\pm$ 0.32 & 8.53 $\pm$ 0.36
& 5.03 $\pm$ 0.23 & 2.29 $\pm$ 0.17 
& 2.49 $\pm$ 0.21 & 1.96 $\pm$ 0.14 \\
3.56\,SE 
& 5.89 $\pm$ 0.33 & 2.44 $\pm$ 0.32
& 3.77 $\pm$ 0.35 & 8.80 $\pm$ 0.40
& 4.37 $\pm$ 0.25 & 2.55 $\pm$ 0.19
& 2.20 $\pm$ 0.23 & 1.82 $\pm$ 0.16 \\
6.84\,SE
& 5.72 $\pm$ 0.40 & 2.42 $\pm$ 0.35
& 3.07 $\pm$ 0.39 & 8.18 $\pm$ 0.46
& 5.09 $\pm$ 0.27 & 2.67 $\pm$ 0.23 
& 2.05 $\pm$ 0.28 & 1.73 $\pm$ 0.20 \\
11.80\,SE
& 6.11 $\pm$ 0.44 & 2.38 $\pm$ 0.37
& 3.50 $\pm$ 0.42 & 6.86 $\pm$ 0.50
& 4.75 $\pm$ 0.29 & 2.58 $\pm$ 0.25
& 2.74 $\pm$ 0.33 & 2.12 $\pm$ 0.23 \\
20.21\,SE
& 4.62 $\pm$ 0.47 & 2.91 $\pm$ 0.43
& 4.42 $\pm$ 0.47 & 8.39 $\pm$ 0.55
& 4.10 $\pm$ 0.30 & 2.84 $\pm$ 0.27
& 2.49 $\pm$ 0.35 & 2.30 $\pm$ 0.25 \\
\hline
1.08\,NW 
& 5.98 $\pm$ 0.30 & 2.43 $\pm$ 0.30
& 3.78 $\pm$ 0.32 & 8.93 $\pm$ 0.36
& 4.94 $\pm$ 0.28 & 2.59 $\pm$ 0.17
& 2.67 $\pm$ 0.21 & 1.93 $\pm$ 0.15 \\
3.54\,NW 
& 6.79 $\pm$ 0.33 & 2.46 $\pm$ 0.31
& 3.70 $\pm$ 0.35 & 8.22 $\pm$ 0.36
& 5.07 $\pm$ 0.24 & 2.49 $\pm$ 0.19
& 2.71 $\pm$ 0.23 & 1.99 $\pm$ 0.16 \\
6.81\,NW
& 6.44 $\pm$ 0.40 & 2.33 $\pm$ 0.35
& 3.61 $\pm$ 0.39 & 7.77 $\pm$ 0.45
& 5.16 $\pm$ 0.26 & 3.01 $\pm$ 0.22
& 2.62 $\pm$ 0.28 & 1.92 $\pm$ 0.19 \\
11.78\,NW
& 4.86 $\pm$ 0.42 & 2.28 $\pm$ 0.36
& 2.85 $\pm$ 0.40 & 8.12 $\pm$ 0.50
& 4.54 $\pm$ 0.28 & 2.80 $\pm$ 0.25
& 2.48 $\pm$ 0.31 & 1.62 $\pm$ 0.21 \\
20.19\,NW
& 5.62 $\pm$ 0.45 & 2.36 $\pm$ 0.36
& 2.83 $\pm$ 0.43 & 5.55 $\pm$ 0.51
& 4.55 $\pm$ 0.28 & 3.03 $\pm$ 0.26
& 1.76 $\pm$ 0.30 & 1.39 $\pm$ 0.21 \\
\hline
\end{tabular}
\label{indices1}
\end{table*}
%

%
\renewcommand{\tabcolsep}{4mm} 
\begin{table*}[htbp]
\caption{Lick indices as a function of the distance to the galaxy center -
NGC\,1052, major axis.}
\begin{tabular} {l|c c c c c c }
\hline
\hline
\multicolumn{7}{c}{$EW$ \,(\AA) \,and \,$MAG$ \,(mag) \,indices} \\
\hline
R (arcsec)
& Fe5709 & Fe5782
& Na\, D & H$\beta$
& Mg$_1$ & Mg$_2$ \\ 
\hline
0.00
& 1.09 $\pm$ 0.16   & 0.98 $\pm$ 0.17
& 6.46 $\pm$ 0.11   & --
& 0.181 $\pm$ 0.004 & 0.331 $\pm$ 0.004 \\
\hline
1.10\,SE
& 0.94 $\pm$ 0.16   & 0.91 $\pm$ 0.17
& 6.03 $\pm$ 0.11   & --
& 0.172 $\pm$ 0.004 & 0.303 $\pm$ 0.004 \\
3.56\,SE 
& 0.91 $\pm$ 0.17   & 0.86 $\pm$ 0.18
& 4.59 $\pm$ 0.12   & --
& 0.148 $\pm$ 0.004 & 0.304 $\pm$ 0.004 \\
6.84\,SE
& 1.20 $\pm$ 0.20   & 0.39 $\pm$ 0.20
& 3.84 $\pm$ 0.14   & 0.68 $\pm$ 0.19
& 0.174 $\pm$ 0.005 & 0.329 $\pm$ 0.005 \\
11.80\,SE
& 1.31 $\pm$ 0.22   & 0.58 $\pm$ 0.22
& 3.33 $\pm$ 0.15   & 1.28 $\pm$ 0.22
& 0.138 $\pm$ 0.006 & 0.298 $\pm$ 0.006 \\
20.21\,SE
& 1.22 $\pm$ 0.23   & 0.26 $\pm$ 0.22
& 3.36 $\pm$ 0.16   & 1.08 $\pm$ 0.23
& 0.148 $\pm$ 0.006 & 0.294 $\pm$ 0.006 \\
\hline
1.08\,NW 
& 1.11 $\pm$ 0.16   & 0.91 $\pm$ 0.17
& 6.31 $\pm$ 0.11   & -- 
& 0.178 $\pm$ 0.004 & 0.322 $\pm$ 0.004 \\
3.54\,NW 
& 1.02 $\pm$ 0.17   & 0.73 $\pm$ 0.18
& 5.41 $\pm$ 0.12   & --
& 0.164 $\pm$ 0.004 & 0.317 $\pm$ 0.004 \\
6.81\,NW
& 1.23 $\pm$ 0.20   & 1.05 $\pm$ 0.21
& 4.51 $\pm$ 0.14   & 0.61 $\pm$ 0.18
& 0.168 $\pm$ 0.005 & 0.330 $\pm$ 0.005 \\
11.78\,NW
& 0.94 $\pm$ 0.20   & 0.62 $\pm$ 0.22
& 3.91 $\pm$ 0.15   & 1.19 $\pm$ 0.21
& 0.147 $\pm$ 0.006 & 0.304 $\pm$ 0.006 \\
20.19\,NW
& 1.06 $\pm$ 0.21   & 0.74 $\pm$ 0.22
& 3.32 $\pm$ 0.16   & 1.28 $\pm$ 0.23
& 0.125 $\pm$ 0.006 & 0.291 $\pm$ 0.006 \\
\hline
\end{tabular}
\begin{list} {Table Notes.}
\item Columns 6 and 7: mag unity.
\end{list}
\label{indices2}
\end{table*}
%

%
\renewcommand{\tabcolsep}{2mm} 
\begin{table*}[htbp]
\caption{Lick indices as a function of the distance to the galaxy center -
NGC\,1052, minor axis.}
\begin{tabular} {l|c c c c c c c c }
\hline
\hline
\multicolumn{9}{c}{$EW$ \,(\AA) \,indices} \\
\hline
R (arcsec)
& Fe4383 & Ca4455
& Fe4531 & Fe4668
& Mg\, b & Fe5270 
& Fe5335 & Fe5406 \\
\hline
0.00
& 7.09 $\pm$ 0.63 & 2.39 $\pm$ 0.48
& 3.84 $\pm$ 0.54 & 8.46 $\pm$ 0.66
& 5.44 $\pm$ 0.39 & 2.60 $\pm$ 0.30 
& 2.86 $\pm$ 0.43 & 2.09 $\pm$ 0.29 \\
\hline
2.17\,SW
& 5.80 $\pm$ 0.84 & 2.39 $\pm$ 0.48
& 3.67 $\pm$ 0.69 & 8.84 $\pm$ 0.86
& 4.92 $\pm$ 0.47 & 2.59 $\pm$ 0.38 
& 2.48 $\pm$ 0.54 & 1.98 $\pm$ 0.37 \\
4.90\,SW 
& 4.99 $\pm$ 1.17 & 2.67 $\pm$ 0.71
& 2.17 $\pm$ 0.90 & 6.64 $\pm$ 1.16
& 4.07 $\pm$ 0.55 & 1.88 $\pm$ 0.49
& 2.49 $\pm$ 0.67 & 1.95 $\pm$ 0.47 \\
9.41\,SW
& 6.68 $\pm$ 1.66 & 2.42 $\pm$ 0.95
& 3.61 $\pm$ 1.25 & 7.21 $\pm$ 1.60
& 5.51 $\pm$ 0.76 & 3.40 $\pm$ 0.70 
& 3.03 $\pm$ 0.98 & 2.33 $\pm$ 0.70 \\
\hline
2.19\,NE 
& 6.95 $\pm$ 0.86 & 2.56 $\pm$ 0.61
& 3.27 $\pm$ 0.69 & 8.68 $\pm$ 0.87
& 5.03 $\pm$ 0.48 & 2.61 $\pm$ 0.39
& 3.24 $\pm$ 0.58 & 1.99 $\pm$ 0.38 \\
4.92\,NE 
& 5.32 $\pm$ 1.25 & 2.40 $\pm$ 0.74
& 3.43 $\pm$ 0.97 & 7.56 $\pm$ 1.23
& 5.53 $\pm$ 0.60 & 2.59 $\pm$ 0.53
& 2.66 $\pm$ 0.74 & 2.01 $\pm$ 0.52 \\
9.43\,NE
& 4.27 $\pm$ 1.75 & 1.90 $\pm$ 0.69
& 1.80 $\pm$ 1.16 & 4.12 $\pm$ 1.51
& 4.38 $\pm$ 0.63 & 1.39 $\pm$ 0.63
& 1.56 $\pm$ 0.77 & 1.49 $\pm$ 0.56 \\
\hline
\end{tabular}
\label{indices3}
\end{table*}
%

%
\renewcommand{\tabcolsep}{4mm} 
\begin{table*}[htbp]
\caption{Lick indices as a function of the distance to the galaxy center -
NGC\,1052, minor axis.}
\begin{tabular} {l|c c c c c c }
\hline
\hline
\multicolumn{7}{c}{$EW$ \,(\AA) \,and \,$MAG$ \,(mag) \,indices} \\
\hline
R (arcsec)
& Fe5709 & Fe5782
& Na\, D & H$\beta$
& Mg$_1$ & Mg$_2$ \\ 
\hline
0.00
& 1.08 $\pm$ 0.25   & 0.87 $\pm$ 0.27
& 6.74 $\pm$ 0.21   & --
& 0.183 $\pm$ 0.007 & 0.285 $\pm$ 0.007 \\
\hline
2.17\,SW
& 0.91 $\pm$ 0.29   & 1.15 $\pm$ 0.34
& 5.85 $\pm$ 0.25   & --
& 0.172 $\pm$ 0.008 & 0.306 $\pm$ 0.008 \\
4.90\,SW 
& 0.95 $\pm$ 0.36   & 1.21 $\pm$ 0.42
& 5.41 $\pm$ 0.31   & --
& 0.178 $\pm$ 0.010 & 0.348 $\pm$ 0.011 \\
9.41\,SW
& 0.90 $\pm$ 0.47   & 0.76 $\pm$ 0.55
& 2.04 $\pm$ 0.40   & 0.88 $\pm$ 0.56
& 0.168 $\pm$ 0.012 & 0.333 $\pm$ 0.014 \\
\hline
2.19\,NE 
& 0.95 $\pm$ 0.30   & 1.26 $\pm$ 0.35
& 5.52 $\pm$ 0.25   & -- 
& 0.179 $\pm$ 0.008 & 0.303 $\pm$ 0.009 \\
4.92\,NE 
& 1.25 $\pm$ 0.39   & 1.42 $\pm$ 0.46
& 3.93 $\pm$ 0.32   & --
& 0.183 $\pm$ 0.011 & 0.460 $\pm$ 0.012 \\
9.43\,NE
& 1.18 $\pm$ 0.45   & 1.30 $\pm$ 0.50
& 2.90 $\pm$ 0.38   & 0.73 $\pm$ 0.54
& 0.157 $\pm$ 0.012 & 0.301 $\pm$ 0.014 \\
\hline
\end{tabular}
\begin{list} {Table Notes.}
\item Columns 6 and 7: mag unity.
\end{list}
\label{indices4}
\end{table*}
%

%
\renewcommand{\tabcolsep}{2mm}
\begin{table*}[htbp]
\caption{Lick indices as a function of the distance to the galaxy center -
NGC\,7796, major axis.}
\begin{tabular} {l|c c c c c c c c }
\hline
\hline
\multicolumn{9}{c}{$EW$ \,(\AA) \,indices} \\
\hline
R (arcsec)
& Fe4383 & Ca4455
& Fe4531 & Fe4668
& Mg\, b & Fe5270
& Fe5335 & Fe5406 \\ 
\hline
0.00
& 5.25 $\pm$ 0.35 & 2.65 $\pm$ 0.36
& 3.41 $\pm$ 0.35 & 9.22 $\pm$ 0.41
& 4.93 $\pm$ 0.25 & 2.84 $\pm$ 0.20
& 2.42 $\pm$ 0.26 & 1.83 $\pm$ 0.18 \\
\hline
1.09\,S 
& 5.28 $\pm$ 0.38 & 2.82 $\pm$ 0.41
& 3.28 $\pm$ 0.36 & 8.37 $\pm$ 0.44
& 5.16 $\pm$ 0.27 & 2.68 $\pm$ 0.22
& 2.28 $\pm$ 0.30 & 1.60 $\pm$ 0.19 \\
3.55\,S 
& 5.22 $\pm$ 0.48 & 2.71 $\pm$ 0.47
& 3.59 $\pm$ 0.44 & 7.50 $\pm$ 0.53
& 5.47 $\pm$ 0.34 & 2.80 $\pm$ 0.27
& 2.87 $\pm$ 0.39 & 1.30 $\pm$ 0.24 \\
7.10\,S
& 5.05 $\pm$ 0.58 & 2.70 $\pm$ 0.55
& 4.11 $\pm$ 0.53 & 7.38 $\pm$ 0.65
& 5.45 $\pm$ 0.41 & 3.35 $\pm$ 0.33
& 3.14 $\pm$ 0.51 & 1.14 $\pm$ 0.29 \\
13.21\,S
& 5.26 $\pm$ 0.68 & 2.72 $\pm$ 0.59
& 3.59 $\pm$ 0.59 & 5.82 $\pm$ 0.73
& 5.42 $\pm$ 0.44 & 3.01 $\pm$ 0.37
& 2.25 $\pm$ 0.52 & 0.51 $\pm$ 0.30 \\
\hline
1.09\,N 
& 5.27 $\pm$ 0.36 & 2.50 $\pm$ 0.35
& 3.47 $\pm$ 0.36 & 9.12 $\pm$ 0.43
& 5.06 $\pm$ 0.26 & 3.12 $\pm$ 0.21
& 2.65 $\pm$ 0.28 & 1.84 $\pm$ 0.18 \\
3.55\,N 
& 4.60 $\pm$ 0.45 & 2.63 $\pm$ 0.44
& 3.76 $\pm$ 0.43 & 8.42 $\pm$ 0.52
& 4.99 $\pm$ 0.32 & 3.10 $\pm$ 0.26
& 2.56 $\pm$ 0.36 & 1.81 $\pm$ 0.24 \\
7.10\,N
& 4.68 $\pm$ 0.53 & 2.23 $\pm$ 0.45
& 3.92 $\pm$ 0.51 & 7.95 $\pm$ 0.61
& 4.71 $\pm$ 0.35 & 3.06 $\pm$ 0.30
& 3.42 $\pm$ 0.45 & 1.45 $\pm$ 0.27 \\
13.21\,N
& 4.49 $\pm$ 0.66 & 2.53 $\pm$ 0.55
& 4.70 $\pm$ 0.60 & 7.36 $\pm$ 0.72
& 4.35 $\pm$ 0.41 & 3.60 $\pm$ 0.37
& 3.26 $\pm$ 0.55 & 1.35 $\pm$ 0.32 \\
\hline
\end{tabular}
\label{indices5}
\end{table*}
%

%
\renewcommand{\tabcolsep}{4mm} 
\begin{table*}[htbp]
\caption{Lick indices as a function of the distance to the galaxy center -
NGC\,7796, major axis.}
\begin{tabular} {l|c c c c c c }
\hline
\hline
\multicolumn{7}{c}{$EW$ \,(\AA) \,and \,$MAG$ \,(mag) \,indices} \\
\hline
R (arcsec)
& Fe5709 & Fe5782
& Na\, D & H$\beta$
& Mg$_1$ & Mg$_2$ \\
\hline
0.00
& 0.81 $\pm$ 0.18   & 0.66 $\pm$ 0.19
& 5.06 $\pm$ 0.13   & 1.56 $\pm$ 0.19
& 0.155 $\pm$ 0.005 & 0.345 $\pm$ 0.005 \\
\hline
1.09\,S 
& 0.98 $\pm$ 0.19   & 0.51 $\pm$ 0.20
& 5.11 $\pm$ 0.14   & 1.46 $\pm$ 0.19
& 0.157 $\pm$ 0.005 & 0.338 $\pm$ 0.005 \\
3.55\,S 
& 1.09 $\pm$ 0.23   & 0.58 $\pm$ 0.25
& 4.82 $\pm$ 0.17   & 1.79 $\pm$ 0.24
& 0.191 $\pm$ 0.005 & 0.359 $\pm$ 0.006 \\
7.10\,S
& 0.88 $\pm$ 0.26   & 0.91 $\pm$ 0.30
& 4.47 $\pm$ 0.21   & 1.41 $\pm$ 0.27
& 0.147 $\pm$ 0.006 & 0.334 $\pm$ 0.007 \\
13.21\,S
& 0.05 $\pm$ 0.26   & -0.18 $\pm$ 0.28
& 4.26 $\pm$ 0.23   & 1.43 $\pm$ 0.30
& 0.153 $\pm$ 0.007 & 0.311 $\pm$ 0.008 \\
\hline
1.09\,N 
& 0.94 $\pm$ 0.19   & 0.68 $\pm$ 0.20
& 4.99 $\pm$ 0.13   & 1.54 $\pm$ 0.19
& 0.173 $\pm$ 0.005 & 0.356 $\pm$ 0.005 \\
3.55\,N 
& 0.87 $\pm$ 0.22   & 0.67 $\pm$ 0.24
& 4.39 $\pm$ 0.16   & 1.66 $\pm$ 0.23
& 0.184 $\pm$ 0.006 & 0.353 $\pm$ 0.006 \\
7.10\,N
& 0.80 $\pm$ 0.24   & 0.83 $\pm$ 0.28
& 4.48 $\pm$ 0.19   & 1.52 $\pm$ 0.25
& 0.150 $\pm$ 0.006 & 0.319 $\pm$ 0.007 \\
13.21\,N
& 0.55 $\pm$ 0.27   & 0.77 $\pm$ 0.32
& 3.99 $\pm$ 0.23   & 1.02 $\pm$ 0.28
& 0.140 $\pm$ 0.007 & 0.306 $\pm$ 0.008 \\
\hline
\end{tabular}
\begin{list} {Table Notes.}
\item Columns 6 and 7: mag unity.
\end{list}
\label{indices6}
\end{table*}
%

%
\renewcommand{\tabcolsep}{2mm} 
\begin{table*}[htbp]
\caption{Lick indices as a function of the distance to the galaxy center -
NGC\,7796, minor axis.}
\begin{tabular} {l|c c c c c c c c }
\hline
\hline
\multicolumn{9}{c}{$EW$ \,(\AA) \,indices} \\
\hline
R (arcsec)
& Fe4383 & Ca4455
& Fe4531 & Fe4668
& Mg\, b & Fe5270 
& Fe5335 & Fe5406 \\
\hline
0.00
& 5.33 $\pm$ 0.42 & 2.88 $\pm$ 0.44
& 3.05 $\pm$ 0.39 & 8.33 $\pm$ 0.48
& 5.10 $\pm$ 0.30 & 2.50 $\pm$ 0.24 
& 2.60 $\pm$ 0.33 & 1.67 $\pm$ 0.22 \\
\hline
1.09\,E 
& 5.93 $\pm$ 0.45 & 2.95 $\pm$ 0.46
& 3.11 $\pm$ 0.41 & 7.85 $\pm$ 0.50
& 4.91 $\pm$ 0.31 & 2.51 $\pm$ 0.25 
& 2.57 $\pm$ 0.36 & 1.49 $\pm$ 0.23 \\
3.55\,E 
& 5.43 $\pm$ 0.58 & 2.64 $\pm$ 0.53
& 2.70 $\pm$ 0.49 & 6.72 $\pm$ 0.62
& 4.66 $\pm$ 0.37 & 2.88 $\pm$ 0.32
& 3.05 $\pm$ 0.48 & 1.31 $\pm$ 0.29 \\
6.93\,E
& 4.19 $\pm$ 0.79 & 2.13 $\pm$ 0.60
& 2.70 $\pm$ 0.64 & 2.59 $\pm$ 0.80
& 4.66 $\pm$ 0.49 & 4.03 $\pm$ 0.45 
& 2.32 $\pm$ 0.61 & 1.44 $\pm$ 0.41 \\
12.50\,E
& 2.90 $\pm$ 0.94 & 2.40 $\pm$ 0.72
& 1.90 $\pm$ 0.74 & 8.03 $\pm$ 1.02
& 3.99 $\pm$ 0.55 & 3.54 $\pm$ 0.51
& 2.46 $\pm$ 0.73 & 0.58 $\pm$ 0.43 \\
\hline
1.09\,W 
& 5.42 $\pm$ 0.44 & 2.74 $\pm$ 0.44
& 3.08 $\pm$ 0.41 & 7.96 $\pm$ 0.50
& 5.23 $\pm$ 0.31 & 2.48 $\pm$ 0.25
& 2.68 $\pm$ 0.35 & 1.84 $\pm$ 0.23 \\
3.55\,W 
& 4.77 $\pm$ 0.54 & 2.94 $\pm$ 0.54
& 2.98 $\pm$ 0.48 & 6.41 $\pm$ 0.60
& 5.02 $\pm$ 0.38 & 2.69 $\pm$ 0.31
& 2.80 $\pm$ 0.46 & 1.54 $\pm$ 0.29 \\
6.93\,W
& 4.66 $\pm$ 0.72 & 2.56 $\pm$ 0.56
& 2.24 $\pm$ 0.61 & 3.83 $\pm$ 0.79
& 5.01 $\pm$ 0.45 & 2.09 $\pm$ 0.39
& 2.69 $\pm$ 0.56 & 1.42 $\pm$ 0.36 \\
12.50\,W
& 2.57 $\pm$ 0.88 & 1.45 $\pm$ 0.51
& 2.02 $\pm$ 0.73 &-4.62 $\pm$ 0.94
& 4.33 $\pm$ 0.49 & 3.10 $\pm$ 0.47
& 2.92 $\pm$ 0.64 & 0.84 $\pm$ 0.40 \\
\hline
\end{tabular}
\label{indices7}
\end{table*}
%

%
\renewcommand{\tabcolsep}{4mm} 
\begin{table*}[htbp]
\caption{Lick indices as a function of the distance to the galaxy center -
NGC\,7796, minor axis.}
\begin{tabular} {l|c c c c c c }
\hline
\hline
\multicolumn{7}{c}{$EW$ \,(\AA) \,and \,$MAG$ \,(mag) \,indices} \\
\hline
R (arcsec)
& Fe5709 & Fe5782
& Na\, D & H$\beta$
& Mg$_1$ & Mg$_2$ \\ 
\hline
0.00
& 0.68 $\pm$ 0.20   & 0.50 $\pm$ 0.22
& 4.74 $\pm$ 0.15   & 1.11 $\pm$ 0.20
& 0.178 $\pm$ 0.005 & 0.366 $\pm$ 0.006 \\
\hline
1.09\,E
& 0.75 $\pm$ 0.21   & 0.54 $\pm$ 0.23
& 4.85 $\pm$ 0.16   & 1.19 $\pm$ 0.21
& 0.180 $\pm$ 0.006 & 0.360 $\pm$ 0.006 \\
3.55\,E 
& 1.17 $\pm$ 0.27   & 0.31 $\pm$ 0.28
& 4.55 $\pm$ 0.21   & 1.08 $\pm$ 0.25
& 0.169 $\pm$ 0.007 & 0.326 $\pm$ 0.007 \\
6.93\,E
& 0.76 $\pm$ 0.32   & 0.83 $\pm$ 0.38
& 4.61 $\pm$ 0.28   & 2.27 $\pm$ 0.36
& 0.148 $\pm$ 0.008 & 0.331 $\pm$ 0.009 \\
12.50\,E
& 0.17 $\pm$ 0.35   &-0.09 $\pm$ 0.39
& 4.53 $\pm$ 0.32   & 1.52 $\pm$ 0.40
& 0.158 $\pm$ 0.009 & 0.341 $\pm$ 0.010 \\
\hline
1.09\,W
& 0.90 $\pm$ 0.16   & 0.62 $\pm$ 0.23
& 4.46 $\pm$ 0.08   & 1.26 $\pm$ 0.21
& 0.172 $\pm$ 0.006 & 0.363 $\pm$ 0.006 \\
3.55\,W 
& 0.84 $\pm$ 0.20   & 0.70 $\pm$ 0.28
& 4.32 $\pm$ 0.10   & 1.31 $\pm$ 0.25
& 0.163 $\pm$ 0.007 & 0.345 $\pm$ 0.007 \\
6.93\,W
& 0.08 $\pm$ 0.28   & 0.65 $\pm$ 0.34
& 3.95 $\pm$ 0.25   & 1.57 $\pm$ 0.33
& 0.159 $\pm$ 0.008 & 0.340 $\pm$ 0.009 \\
12.50\,W
& 0.89 $\pm$ 0.35   &-0.02 $\pm$ 0.36
& 4.43 $\pm$ 0.29   & 1.40 $\pm$ 0.38
& 0.148 $\pm$ 0.009 & 0.283 $\pm$ 0.0010 \\
\hline
\end{tabular}
\begin{list} {Table Notes.}
\item Columns 6 and 7: mag unity.
\end{list}
\label{indices8}
\end{table*}
%

%
%
\subsection {External comparisons of some Lick indices}

In order to compare our measurements of the nuclear Lick indices,
they must be corrected by the aperture effect
as the central velocity dispersion was (see Subsection 2.2 and
Jorgensen 1997).
The adopted logarithmic radial gradients of the indices
($\nabla$ INDEX = ${\Delta \log EW \over \Delta \log r}$
or ${\Delta MAG \over \Delta \log r}$)
were
$\nabla(Fe \,index)$ = -0.050
including $<$Fe$>$ except $\nabla(Fe4668)$ = -0.080,
$\nabla(Ca4455)$ = 0,
$\nabla(Na \,D)$ = -0.090,
and $\nabla(Mg_1)$ = $\nabla(Mg_2)$ = -0.038 mag.dex$^{-1}$.

The central Lick indices of both galaxies from the current work
are shown in Table 17, in which the mean literature values of some indices
are also presented from several sources:
Terlevich et al. (1981),
Davies et al. (1987),
Burstein et al. (1988),
Couture \& Hardy (1988),
Faber et al. (1989),
Worthey, Faber \& Gonzalez (1992),
Carollo, Danziger \& Buson (1993),
Huchra et al. (1996) (all for Mg$_2^0$ only),
Trager et al. (1998) (not for Mg$_2$),
Beuing et al. (2002) and
Thomas et al. (2005) (only for Mg b, $<$Fe$>$ and H$\beta$)
for NGC\,1052
and
Bertin et al. (1994),
Golev \& Prugniel (1998) (both for Mg$_2^0$ only),
Beuing et al. (2002) and
Thomas et al. (2005) (only for Mg b, $<$Fe$>$ and H$\beta$)
for NGC\,7796.

The estimate of the nuclear Mg$_{2}$
based on the Mg$_{2}^{0}$-$\sigma_{v}^{0}$ relation
was made with the results of
Bender, Burstein \& Faber (1993).
The predictions,
using $\sigma_{v}^{0}$ (NGC\,1052) = 204 km.s$^{-1}$ and
$\sigma_{v}^{0}$ (NGC\,7796) = 253 km.s$^{-1}$,
are very near to our measurements:
Mg$_{2}^{0}$ = 0.296 mag and
Mg$_{2}^{0}$ = 0.315 mag,
respectively for NGC\,1052 and NGC\,7796.
Note that the literature values of Mg$_{2}^{0}$ of
NGC\,7796 are smaller than our value and predicted one
(see Tab. 1 and Tab. 17).

For all indices, the agreements with the literature values
are good considering both errors.
Note that the published Lick indices are non fully corrected
(by the aperture effect and emission lines).

%
\renewcommand{\tabcolsep}{5.3mm} 
\begin{table*}[htbp]
\caption{Lick index external comparisons:
galaxy central values of Lick indices (major axis only).}
\begin{tabular} {l|c c c c }
\hline
\hline
\multicolumn{5}{c}{$EW$ \,(\AA) \,and \,$MAG$ \,(mag) \,indices} \\
\hline
       & NGC 1052 &                 & NGC 7796        &            \\
Index  & our      & literature      & our             & literature \\
\hline
Fe4383$^0$
& 5.62 $\pm$ 0.30 & 6.11 $\pm$ 0.45 & 4.97 $\pm$ 0.35 & -- \\
Ca4455$^0$
& 2.36 $\pm$ 0.30 & 2.10 $\pm$ 0.21 & 2.66 $\pm$ 0.36 & -- \\
Fe4531$^0$
& 3.36 $\pm$ 0.47 & 3.62 $\pm$ 0.30 & 3.22 $\pm$ 0.35 & -- \\
Fe4668$^0$
& 7.65 $\pm$ 0.35 & 8.50 $\pm$ 0.44 & 8.45 $\pm$ 0.41 & -- \\
Mg\, b$^0$
& 3.83 $\pm$ 0.23 & 5.63 $\pm$ 0.14 & 4.67 $\pm$ 0.25 & 5.20 $\pm$ 0.25 \\
Fe5270$^0$
& 2.25 $\pm$ 0.17 & 2.81 $\pm$ 0.04 & 2.69 $\pm$ 0.20 & 3.63 $\pm$ 0.05 \\
Fe5335$^0$
& 2.39 $\pm$ 0.21 & 2.74 $\pm$ 0.07 & 2.30 $\pm$ 0.26 & 2.98 $\pm$ 0.06 \\
Fe5406$^0$
& 1.85 $\pm$ 0.15 & 1.82 $\pm$ 0.02 & 1.73 $\pm$ 0.18 & 1.87 $\pm$ 0.05 \\
Fe5709$^0$
& 1.00 $\pm$ 0.16 & 0.98 $\pm$ 0.02 & 0.76 $\pm$ 0.18 & 1.01 $\pm$ 0.04 \\
Fe5782$^0$
& 0.90 $\pm$ 0.17 & 0.93 $\pm$ 0.02 & 0.62 $\pm$ 0.19 & 1.02 $\pm$ 0.04 \\
Na\, D$^0$
& 5.52 $\pm$ 0.11 & 6.10 $\pm$ 0.03 & 4.58 $\pm$ 0.13 & 5.48 $\pm$ 0.06 \\
$<$Fe$>^0$
& 2.32 $\pm$ 0.21 & 2.78 $\pm$ 0.02 & 2.49 $\pm$ 0.33 & 3.04 $\pm$ 0.37 \\
H$\beta$$^0$
& --              & 1.22 $\pm$ ?    & 1.57 $\pm$ 0.19 & 1.64 $\pm$ 0.16 \\
Mg$_1^0$
& 0.149$\pm$0.004 & 0.193$\pm$0.001 & 0.137$\pm$0.005 & 0.165$\pm$0.002 \\
Mg$_2^0$
& 0.293$\pm$0.004 & 0.303$\pm$0.017 & 0.326$\pm$0.005 & 0.267$\pm$0.051 \\
\hline
\end{tabular}
\label{centralinds}
\end{table*}
%

%
%
\section{Radial gradients of the Lick indices}

The calibrated/corrected Lick indices (except Fe5015) 
of the aperture spectra along both axes
are shown in Tables 9 to 12 for NGC\,1052
and Tables 13 to 16 for NGC\,7796.

The Lick indices for which we have measured their radial gradients
along both axes of two galaxies
are Fe4383, Ca4455, Fe4531, Fe4668, Mg b, Fe5270, Fe5335, $<$Fe$>$,
Fe5406, Fe5709, Fe5782 and Na D
(including H$\beta$, Mg$_1$ and Mg$_2$ for NGC\,7796).
The $<$Fe$>$ index is an average value of two iron indices:
$<$Fe$>$ = (Fe5270+Fe5335) $\div$ 2.

For the major axis of NGC\,1052,
we have used the aperture spectra up to r = 20.2 arcsec
and for its minor axis up to 9.4 arcsec.
For the major and minor axes of NGC\,7796,
we have used the spectra up to r = 13.2 and 12.5 arcsec
respectively (all data). 

Some atomic line-strengths are contaminated
by molecular absorption,
such as Fe4383 that is influenced by CH lines
and Fe4668 by C$_{2}$ lines of the Swan System
(Tripicco \& Bell 1995).
Thomas, Maraston \& Bender (2003a)
re-called Fe4668 as C$_2$4668.
However they affirmed that
Fe4383 index is still very sensitive to the iron abundance.

The blue continuum of Na D index is partially contaminated
by the same Sodium absorption of the interestellar medium of the Galaxy, 
but this influence performs equally for all spectra.

The distances to the galaxy center are normalized
by the effective radius $r_{e}$
that is corrected considering the apparent ellipticity $\epsilon$,
so that $r_{e}^{corr}$ = $r_{e}$(1-$\epsilon$)$^{-1/2}$
at the major axis direction
or $r_{e}^{corr}$ = $r_{e}$(1-$\epsilon$)$^{1/2}$
at the minor one
(Davies, Sadler \& Peletier 1993 and
Kobayashi \& Arimoto 1999).

We have computed the radial gradients of the Lick indices
as a function of the logarithm of the normalized radius
through a linear regression fit,

$$ Index(r) = Index_{e} + 
\frac{\Delta Index}{\Delta \log r} . \log \frac{r}{r_{e}^{corr}} $$

where $Index_{e}$ = $Index(\log \frac{r}{r_{e}^{corr}} = 0)$
= $Index(r = r_{e}^{corr})$
is the linear regression constant.
The linear fits are computed by the least square method
that takes into account the data errors only on the Lick indices
(like Kobayashi \& Arimoto 1999).

In Table 18, we have shown the respective results of the fits:
the own radial gradients and the index values at one $r_{e}^{corr}$.

In Figures 5, 6 and 7,
the Lick indices as a function of the logarithm of the normalized radius
along both axes
are shown with their respective linear fits for NGC\,1052.
Figure 5 has the gradients of Fe4383, Fe4531, Fe4668,
Fe5270, Fe5335 and $<$Fe$>$.
Figure 6 shows the gradients of Fe5406, Fe5709, Fe5782, 
Ca4455, Mg b and Na D.
Figure 7 presents the radial gradients of Mg$_1$ and Mg$_2$.

In Figures 8, 9 and 10,
the Lick indices as a function of the logarithm of the normalized radius
along both axes
are presented with their respective linear fits for NGC\,7796.
Figure 8 has the gradients of Fe4383, Fe4531, Fe4668,
Fe5270, Fe5335 and $<$Fe$>$.
Figure 9 shows the gradients of Fe5406, Fe5709, Fe5782,
Ca4455, Mg b and Na D.
Figure 10 presents the radial gradients of H$\beta$, Mg$_1$ and Mg$_2$.

%
\begin{figure}[htbp]
\resizebox{\hsize}{!}{\includegraphics{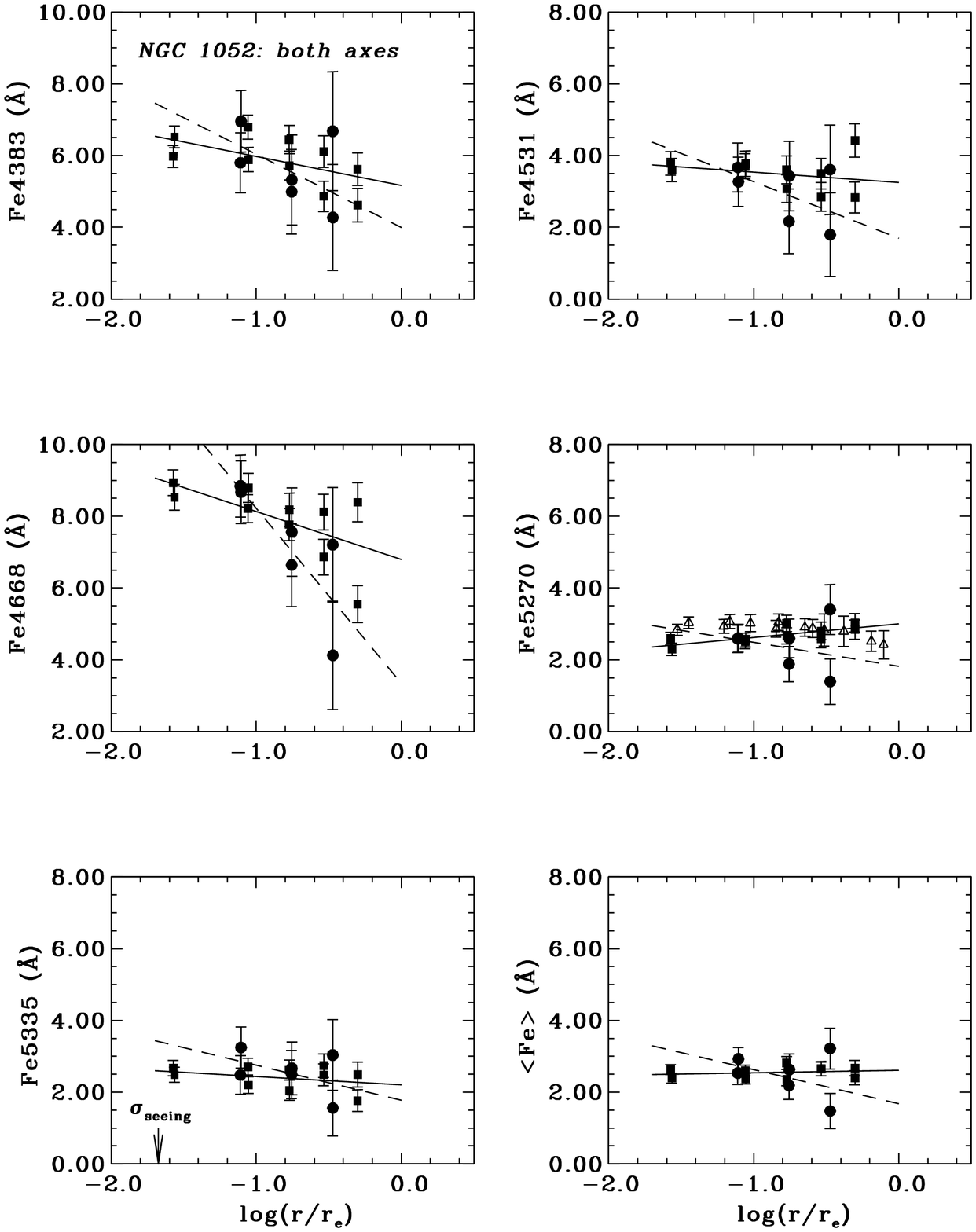}}
\caption{Radial gradients of
Fe4383, Fe4531, Fe4668, Fe5270, Fe5335 and $<$Fe$>$
along both axes of NGC\,1052
whose respective linear regressions are shown by solid and dashed lines
for the major (solid squares) and minor axis (solid circles) data respectively.
The seeing size $\sigma_{seeing}$ is given on the abscissa of the bottom-left plot
($\sigma_{seeing}$ = FWHM$_{seeing}$/2.355).
The Fe5270 data of Carollo, Danziger \& Buson (1993) along the E-W direction, 
as open triangles, are also plotted.}
\label{grads1}
\end{figure}
%

%
\begin{figure}[htbp]
\resizebox{\hsize}{!}{\includegraphics{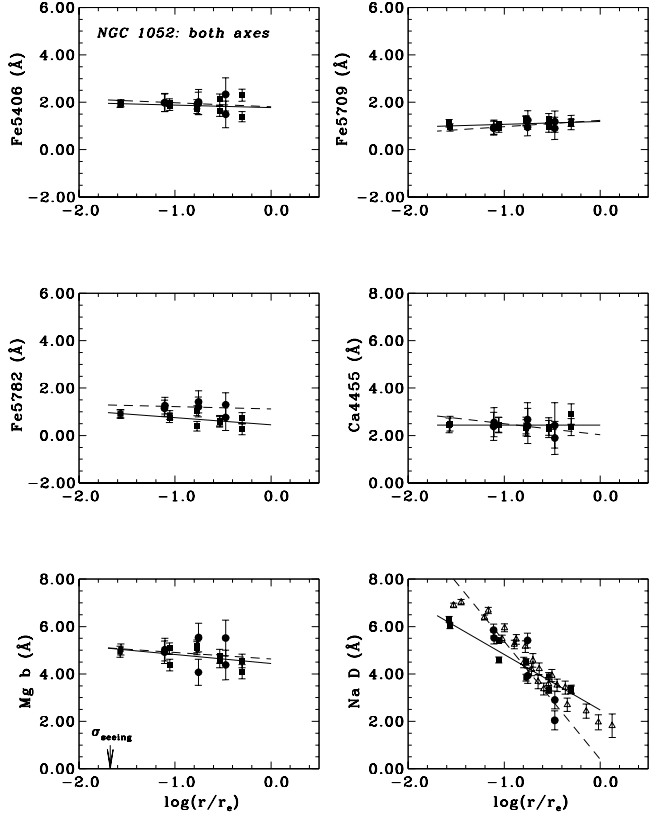}}
\caption{Radial gradients of
Fe5406, Fe5709, Fe5782, Ca4455, Mg b and Na D
along both axes of NGC\,1052.
The notation is the same of Fig. 5.
The Na D data of
Carollo, Danziger \& Buson (1993) along the E-W direction, 
as open triangles, are also plotted.}
\label{grads2}
\end{figure}
%

%
\begin{figure}[htbp]
\resizebox{\hsize}{!}{\includegraphics{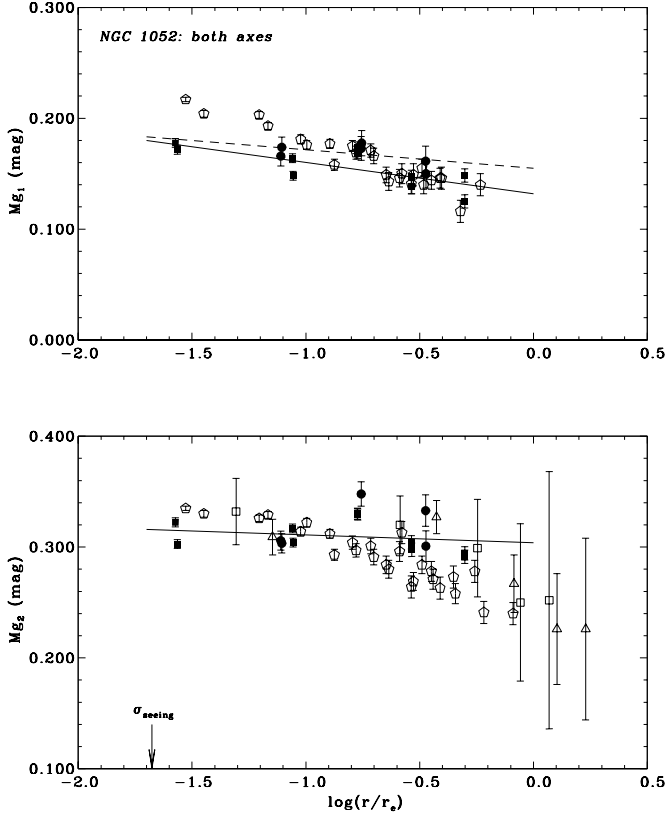}}
\caption{Radial gradients of
Mg$_1$ and Mg$_2$
along both axes of NGC\,1052.
The notation is the same of Fig. 5.
The fit for the minor axis Mg$_2$ data is not plotted.
The Mg$_2$ data of
Couture \& Hardy (1988) along the major and minor axis, 
as open squares and open triangles respectively,
are also plotted (bottom panel).
The Mg$_1$ and Mg$_2$ data of
Carollo, Danziger \& Buson (1993)
along the E-W direction, as open pentagons, are plotted as well (both panels).}
\label{grads3}
\end{figure}
%

%
\begin{figure}[htbp]
\resizebox{\hsize}{!}{\includegraphics{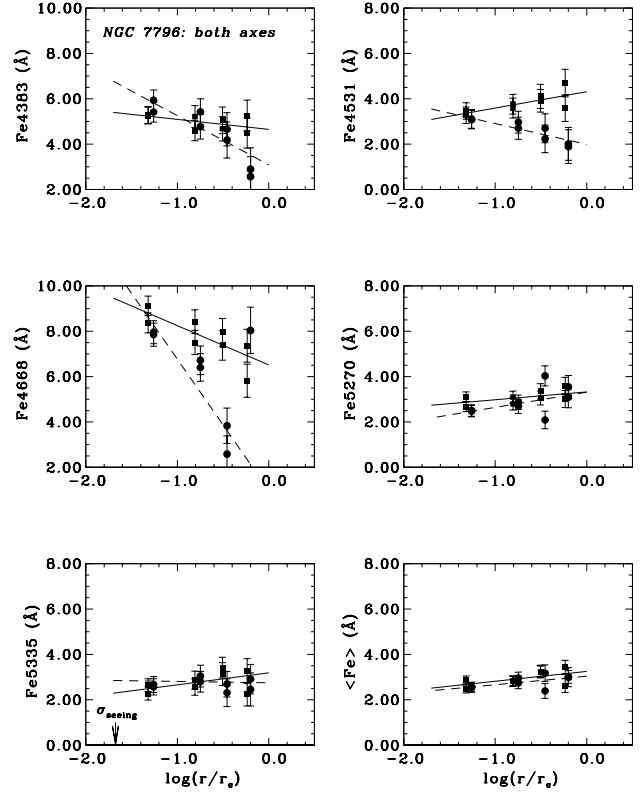}}
\caption{Radial gradients of
Fe4383, Fe4531, Fe4668, Fe5270, Fe5335 and $<$Fe$>$
along both axes of NGC\,7796.
The notation is the same of Fig. 5.}
\label{grads4}
\end{figure}
%

%
\begin{figure}[htbp]
\resizebox{\hsize}{!}{\includegraphics{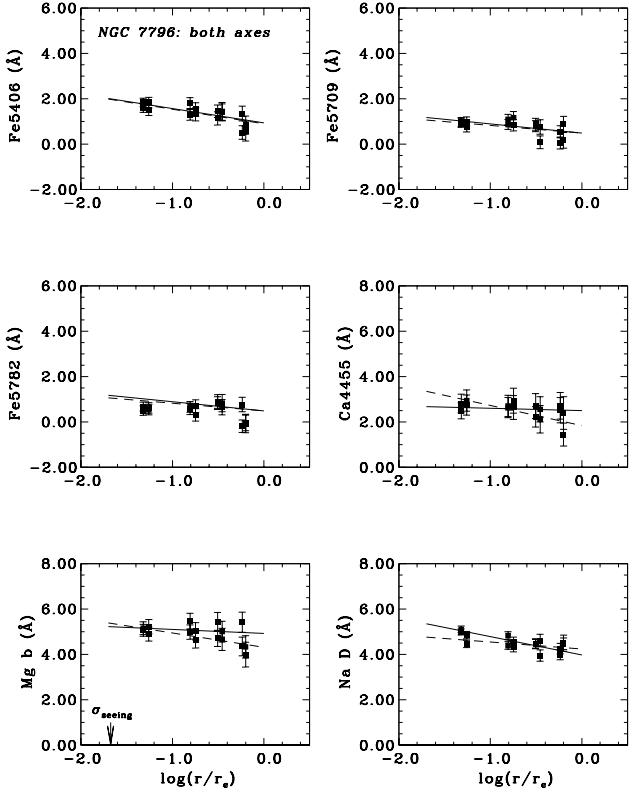}}
\caption{Radial gradients of
Fe5406, Fe5709, Fe5782, Ca4455, Mg b and Na D
along both axes of NGC\,7796.
The notation is the same of Fig. 5.}
\label{grads5}
\end{figure}
%

%
\begin{figure}[htbp]
\resizebox{\hsize}{!}{\includegraphics{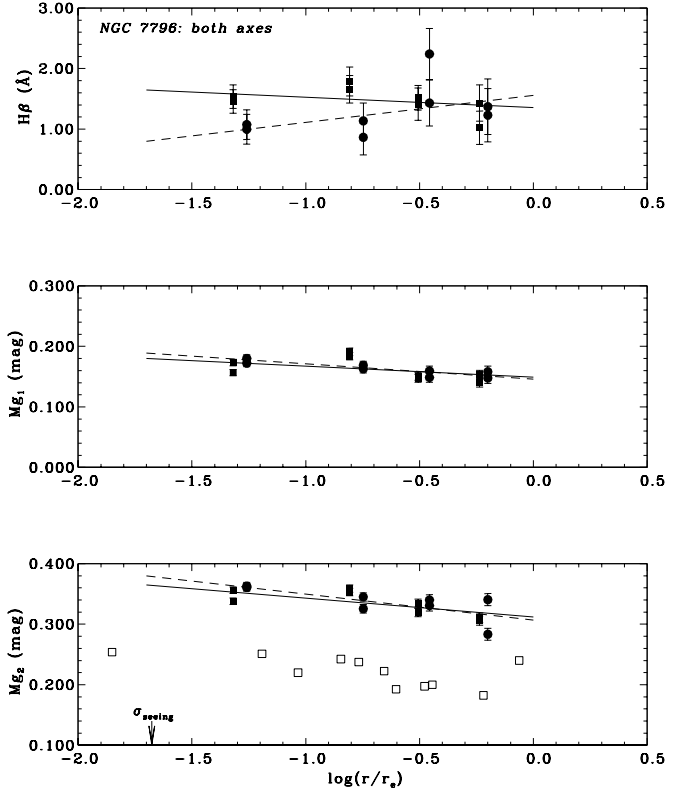}}
\caption{Radial gradients of
H$\beta$, Mg$_1$ and Mg$_2$ along both axes of NGC\,7796.
The notation is the same of Fig. 5.
The Mg$_2$ data of Bertin et al. (1994) along the major, as open squares,
are also plotted (bottom panel).}
\label{grads6}
\end{figure}
%

%
\renewcommand{\tabcolsep}{5.3mm} 
\begin{table*}[htbp]
\caption{Radial gradients along the major and minor axes of both galaxies
respectively for each Lick index.}
\begin{tabular} {l|c c c c }
\hline
\hline
\multicolumn{5}{c}{{\it Gradients  \,and \,Linear constants}} \\
\hline
Lick Index & NGC 1052 &      & NGC 7796 &  \\
\hline
           & (\AA.dex$^{-1}$) & (\AA)
           & (\AA.dex$^{-1}$) & (\AA) \\
Fe4383
& -0.81 $\pm$ 0.24 & 5.16 $\pm$ 0.26 & -0.44 $\pm$ 0.39 & 4.65 $\pm$ 0.38 \\
& -2.04 $\pm$ 0.64 & 3.99 $\pm$ 0.59 & -2.17 $\pm$ 0.50 & 3.09 $\pm$ 0.47 \\
Fe4531
& -0.29 $\pm$ 0.24 & 3.25 $\pm$ 0.26 & +0.72 $\pm$ 0.39 & 4.31 $\pm$ 0.38 \\
& -1.57 $\pm$ 0.64 & 1.70 $\pm$ 0.59 & -0.92 $\pm$ 0.50 & 1.99 $\pm$ 0.47 \\
Fe4668
& -1.34 $\pm$ 0.24 & 6.80 $\pm$ 0.26 & -1.73 $\pm$ 0.39 & 6.52 $\pm$ 0.38  \\
& -5.15 $\pm$ 0.64 & 3.08 $\pm$ 0.59 & -5.80 $\pm$ 0.50 & 0.97 $\pm$ 0.47 \\
Fe5270
& +0.38 $\pm$ 0.24 & 3.00 $\pm$ 0.26 & +0.34 $\pm$ 0.39 & 3.33 $\pm$ 0.38 \\
& -0.67 $\pm$ 0.64 & 1.82 $\pm$ 0.59 & +0.67 $\pm$ 0.50 & 3.32 $\pm$ 0.47 \\
& -0.56 $\pm$ 0.27 & --             & --              & --             \\
Fe5335
& -0.23 $\pm$ 0.24 & 2.20 $\pm$ 0.26 & +0.53 $\pm$ 0.39 & 3.19 $\pm$ 0.38 \\
& -0.98 $\pm$ 0.64 & 1.78 $\pm$ 0.59 & +0.06 $\pm$ 0.50 & 2.75 $\pm$ 0.47 \\
$<$Fe$>$
& +0.07 $\pm$ 0.23 & 2.61 $\pm$ 0.25 & +0.44 $\pm$ 0.39 & 3.26 $\pm$ 0.38 \\
& -0.82 $\pm$ 0.64 & 1.80 $\pm$ 0.59 & +0.37 $\pm$ 0.73 & 3.04 $\pm$ 0.69 \\
Fe5406
& -0.11 $\pm$ 0.24 & 1.77 $\pm$ 0.26 & -0.63 $\pm$ 0.39 & 0.94 $\pm$ 0.38 \\
& -0.17 $\pm$ 0.64 & 1.81 $\pm$ 0.59 & -0.64 $\pm$ 0.50 & 0.90 $\pm$ 0.47 \\
Fe5709
& +0.12 $\pm$ 0.24 & 1.19 $\pm$ 0.26 & -0.50 $\pm$ 0.50 & 0.41 $\pm$ 0.53 \\
& +0.26 $\pm$ 0.64 & 1.23 $\pm$ 0.59 & -0.34 $\pm$ 0.50 & 0.49 $\pm$ 0.47 \\
Fe5782
& -0.30 $\pm$ 0.24 & 0.45 $\pm$ 0.26 & -0.41 $\pm$ 0.39 & 0.49 $\pm$ 0.38 \\
& -0.10 $\pm$ 0.64 & 1.12 $\pm$ 0.59 & -0.34 $\pm$ 0.50 & 0.49 $\pm$ 0.47 \\
Ca4455
&  0.00 $\pm$ 0.49 & 2.44 $\pm$ 0.50 & -0.10 $\pm$ 0.80 & 2.51 $\pm$ 0.74 \\
& -0.51 $\pm$ 1.08 & 1.97 $\pm$ 0.95 & -0.89 $\pm$ 0.91 & 1.84 $\pm$ 0.78 \\
Mg\, b
& -0.38 $\pm$ 0.18 & 4.44 $\pm$ 0.19 & -0.18 $\pm$ 0.58 & 4.93 $\pm$ 0.55 \\
& -0.28 $\pm$ 0.89 & 4.63 $\pm$ 0.80 & -0.63 $\pm$ 0.70 & 4.32 $\pm$ 0.64 \\
Na\, D
& -2.34 $\pm$ 0.19 & 2.47 $\pm$ 0.20 & -0.81 $\pm$ 0.31 & 3.98 $\pm$ 0.30 \\
& -4.60 $\pm$ 0.49 & 0.73 $\pm$ 0.45 & -0.31 $\pm$ 0.39 & 4.24 $\pm$ 0.36 \\
& -4.01 $\pm$ 0.14 & --             & --              & --             \\
H$\beta$
& --              & --             & -0.17 $\pm$ 0.40 & 1.35 $\pm$ 0.38 \\
& --              & --             & +0.38 $\pm$ 0.50 & 1.68 $\pm$ 0.47 \\
           & (mag.dex$^{-1}$) & (mag)& (mag.dex$^{-1}$) & (mag) \\
Mg$_1$
& -0.028$\pm$0.004 & 0.132$\pm$0.004 & -0.018$\pm$0.011 & 0.149$\pm$0.010 \\
& -0.015$\pm$0.016 & 0.162$\pm$0.014 & -0.024$\pm$0.013 & 0.146$\pm$0.011 \\
& -0.074$\pm$0.004 & --             & --              & --             \\
Mg$_2$
& -0.007$\pm$0.004 & 0.304$\pm$0.004 & -0.031$\pm$0.011 & 0.312$\pm$0.010 \\
& +0.073$\pm$0.017 & 0.400$\pm$0.016 & -0.043$\pm$0.013 & 0.307$\pm$0.012 \\
& -0.074$\pm$0.004 & --             & --              & --             \\
& -0.061$\pm$0.019 & --             & --              & --             \\
& -0.068$\pm$0.029 & --             & --              & --             \\
\hline
\end{tabular}
\begin{list} {}
\item Columns 3 and 5: the linear regression constant of the fitting
that represents the Lick index at one effective radius.
\item For Fe5270, Na D, Mg$_1$ and Mg$_2$, the third line shows
their gradients measured in NGC\,1052 by
Carollo, Danziger \& Buson (1993).
\item For Mg$_2$ only, the fourth and fifth lines show the gradients
measured respectively along the major and minor axis of NGC\,1052
by Couture \& Hardy (1988).
\end{list}
\label{grads}
\end{table*}

For {\bf NGC\,1052},
the majority of the gradients of the iron indices
are practically zero considering their errors:
Fe4531, Fe5335, $<$Fe$>$, Fe5406 and Fe5709 along the major axis
and Fe5270, Fe5406, Fe5709 and Fe5782 along the minor axis.
One of them is positive: that of Fe5270 at major axis.
The Mg b, Mg$_1$ and Mg$_2$ gradients are actually negative.
The Mg$_2$ minor axis gradient
was seriously affected by the emission line corrections
for the aperture spectrum at r = 4.92 arcsec 
(see Tables 12 and 18).
However, the Mg$_1$ and Mg$_2$ gradients are less stronger than
the average ones for ellipticals
(-0.038 mag.dex$^{-1}$ as denoted by
Jorgensen 1997) and
the gradients of other Fe indices are still negative.
The Ca4455 gradient is also zero along both axes.
The Na D gradient is strongly negative at both axes
and it is greater than the Mg b one.
The Na D is probably contaminated
by the interstellar absorption of NGC\,1052
which should increase inwards.
If Fe4383 and Fe4668
are actually contaminated by CH and C$_{2}$ lines respectively,
we would propose that might exist a negative radial gradient
of the Carbon stellar abundance in NGC\,1052.

For {\bf NGC\,7796},
the gradients of the Lick indices along the major axis are equal
to the other ones along the minor axis for the majority of them.
The exceptions are for Fe4383, Fe4531, Fe4668 and marginally H$\beta$.
Considering both directions and the uncertainties of the gradients,
the majority of them for the iron indices are zero or positive
(Fe4531, Fe5270, Fe5335, $<$Fe$>$, Fe5709 and Fe5782),
while the Mg b gradient is zero
and the Mg$_1$ and Mg$_2$ gradients are negative.
However, the gradients of Fe4383, Fe4668 and Fe5406 are still negative.
The Ca4455 gradient is zero.
The Na D gradient is negative.
The Na D can be contaminated by own internal interstellar absorption
of NGC\,7796 like NGC\,1052.
Due to the contribution of molecular lines in Fe4383 and Fe4668,
a negative gradient of the Carbon stellar abundance can also be proposed for NGC\,7796.
The mean H$\beta$ gradient is also zero.

In order to derive the radial variations of the abundance ratios,
we have analyzed the index variations directly on the index-index planes,
where the direction of the vector age-metallicity-abundance ratio can be modeled properly
(see Sect. 5).
The stellar population synthesis helps to disentangle
the age-metallicity degeneracy of the Lick indices as well
(see Sect. 6).

%
%
\subsection {External comparisons of some radial gradients}

Couture \& Hardy (1988) have measured the Mg$_2$ gradient
along both axes of NGC\,1052 and their data are plotted
in comparison with ours in Figure 7.
The agreement is marginal because the errors of the
Couture \& Hardy (1988)'s
measurements are greater.
The Mg$_2$ gradients measured by them are different from ours (see Tab. 18).
The same exists when a comparison is made with the Mg$_2$ gradient measured by
Carollo, Danziger \& Buson (1993)
along the E-W direction.
Carollo et al. (1993)
have also computed the gradients
of Fe5270, Na D and Mg$_1$ along the same direction.
Despite some agreement for the radial variation of them in the plots of
Fig. 5 (Fe5270), Fig. 6 (Na D) and Fig. 7 (Mg$_1$ and Mg$_2$),
the computed gradients are different of ours (Tab. 18).
For Mg$_1$ and Mg$_2$ only,
Carollo et al. (1993) should not have applied the emission line corrections.

No radial gradient of Lick index has already been quantified for NGC\,7796.
Bertin et al. (1994)
have measured the radial profile of Mg$_2$
along the major axis whose direct comparison with our data
is presented in Figure 10.
There is a systematic difference between the
Bertin et al. (1994)'s
data and our measurements ($\sim$ -0.113 mag)
although the Mg$_2$ computed gradient using their data
is very near to our result: -0.031 $\pm$ 0.013 mag.dex$^{-1}$.
Note that the predicted Mg$_{2}^{0}$ for NGC\,7796
by the Mg$_{2}^{0}$-$\sigma_{v}^{0}$ relation of
Bender, Burstein \& Faber (1993)
is greater than that one using the observations of
Bertin et al. (1994).

%
%
\section{Comparisons with single-aged stellar population models}

The single-aged stellar population (SSP) models of
Thomas, Maraston \& Bender (2003a), hereafter as TMB (2003a),
have been adopted in order to make predictions about the spatial distribution
of the stellar populations inside the observed regions of each galaxy
in terms of chemical abundances and age.
We have made direct comparisons
of the measured Lick indices of each aperture spectrum
with the theoretical ones of these SSP models.

The SSP models of
TMB (2003a)
are built using the evolutive stellar population synthesis of
Maraston (1998).
The main characteristic of the code of
TMB (2003a)
is to compute the influence of the abundance variations on the Lick indices,
although this code uses the evolutive models of
Maraston (1998)
which are essentially based on the solar abundance ratios.
The Lick indices of these SSP models are computed
adopting the stellar fitting functions of
Worthey (1994)
and the response functions of
Tripicco \& Bell (1995).
The abundance variations are only considered
for the alpha-elements like Magnesium.

The TMB (2003a)'s SSP models assuming the IMF of Salpeter (1955)
have been employed.
We have used the SSP models with five ages
(2, 6, 10, 12 and 15 Gyr),
six global metallicities
([Z/Z$_{\odot}$] = -2.25, -1.35, -0.33, 0.00, +0.35, +0.67 dex)
and four alpha-element to iron ratios
([$\alpha$/Fe] = -0.3, 0.0, +0.3, +0.5 dex).

We have plotted, on planes Lick index versus Lick index,
our data of all aperture spectra of NGC\,1052
and NGC\,7796 with the theoretical ones of TMB's SSP models.

For NGC\,1052, Figs. 11a/b, 12a/b and 13a/b
show the Mg b index plotted versus some iron index:
Mg b versus Fe4531 in Fig. 11a, Mg b versus Fe5270 in Fig. 11b,
Mg b versus Fe5335 in Fig. 12a, Mg b versus Fe5406 in Fig. 12b,
Mg b versus Fe5709 in Fig. 13a and Mg b versus Fe5782 in Fig. 13b.
Figures 14a and 14b show Mg$_1$ versus $<$Fe$>$ and
Mg$_2$ versus $<$Fe$>$ respectively.
Figure 15 shows Na D versus $<$Fe$>$.

The Ca4455, Fe4383 and Fe4668 indices were not adopted because
they are blended with other absorption lines:
Ca4455 with atomic lines of Fe and Cr as
TMB (2003a),
and Fe4383 and Fe4668 with molecular lines
of C$_2$ and CH respectively.

For NGC\,7796, Figs. 17a/b, 18a/b and 19a/b
show respectively the same plots of Mg b index versus an iron index
as for NGC\,1052.
Figures 20a and 20b show Mg$_1$ versus $<$Fe$>$ and
Mg$_2$ versus $<$Fe$>$ respectively.
Figure 21 shows Na D versus $<$Fe$>$ respectively.

The plots present an element abundance indicator
(Mg b, Mg$_1$ and Mg$_2$ for Magnesium and Na D for Sodium)
versus an iron abundance indicator
(Fe4531, Fe5270, Fe5335, Fe5408, Fe5709, Fe5782 and $<$Fe$>$).
In all these plots,
the predicted Lick indices of the SSPs are localized in four curve families,
each one with a different $\alpha$-element/Fe ratio.
Each curve family presents Lick indices of the SSPs
with five ages and six metallicities
that are distributed with very small dispersion
in some restricted area
(the age-metallicity degeneracy of Lick indices).
The present index errors correspond to an mean imprecision of $\pm$ 0.1 dex
for the [$\alpha$/Fe] determination in that plots.

Figure 16 shows $[MgFe]_{TMB}$ modified Lick
(TMB 2003a)
index versus H$\beta$ for NGC\,1052
using the data of the more external spectra only.
Figure 22 presents the same for NGC\,7796 using all data.
Both figures are commented ahead.
The $[MgFe]_{TMB}$ index of
TMB (2003a)
is a metallicity indicator:
$[MgFe]_{TMB}$ 
= $\sqrt{Mg\,b \times (0.72 \times Fe5270 + 0.28 \times Fe5335)}$.
This index is insensitive to the $\alpha$/Fe ratio.
The H$\beta$ Lick index is a stellar age indicator
for SSPs with 2 Gyr or more.

For {\bf NGC\,1052}, we have obtained the following stellar population analysis
using the Lick indices of all extracted spectra along both photometric axes.

The stellar populations inside the observed region
of {\bf NGC\,1052} present a variable alpha-enhancement
(maybe between the solar ratio and [$\alpha$/Fe] = +0.5 dex).
The luminosity-weighted mean global metallicity
seems to change in the interval
from [Z/Z$_{\odot}$] = 0.0 up to +0.35 dex.
None information about age can be obtained from these plots.
All plots of Mg b vs. an iron index
do not show a monotonic radial dependency for the $\alpha$/Fe ratio.
On the other hand,
the data of the central extracted spectrum
indicate a little oversolar $\alpha$/Fe ratio
(luminosity-averaged [$\alpha$/Fe] $\approx$ +0.2 $\pm$ 0.2 dex)
or nearly solar taking into account its error.

Therefore, the timescale of the star formation in the nucleus was 1 Gyr at least
if we adopt the relation between the $\alpha$/Fe ratio and timescale from   
Thomas et al. (2005); see Section 7.
Nearly the same timescale is obtained for the observed region
along the minor axis direction, i.e. the bulge.
However, the strong spread of the Mg/Fe ratio along the major axis or disc
directly shows a great dispersion of the star formation timescale as well
based on an unknown galaxy formation process.
Therefore, in the disc of NGC\,1052 the chemical enrichment
of $\alpha$-elements by SN-II relative to the iron elements by SN-Ia
should has happened in a heterogeneous way.
None information about the star ages can be obtained 
because the H$\beta$ index is completely modified by nebular emission of this Liner.

The relation between the global metallicity, the iron abundance and the $\alpha$/Fe ratio by
Salaris, Chieffi \& Straniero (1993),
immediately below,
can be used to estimate one of them as a function of the other ones
and to understand the interconnected variations of these quantities.
For example, a constant value of the global metallicity with
the increasing of the $\alpha$/Fe ratio together with the decreasing of the iron abundance,
and a decreasing of Z due basically to a decreasing of $\alpha$/Fe
are explained for this relation.

$$ [Z/Z_{\odot}] = [Fe/H] + log(0.638 \times 10^{[\alpha/Fe]} + 0.362) $$

In the next paragraphs,
we show the analysis in each index versus index plot for NGC\,1052.

%
\begin{figure}[htbp]
\resizebox{\hsize}{!}{\includegraphics{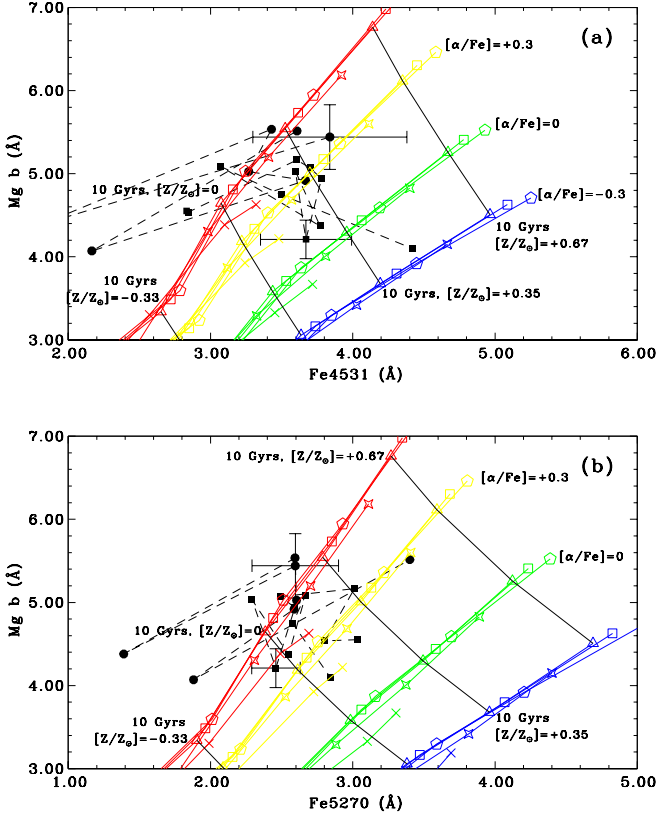}}
\caption{{\bf a)} Mg b versus Fe4531 plot (top panel);
{\bf b)} Mg b versus Fe5270 plot (bottom panel):
comparisons of Lick indices of the long slit spectra of NGC\,1052
(both photometric axes)
with the theoretical ones of TMB (2003a)'s SSPs.
The index values of the spectra are represented
by solid black squares for the major axis data
and solid black circles for the minor axis data,
which are both connected by traced black lines.
The index erros are only plotted for the central spectrum data.
The SSP predicted index values are represented by open symbols
whose colour are indicating the curve families with different [$\alpha$/Fe]:
-0.3 (blue), 0.0 (green), +0.3 (yellow) and +0.5 dex (red).
The SSP ages are represented by open symbols:
2 (crosses), 6 (four-points stars),
10 (triangles), 12 (squares) and 15 Gyr (pentagons).
The global metallicities,
[Z/Z$_{\odot}$] =  -2.25, -1.35, -0.33, 0.00, +0.35, +0.67 dex,
increase from the left-bottom to the right-up
for a given [$\alpha$/Fe] family curve.
The predicted Lick indices of the SSPs with 10 Gyr
and different [$\alpha$/Fe] are connected by solid black lines
for an specific metallicity.}
\label{indxindN1052_1}
\end{figure}
%

%
\begin{figure}[htbp]
\resizebox{\hsize}{!}{\includegraphics{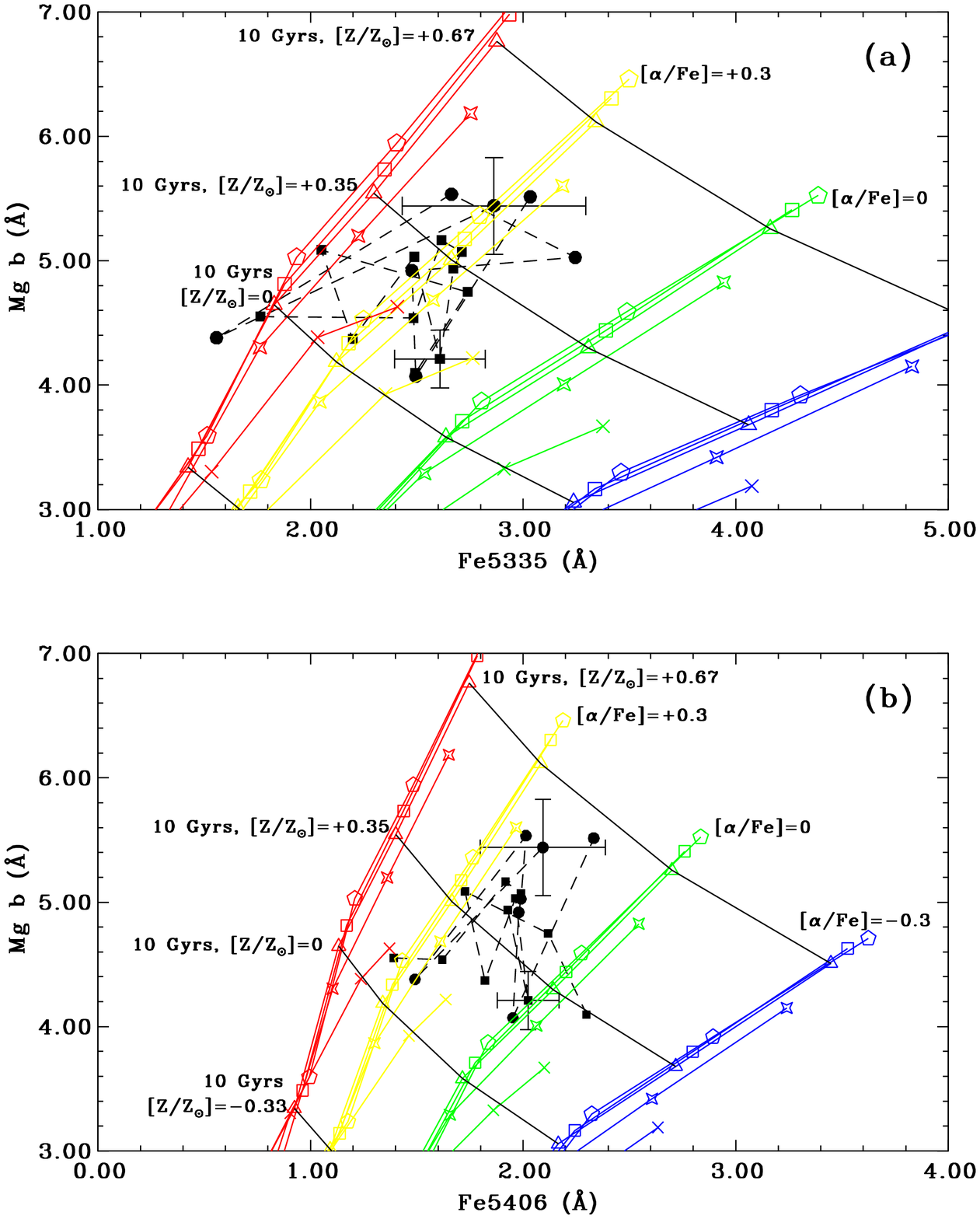}}
\caption{{\bf a)} Mg b versus Fe5335; Mg b versus Fe5406 {\bf b)}:
comparisons of the observed Lick indices of the aperture spectra of NGC\,1052
with the theoretical ones of TMB (2003a)'s SSPs.
The notation is the same of Fig. 11.}
\label{indxindN1052_2}
\end{figure}
%

%
\begin{figure}[htbp]
\resizebox{\hsize}{!}{\includegraphics{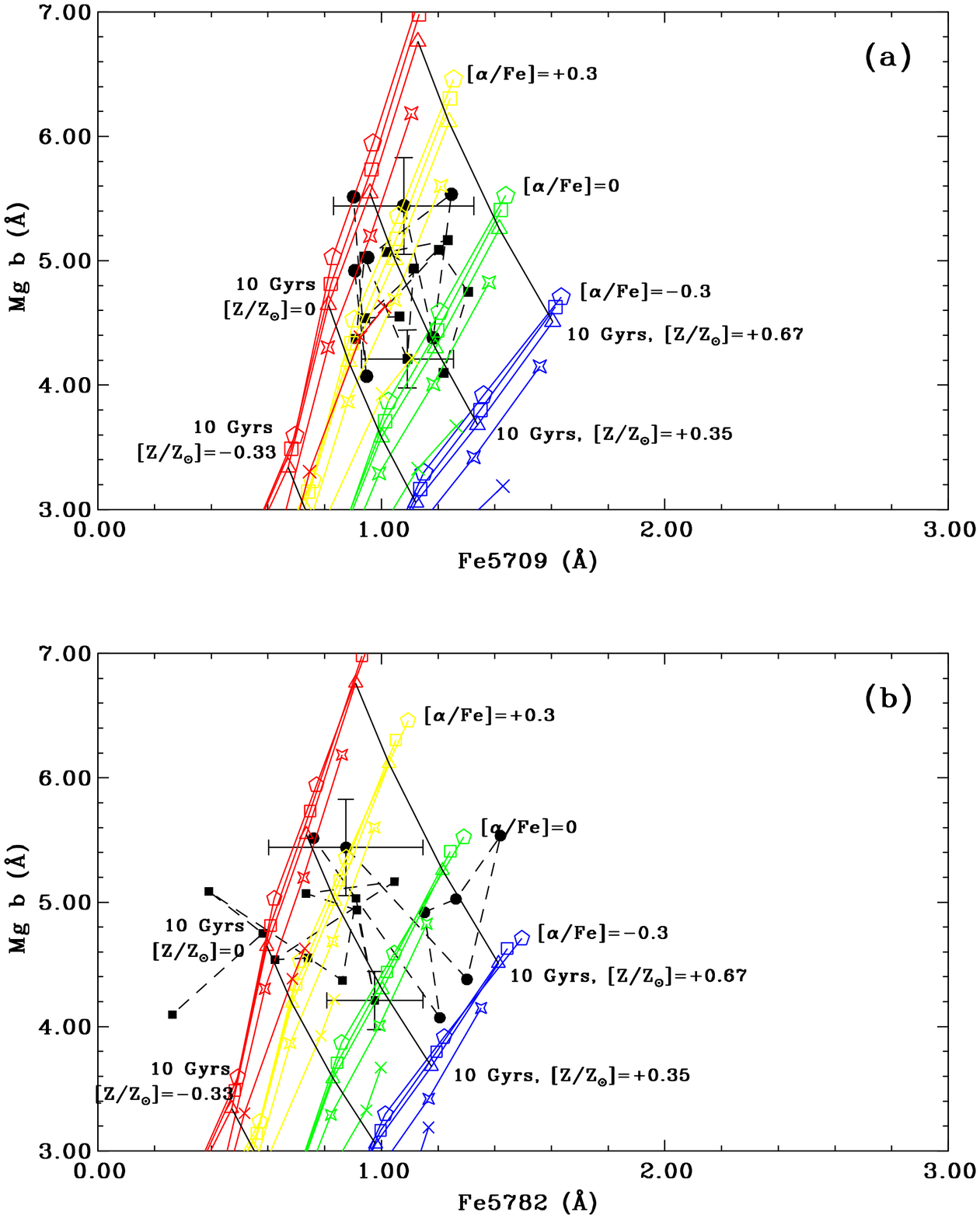}}
\caption{{\bf a)} Mg b versus Fe5709; {\bf b)} Mg b versus Fe5782:
comparisons of the observed Lick indices of the aperture spectra of NGC\,1052
with the theoretical ones of TMB (2003a)'s SSPs.
The notation is the same of Fig. 11.}
\label{indxindN1052_3}
\end{figure}
%

%
\begin{figure}[htbp]
\resizebox{\hsize}{!}{\includegraphics{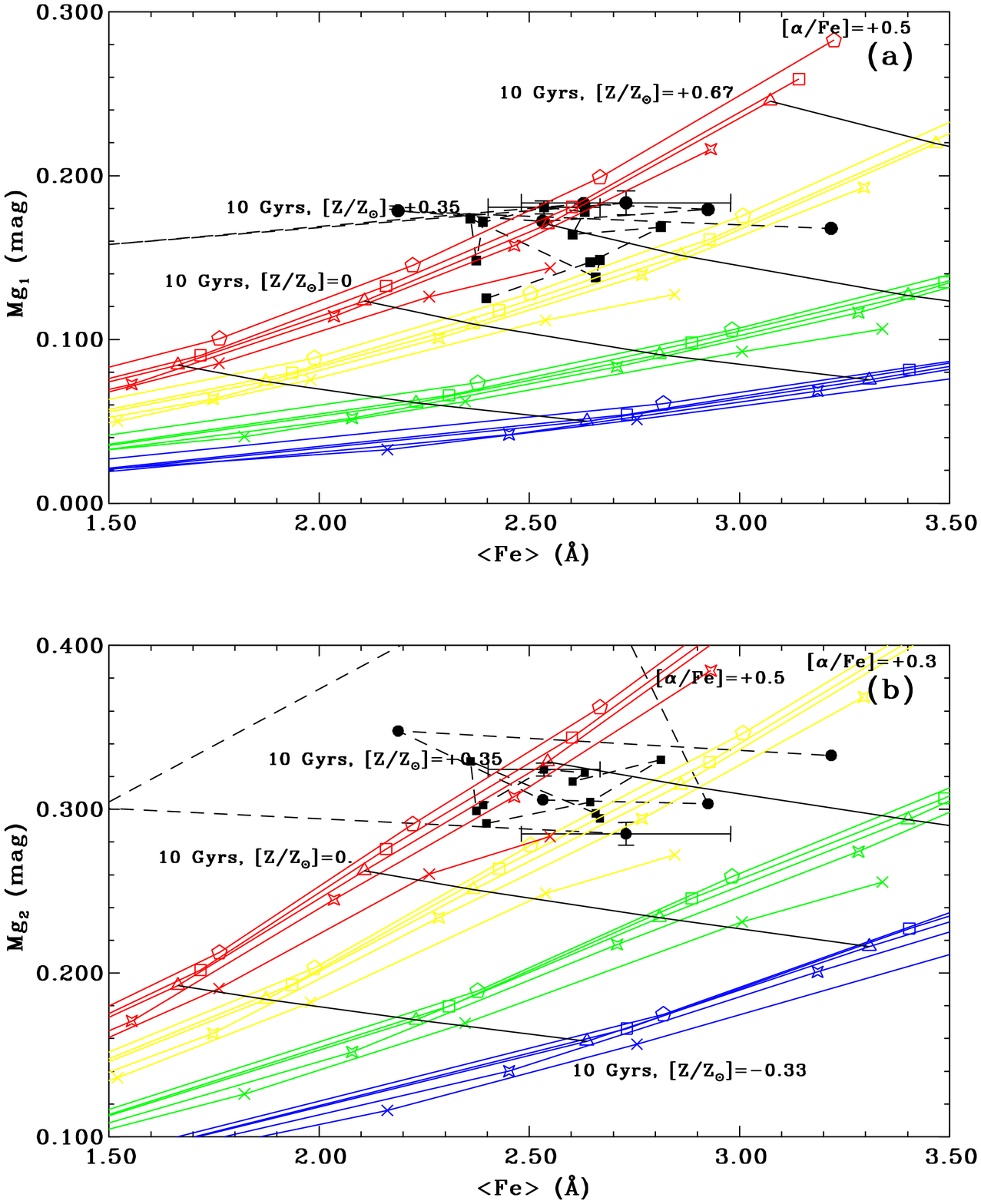}}
\caption{{\bf a)} Mg$_1$ versus $<$Fe$>$; {\bf b)} Mg$_2$ versus $<$Fe$>$:
comparisons of the observed Lick indices of the aperture spectra of NGC\,1052
with the theoretical ones of TMB (2003a)'s SSPs.
The notation is the same of Fig. 11.}
\label{indxindN1052_4}
\end{figure}
%

%
\begin{figure}[htbp]
\resizebox{\hsize}{!}{\includegraphics[angle=-90]{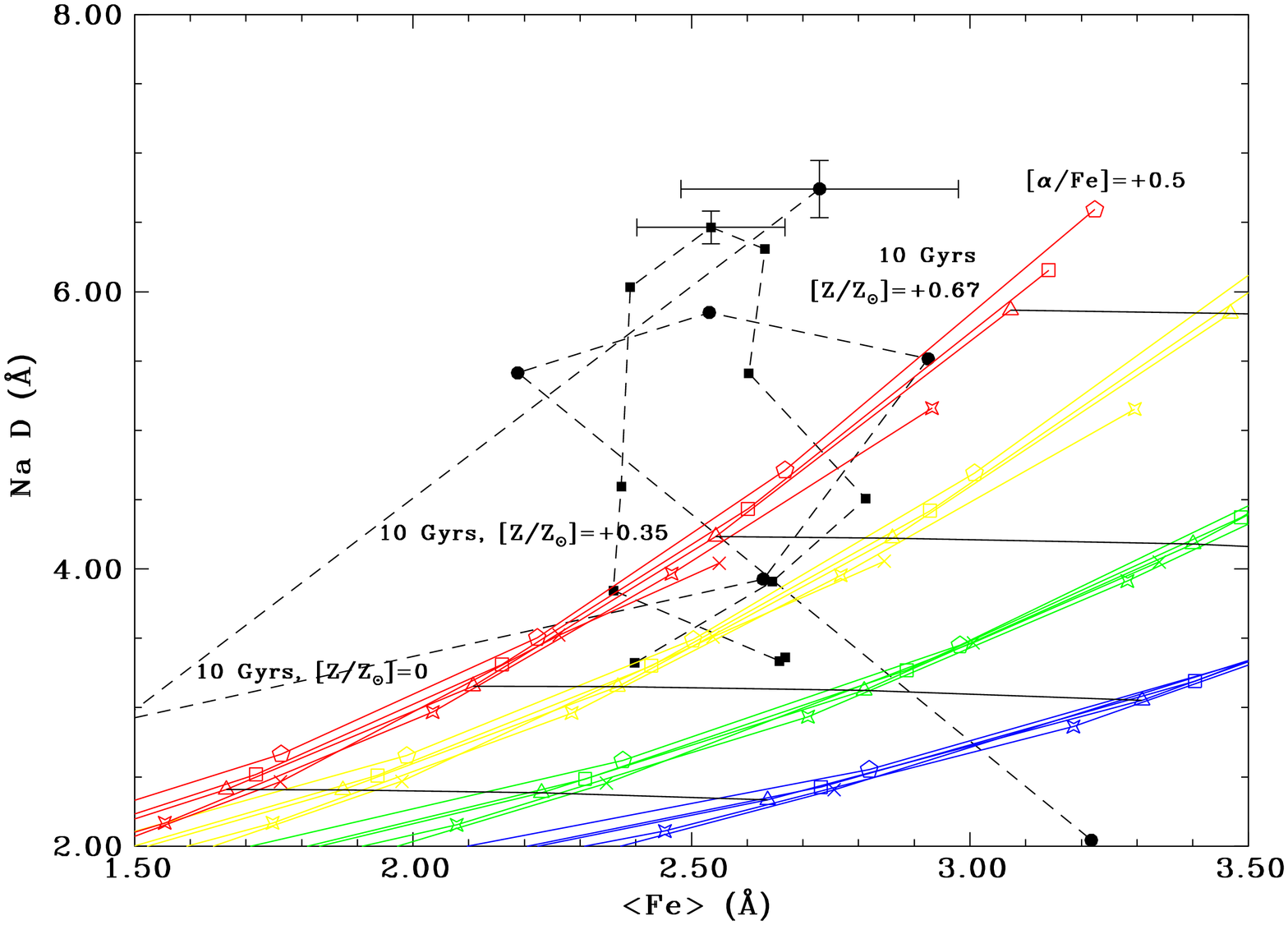}}
\caption{Na D versus $<$Fe$>$:
comparisons of the observed Lick indices of the aperture spectra of NGC\,1052
with the theoretical ones of TMB (2003a)'s SSPs.
The notation is the same of Fig. 11.}
\label{indxindN1052_5}
\end{figure}
%

%
\begin{figure}[htbp]
\resizebox{\hsize}{!}
{\includegraphics[angle=-90]{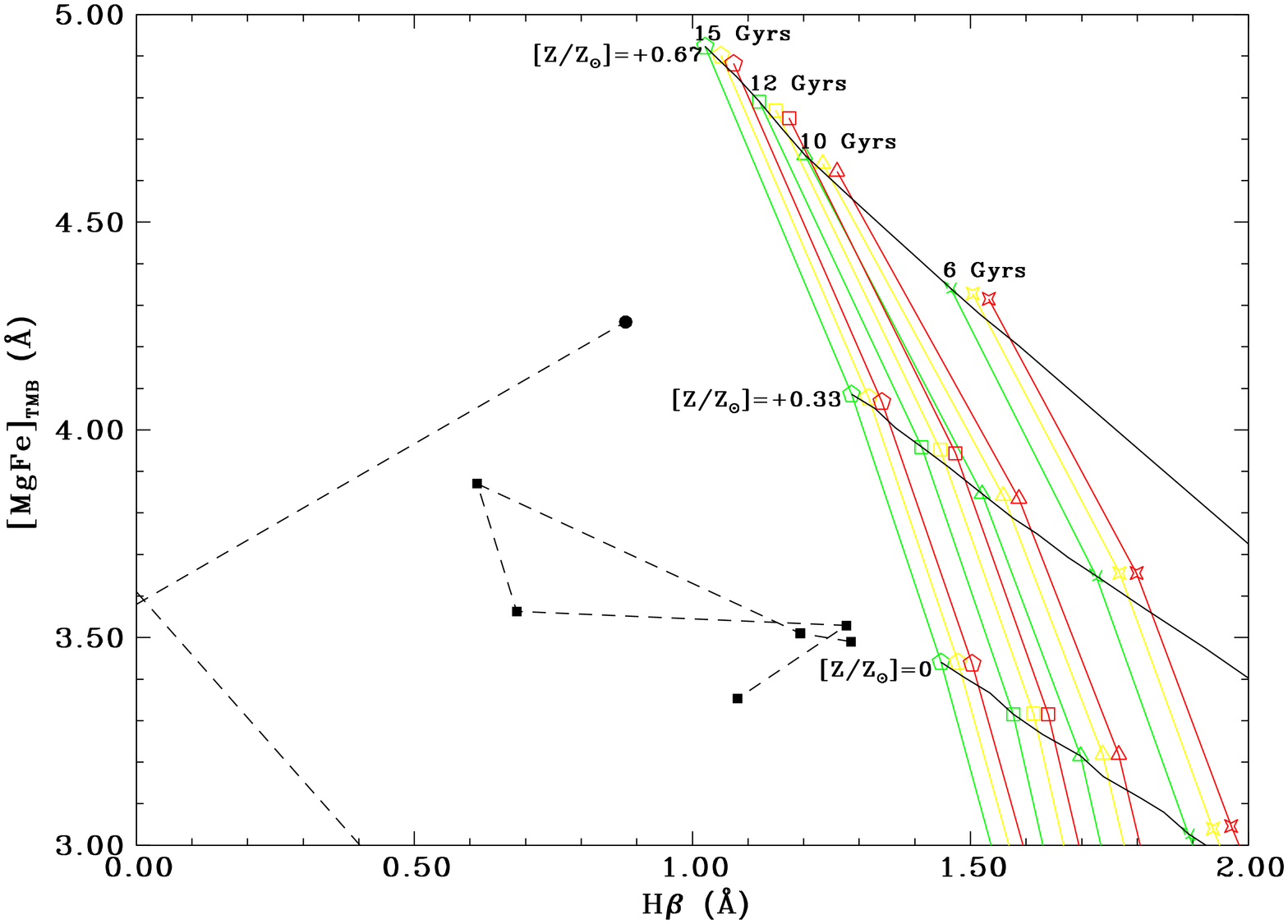}}
\caption{[MgFe]$_{TMB}$ modified Lick index versus H$\beta$ Lick index:
comparison of Lick indices of the long slit spectra of NGC\,1052
with the theoretical ones of TMB (2003a)'s SSPs.
The indices of the aperture spectra are represented by
solid black squares
(major axis data for 6.8 arcsec $\geq$ r $\geq$ 20.2 arcsec) and
solid black circles
(minor axis data for 9.4 arcsec $\geq$ r $\geq$ 18.0 arcsec) 
which are connected by traced black lines like in Fig. 15.
The predicted indices of the SSPs are represented
by open colour symbols like Fig. 15 too
(ages by symbols and [$\alpha$/Fe] by colours),
however the solid black lines are connecting Lick indices of the SSPs
with [$\alpha$/Fe] = 0.0 for each global metallicity covering different ages.
In this plot, the global metallicities increase from the bottom to the top.}
\label{indxindN1052_6}
\end{figure}

The Lick index data obtained along both axes are reasonably distributed
in the same regions of the index vs. index planes.

The Mg b versus Fe4531 plot (Fig. 11a) and
the Mg b versus Fe5782 plot (Fig. 13b)
show the observed Lick indices
of the extracted spectra with distribution basically
between the curve families of the SSPs with [$\alpha$/Fe] = 0.0 and +0.5 dex
at the region of the metallicities [Z/Z$_{\odot}$] = 0.00 and +0.35 dex.

The Mg b versus Fe5270 plot (Fig. 11b) and
the Mg b versus Fe5335 plot (Fig. 12a)
show the indices
of the spectra with distribution between the curves
of the SSPs with [$\alpha$/Fe] = +0.3 (or below) and +0.5 dex
at the region of [Z/Z$_{\odot}$] = 0.00 and +0.35 dex.

The Mg b versus Fe5406 plot (Fig. 12b) and
the Mg b versus Fe5709 plot (Fig. 13a)
show the indices
of the spectra with distribution between the curves
of the SSPs with [$\alpha$/Fe] = 0.0 and +0.3 dex
at the region of [Z/Z$_{\odot}$] = +0.35 dex.

The Mg$_1$ versus $<$Fe$>$ plot (Fig. 14a) and
the Mg$_2$ versus $<$Fe$>$ plot (Fig. 14b)
show the indices
of the spectra with distribution between the curves
of the SSPs with [$\alpha$/Fe] = +0.3 and +0.5 dex
at the region of [Z/Z$_{\odot}$] = 0.00 and +0.35 dex.

The Na D versus $<$Fe$>$ plot (Fig. 15) shows the indices
of the spectra with distribution between the curves
of the SSPs with [$\alpha$/Fe] = +0.3 and +0.5 dex (or higher!)
basically at the region of [Z/Z$_{\odot}$] = +0.35 dex;
where the data of more central spectra are localized
above the curve families with highest $\alpha$/Fe ratio.
However, Na is not an $\alpha$-element.
Moreover, the interstellar medium can increase
the values of the Na D index.
Therefore, the overabundance of Sodium relative to Iron
is still an open question.

The $[MgFe]_{TMB}$ versus H$\beta$ plot for NGC\,1052 (Fig. 16)
was made using the data basically along the major axis
(6.8 arcsec $\geq$ r $\geq$ 20.2 arcsec).
The data of NGC\,1052 (r $\geq$ 11.8 arcsce) are concentrated
outside of the region of the theoretical Lick indices of the SSPs,
i.e. in the region of lower values of H$\beta$ than 1 {\AA},
however nearby to the curves with [Z/Z$_{\odot}$] = 0.0 
and the highest ages.
This indicates that there still is a gas emission contribution
in the outer regions on the H$\beta$ line at least.
The stellar ages in these regions are undetermined
through the analysis of this Balmer line.

For {\bf NGC\,7796}, we have made the following stellar population analysis
using the Lick indices of all extracted spectra
along both photometric axes.

The stellar populations inside the observed region
of {\bf NGC\,7796} present a variable high alpha-enhancement
(maybe between the solar ratio and [$\alpha$/Fe] = +0.5 dex).
However, differently from NGC\,1052,
the plots of Mg b versus Fe5270 (Fig. 17b), 
maybe Mg b vs. Fe5335 (Fig. 18a),
Mg$_1$ vs. $<$Fe$>$ (Fig. 20a)
and Mg$_2$ vs. $<$Fe$>$ (Fig. 20b)
show a radial dependency
for the luminosity-weighted mean $\alpha$/Fe ratio
that seems to be greater in the nucleus 
([$\alpha$/Fe] $\approx$ +0.44 $\pm$ 0.09 dex)
than in the outer regions.
The stellar populations in the observed region of NGC\,7796 
have a luminosity-averaged global metallicity between
[Z/Z$_{\odot}$] = 0.0 and +0.35 dex,
or almost around +0.35 dex,
with ages from 6 up to 12 Gyr (see Fig. 22).

Therefore, the timescale of the stellar formation in the nucleus was nearly 0.04 Gyr
adopting the relation of
Thomas et al. (2005); Section 7.
Along both axis directions, there is strong spread of the Mg/Fe ratio 
that seems to have a monotonic radial dependency.
This can be based on some inside-out galaxy formation process
associated together with a possible outwards radial increasing
of the star formation timescale.
Therefore, in NGC\,7796 the chemical enrichment
of $\alpha$-elements (SN-II) relative to the iron elements (SN-Ia)
should be occurred with a radial dependency.

In the following paragraphs,
the analysis in each index-index plane is given.

%
\begin{figure}[htbp]
\resizebox{\hsize}{!}
{\includegraphics{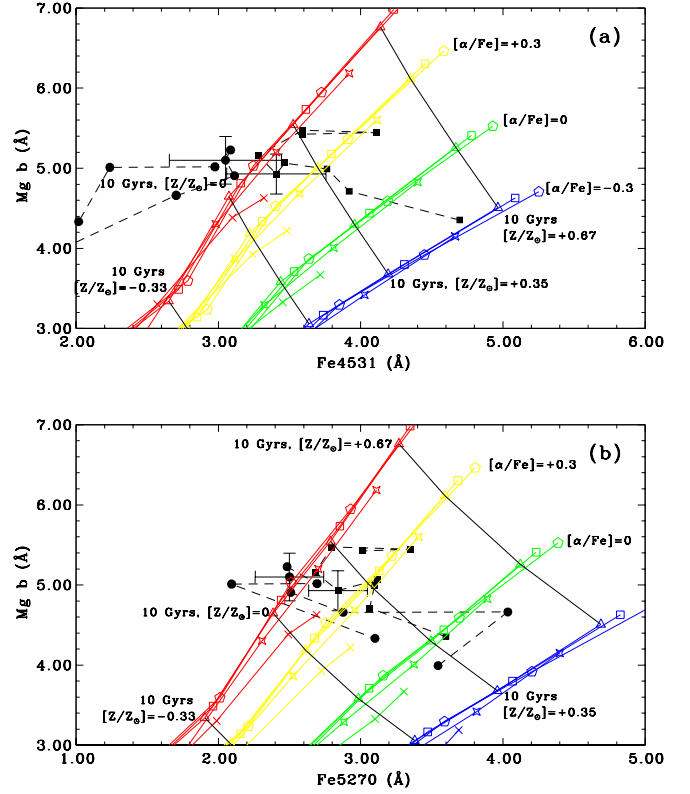}}
\caption{{\bf a)} Mg b versus Fe4531; {\bf b)} Mg b versus Fe5270:
comparisons of the observed Lick indices of the aperture spectra of NGC\,7796
with the theoretical ones of TMB (2003a)'s SSPs.
The notation is the same of Fig. 11.}
\label{indxindN7796_1}
\end{figure}
%

%
\begin{figure}[htbp]
\resizebox{\hsize}{!}
{\includegraphics{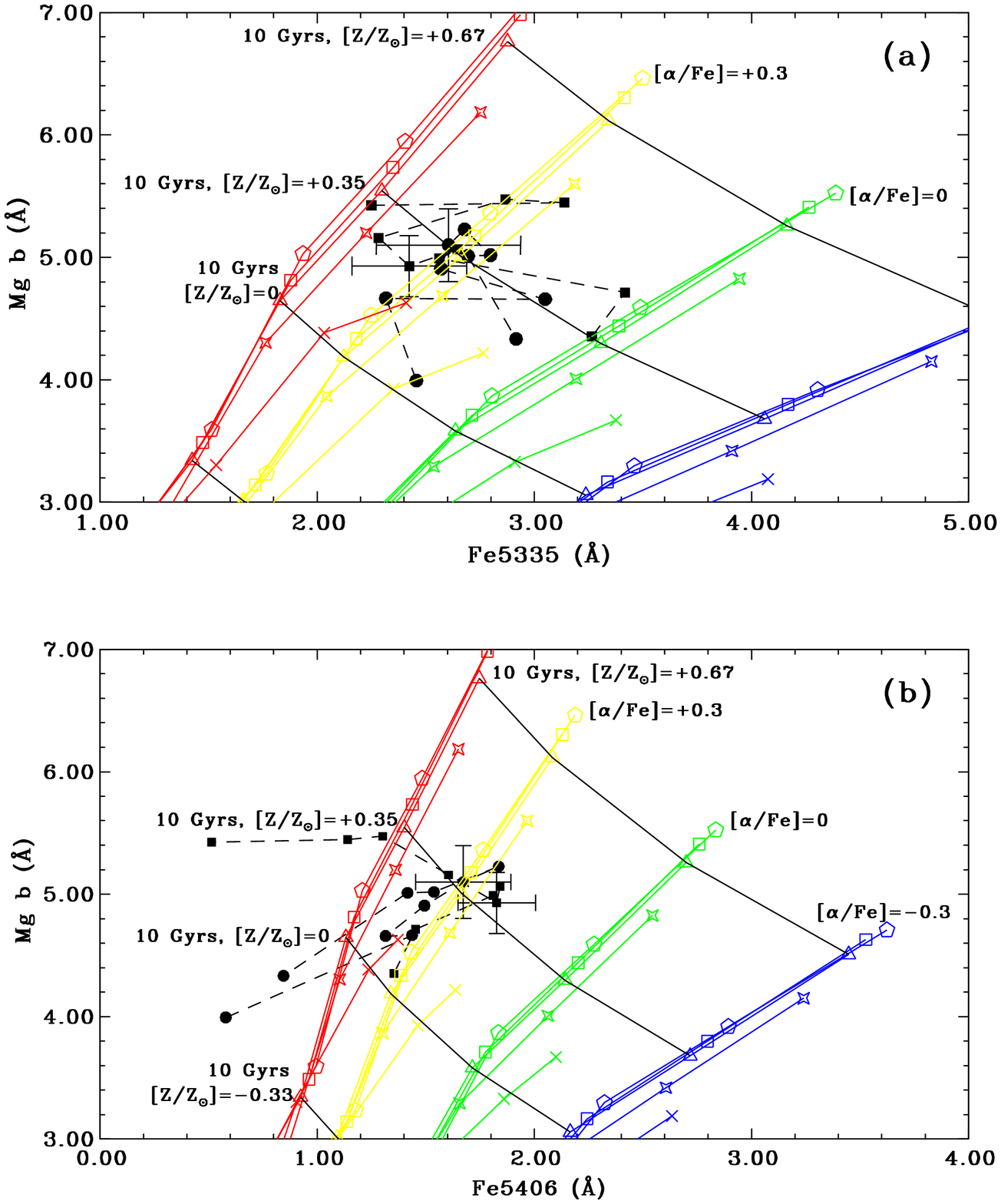}}
\caption{{\bf a)} Mg b versus Fe5335; {\bf b)} Mg b versus Fe5406:
comparisons of the observed Lick indices of the aperture spectra of NGC\,7796
with the theoretical ones of TMB (2003a)'s SSPs.
The notation is the same of Fig. 11.}
\label{indxindN7796_2}
\end{figure}
%

%
\begin{figure}[htbp]
\resizebox{\hsize}{!}
{\includegraphics{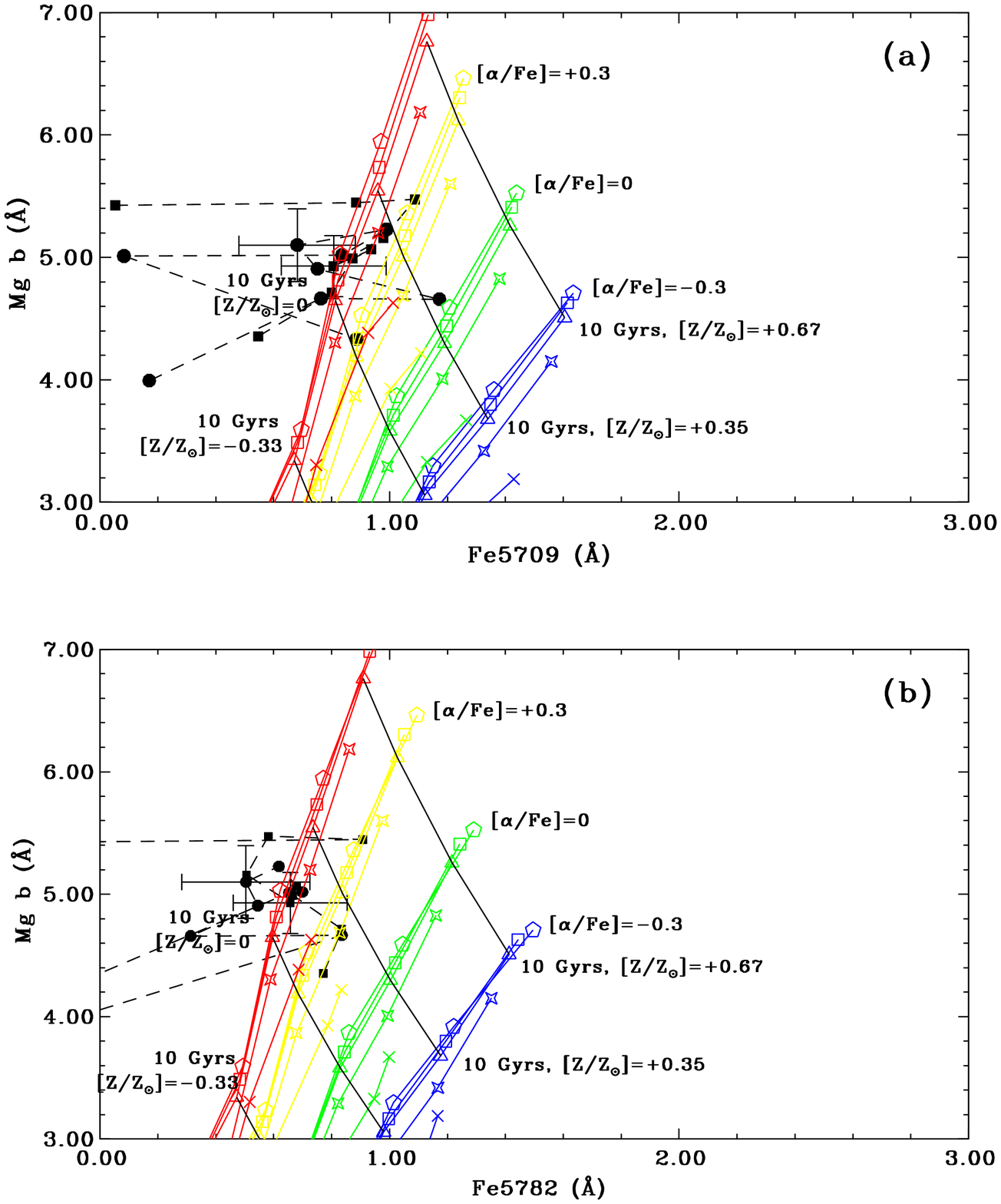}}
\caption{{\bf a)} Mg b versus Fe5709; {\bf b)} Mg b versus Fe5782:
comparisons of the observed Lick indices of the aperture spectra of NGC\,7796
with the theoretical ones
of TMB (2003a)'s SSPs.
The notation is the same of Fig. 11.}
\label{indxindN7796_3}
\end{figure}
%

%
\begin{figure}[htbp]
\resizebox{\hsize}{!}{\includegraphics{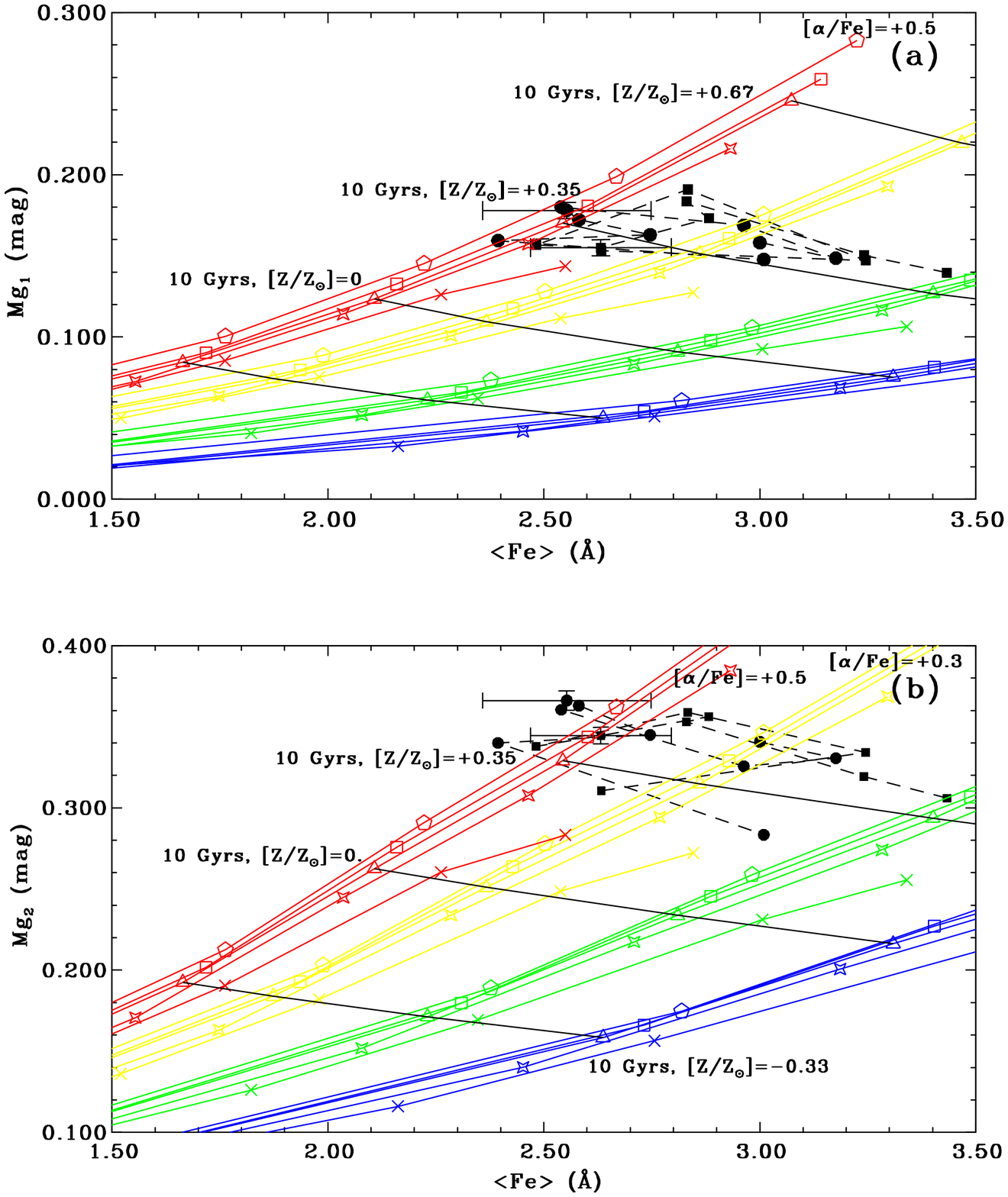}}
\caption{{\bf a)} Mg$_1$ versus $<$Fe$>$; {\bf b)} Mg$_2$ versus $<$Fe$>$:
comparisons of the observed Lick indices of the aperture spectra of NGC\,7796
with the theoretical ones of
TMB (2003a)'s SSPs.
The notation is the same of Fig. 11.}
\label{indxindN7796_4}
\end{figure}
%

%
\begin{figure}[htbp]
\resizebox{\hsize}{!}{\includegraphics[angle=-90]{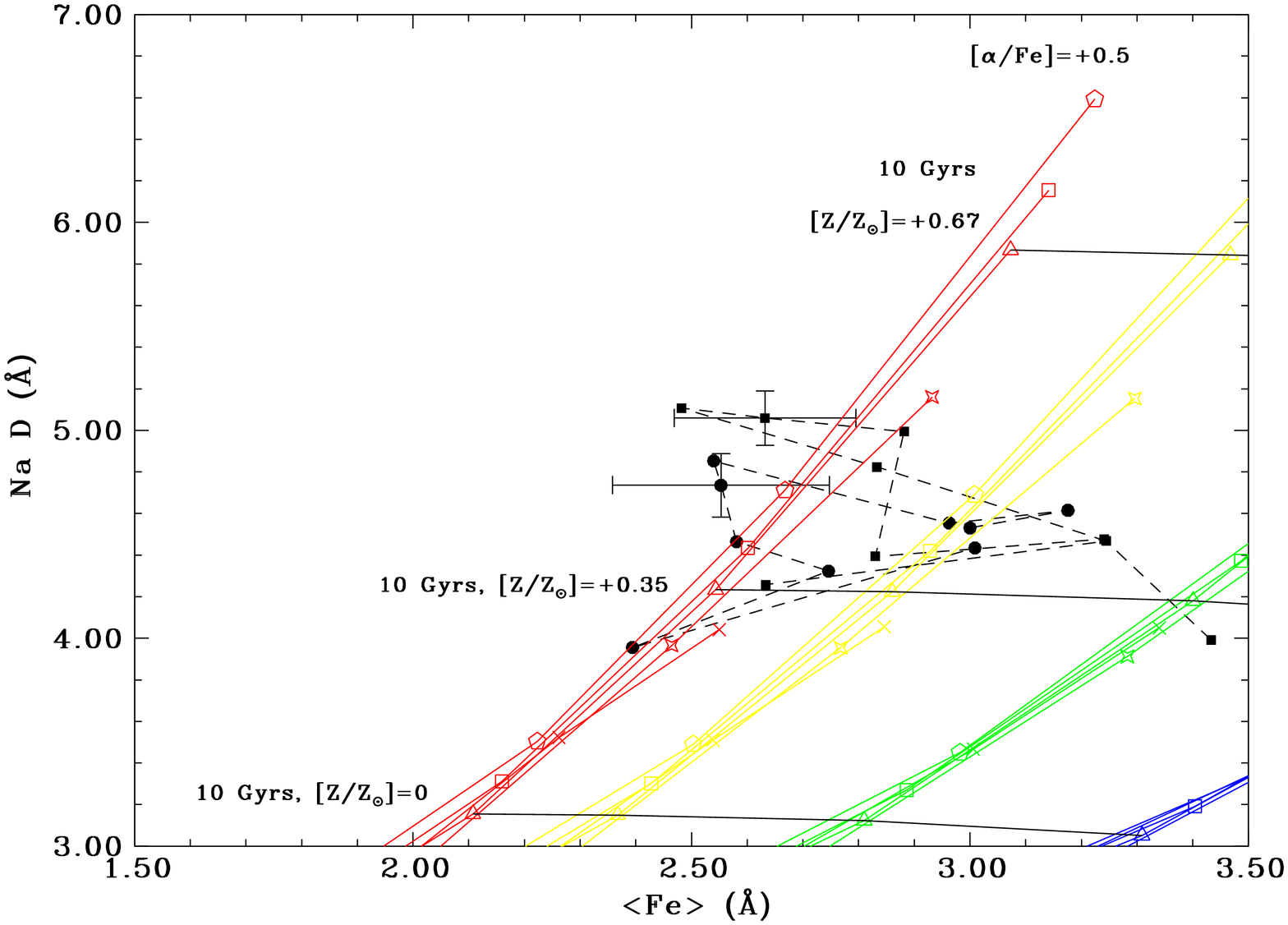}}
\caption{Na D versus $<$Fe$>$:
comparisons of the observed Lick indices of the aperture spectra of NGC\,7796
with the theoretical ones of
TMB (2003a)'s SSPs.
The notation is the same of Fig. 11.}
\label{indxindN7796_5}
\end{figure}
%

%
\begin{figure}[htbp]
\resizebox{\hsize}{!}{\includegraphics[angle=-90]{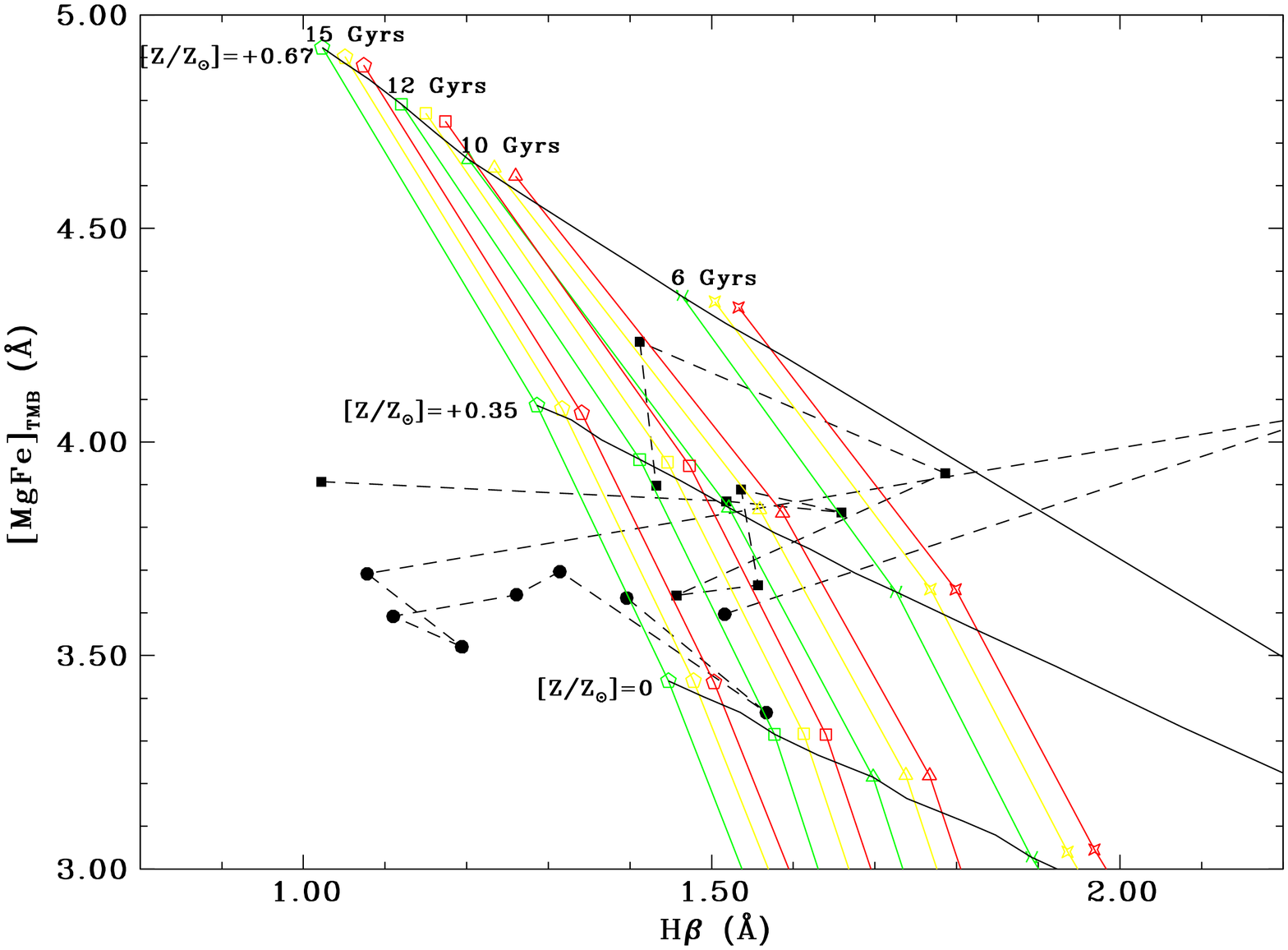}}
\caption{[MgFe]$_{TMB}$ versus H$\beta$:
comparison of Lick indices of the long slit spectra of NGC\,7796
with the theoretical ones of TMB (2003a)'s SSPs.
The notation is the same of Fig 16.
In this plot, the global metallicities increase from the bottom to the top.}
\label{indxindN7796_6}
\end{figure}

In the planes index vs. index,
the data obtained along both axes are reasonably distributed
in the same regions.
For example, the data of both central apertures are close each other.

The Mg b versus Fe4531 plot (Fig. 17a) shows the observed Lick indices
of all extracted spectra with distribution basically
along the curve families of the SSPs with
[$\alpha$/Fe] = 0.00, +0.3 and +0.5 dex
at the region of the metallicities [Z/Z$_{\odot}$] = 0.0 and +0.35 dex;
where the data of the central spectra are localized
around the curve families with highest $\alpha$/Fe ratio
for the major axis data only.
Some minor axis data are outside of the SSP region.

The Mg b versus Fe5270 plot (Fig. 17b) shows the indices
with distribution between the curves of the SSPs
with [$\alpha$/Fe] = 0.0 and +0.5 dex
at the region of [Z/Z$_{\odot}$] = 0.0 and +0.35 dex.
In this plot, the data are spread transversely
to the $\alpha$/Fe family curves so that
the data from the nucleus are located nearby the [$\alpha$/Fe] = +0.5 curve.

The Mg b versus Fe5335 plot (Fig. 18a) shows the indices
with distribution along the curves of the SSPs
with [$\alpha$/Fe] = 0.0, +0.3 and +0.5 dex
basically at [Z/Z$_{\odot}$] = +0.35 dex;
where the data of the more central spectra are quite localized
around the curves with higher $\alpha$/Fe abundance ratios.

The Mg b versus Fe5406 plot (Fig. 18b) shows the indices
with distribution between the curves
of the SSPs with [$\alpha$/Fe] = +0.3 and +0.5 dex
at the region of [Z/Z$_{\odot}$] = 0.00 and +0.35 dex.

The plots of Mg b vs. Fe5709 (Fig. 19a)
Mg b vs. Fe5782 (Fig. 19b)
show the indices with distribution along the curves
of the SSPs with [$\alpha$/Fe] = +0.3 and +0.5
basically in the region of [Z/Z$_{\odot}$] = 0.00 and +0.35 dex.
Some data are outside of the SSP region in both plots.

The Mg$_1$ versus $<$Fe$>$ plot (Fig. 20a) shows the indices
with distribution between the curves
of the SSPs with [$\alpha$/Fe] = 0.0 and +0.5 dex
basically along the [Z/Z$_{\odot}$] = +0.35 dex region.

The Mg$_2$ versus $<$Fe$>$ plot (Fig. 24b) shows the indices
with distribution between the curves
of the SSPs with [$\alpha$/Fe] = 0.0 and +0.5 dex
above and along the [Z/Z$_{\odot}$] = +0.35 dex region.

Both Mg$_1$ and Mg$_2$ (vs. $<$Fe$>$) plots
present the data distributed transversely to the
$\alpha$/Fe curves so that
the data from nucleus are located nearby the
curve of the highest $\alpha$/Fe ratio
and the data of the more external apertures are
near to curves with smaller [$\alpha$/Fe].

The Na D versus $<$Fe$>$ plot (Fig. 21) shows the indices
with distribution between the curves
of the SSPs with [$\alpha$/Fe] = 0.0 and +0.5 dex
basically at [Z/Z$_{\odot}$] = +0.35 dex;
where the data of more spectra are localized
around the curves with the highest $\alpha$/Fe ratio.
However, Na is not an $\alpha$-element.
Therefore, the overabundance of Sodium relative to Iron 
is still questionable.

The $[MgFe]_{TMB}$ versus H$\beta$ plot (Fig. 22)
was made using all data of NGC\,7796.
The data of NGC\,7796 are concentrated around the region of the theoretical
Lick indices of the SSPs with global metallicity [Z/Z$_{\odot}$] = +0.35 dex,
basically including the ages 6 Gyr, 10 Gyr and 12 Gyr
and different [$\alpha$/Fe].

We have noted that the average dispersion of [$\alpha$/Fe] is nearly 0.35 dex
inside the observed regions of both galaxies
(the rotating disc or major axis of NGC\,1052
and both main photometric directions of NGC\,7796). 
This dispersion is still greater than
the error of a single estimation of [$\alpha$/Fe],
which is nearly $\pm$ 0.1 dex,
when only one Lick index versus Lick index plot is adopted.
Few plots (mainly Fig. 11a and Fig. 17a)
have shown part of the data outside the curves grid
whose reason might be basically due to the accuracy of the Lick calibration
(see Tab. 6 for Fe4531 and Proctor et al. 2004).

%
%
\section{Stellar Population Synthesis}

For a good understanding of some properties of NGC\,1052 and NGC\,7796,
such as the presence of central gas emission,
a precise determination of the star formation process is necessary.
Therefore, the age of the constituent stars in these galaxies
is an important parameter to be determined.

The integrated spectrum of a given galaxy contains significant information
on its stellar content and chemical enrichment
(Bica \& Alloin, 1986a).
This information together with a stellar population synthesis method
is used to determine the galaxy star formation history
(Bica \& Alloin, 1986b).
In this work we have employed
the stellar population synthesis method developed by
Bica (1988)
which is based on integrated spectra of star clusters
and \ion{H}{ii} regions,
characterized by different ages and metallicities.
In the present case, we have used seven components in the spectral base
(G1, G2, G3, Y1, Y2, Y3 and RHII).
The old components (G1, G2 and G3) have ages of 10, 13 and 15 Gyr
and metallicities of the
[Z/Z$_{\odot}$] $\sim$ 0.0, $\sim$ -0.4
and $\sim$ -1.1 dex respectively.
The components Y1, Y2 and Y3 have ages of
$\sim$ 10, $\sim$ 25 and $\sim$ 80 Myr
with metallicities of the [Z/Z$_{\odot}$] $\sim$ -0.25,
$\sim$ -0.4 and $\sim$ -0.5 dex respectively.
The HII region represents a stellar population
of 10$^{6}$ years old and with solar metallicity.

Basically the algorithm uses some {\it EW}s and
continuum points measured in a given galaxy spectrum
and compares them to those of a model computed from
a base of simple stellar population elements (see in
Rickes, Pastoriza \& Bonatto 2004).

The corresponding selected {\it EW}s and continuum points for the base elements
have been measured similarly as those for the observed galaxies,
and are given in Tables 19 and 20.
The base spectra are presented in Figure 23.

%
\begin{figure}[htbp]
\resizebox{\hsize}{!}{\includegraphics{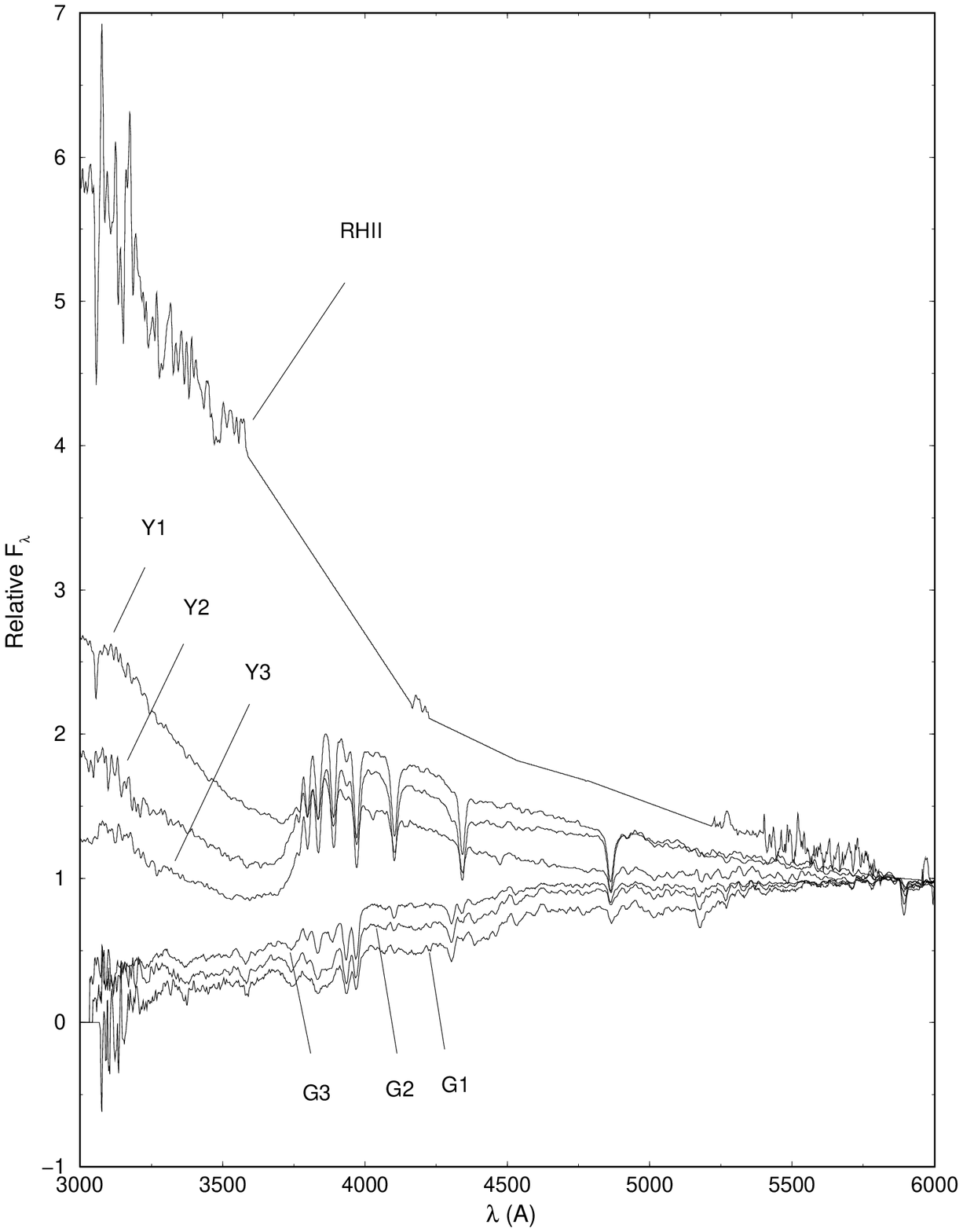}}
\caption{Spectra of the stellar population templates used in the synthesis,
normalized at $\lambda5870$\,\AA.}
\label{base1052}
\end{figure}
%

%
%
\begin{table}[htbp]
\caption{Equivalent widths for the base}
\renewcommand{\tabcolsep}{0.30mm}
\tiny
\begin{tabular} {l|c|c c c c c c c c}
\hline
\hline
\multicolumn{10}{c}{$EW$ (\AA)} \\
\hline
 & Metallicity & & & & & & & & \\
& [Z/Z$_{\odot}$] & Fe4383 & Ca4455 & Fe4531
& Fe5270 & Fe5335 & Fe5406
& Fe5709 & Fe5782  \\
\hline
G1 & 0.00  & 6.00 & 2.76 & 4.06
& 2.77 & 1.65 & 1.50
& 1.36 & 1.14 \\
G2 & -0.4  & 3.28 & 0.94 & 2.94
& 2.99 & 1.70 & 1.46
& 1.30 & 0.80 \\
G3 & -1.1  & 2.42 & 0.40 & 2.10
& 1.59 & 0.95 & 0.77
& 0.57 & 0.52 \\
Y1 & -0.25 & 2.73 & 1.08 & 1.10
& 1.95 & 0.70 & 1.36
& 1.22 & 0.87 \\
Y2 & -0.4  & 2.58 & 0.49 & 0.78
& 1.06 & 0.71 & 0.89
& 0.70 & 0.28 \\
Y3 & -0.5  & 1.78 & 0.57 & 1.13
& 0.88 & 0.76 & 0.53
& 0.43 & 0.33 \\
RHII& 0.00 & 0.00 & 0.00 & 0.00
& 0.00 & 0.00 & 0.00
& 0.00 & 0.00 \\
\hline
\end{tabular}
\label{largbase1052}
\end{table}
%

%
\renewcommand{\tabcolsep}{1.5mm}
\begin{table}[htbp]
\caption{Continuum points for the base}
\tiny
\begin{tabular} {c|c c c c c c c c}
\hline
\hline
\multicolumn{9}{c}{{$C_{\lambda}/C_{5870}$}} \\
\hline
 & $\lambda$4475 & $\lambda$4800 & $\lambda$5100 & $\lambda$5300
 & $\lambda$5546 & $\lambda$5800 & $\lambda$5822 & $\lambda$6200 \\
\hline
G1   & 0.67 & 0.78 & 0.79 & 0.91 & 0.97 & 1.01 & 1.01 & 0.94 \\
G2   & 0.83 & 0.92 & 0.91 & 0.97 & 0.98 & 0.99 & 1.01 & 0.94 \\
G3   & 0.90 & 0.96 & 0.95 & 0.97 & 0.99 & 0.99 & 0.99 & 0.93 \\
Y1   & 1.16 & 1.06 & 1.04 & 1.04 & 1.12 & 1.01 & 1.03 & 0.90 \\
Y2   & 1.52 & 1.34 & 1.22 & 1.17 & 1.04 & 1.04 & 1.03 & 0.90 \\
Y3   & 1.37 & 1.29 & 1.23 & 1.08 & 1.06 & 1.00 & 0.99 & 0.81 \\
RHII & 1.87 & 1.66 & 1.45 & 1.33 & 1.21 & 1.03 & 1.00 & 0.92 \\
\hline
\end{tabular}
\label{contbase1052}
\end{table}
%

%
%
\subsection{Synthesis results}

The synthesis model presents the contribution of each base element
to the flux at $\lambda$5870\,{\AA}.
The computed models along the major and minor axes of NGC\,1052 and NGC\,7796
are given in Tables 21, 22, 23 and 24.
The $E(B-V)_{\rm{i}}$ values are also derived from the synthesis.

For {\bf NGC\,1052},
the old components (G1, G2 and G3)
dominate the $\lambda$5870\,{\AA}\,flux in the more central extractions.
In its nucleus, these components contribute
with 82.\% ($\pm$ 10.\%) of the total flux
(89.\% for the major axis spectral extraction
and 75.\% for the minor axis one),
while at $\sim$ 0.3 r$_{e}^{corr}$ this contribution seems to decrease 
to 71.\% ($\pm$ 8.\%) along the major axis (or the stellar disc)
while it remains the same (80 $\pm$ 10.\%)
along the minor axis (or the bulge).
The young components (Y1, Y2 and Y3)
have a small increasing along the disc from the nucleus outwards:
from 17 $\pm$ 10.\% (center) up to 25 $\pm$ 5.\% (at 0.3 r$_{e}^{corr}$).
Around the bulge the contribution of the young components is constant:
at the 0.3 r$_{e}^{corr}$ it is also 16 $\pm$ 9.\%.
In terms of stellar metallicity,
the synthesis results have been also rebinned
so that the G1+Y1, G2+Y2+Y3 and G3 contributions  
were summed in order to represent respectively the fractions for
[Z/Z$_{\odot}$] $\sim$ 0, $\sim$ -0.4 and $\sim$ -1 dex.
The solar metallicity components dominate the nucleus (68 $\pm$ 3.\%)
and its contribution decreases along the stellar disc
(49 $\pm$ 17.\% at 0.3 r$_{e}^{corr}$)
while it remains constant around the bulge
(65 $\pm$ 6.\% at 0.3 r$_{e}^{corr}$).
The fraction of the G2, Y2 and Y3 components (under-solar metallicities)
is the same in the nucleus and around the bulge (26 $\pm$ 5.\%).
However, it increases along the disc (41 $\pm$ 6.\% at 0.3 r$_{e}^{corr}$).
The extremely metal poor component G3 has a very small contributions
around all observed region:
4 $\pm$ 1.\% at the nucleus,
7 $\pm$ 5.\% at 0.3 r$_{e}^{corr}$ of the minor axis (bulge) and
6 $\pm$ 3.\% at 0.3 r$_{e}^{corr}$ of the disc.
The RHII component does not show any considerable contribution and spread.
The synthesis results of NGC\,1052 can be seen in Tabs. 21-22 (all apertures)
and Figure 24 (nucleus, major axis observation).

The synthesis results for {\bf NGC\,7796} are analogous to the NGC\,1052 one.
Moreover, there is more symmetry of them around the center for both axes
than in NGC\,1052.
The old components dominate in the nucleus
with $\sim$ 82.5 $\pm$ 1.\% of the total flux,
(83.\% for the major axis aperture
and 82.\% for the minor axis one),
while at $\sim$ 0.3 r$_{e}^{corr}$ this contribution decreases 
to $\sim$ 73 $\pm$ 3.\%
(75 $\pm$ 8.\% for the major axis
and 71 $\pm$ 1.\% for the minor one).
The young components Y1, Y2 and Y3
have an outwards radial increasing
(14 $\pm$ 1.\%, at the nucleus, to 23 $\pm$ 4.\% outside).
Along the both axes these components increase
from the central part outwards:
from 13 $\pm$ 3.\% up to 20 $\pm$ 6.\% at 0.3 r$_{e}^{corr}$ (major axis)
and from 15 $\pm$ 8.\% up to 25 $\pm$ 1.\% at 0.3 r$_{e}^{corr}$ (minor axis).
In terms of stellar metallicity,
the synthesis results have been also rebinned
on the same way applied for NGC\, 1052.
The solar metallicity components dominate the nucleus (58 $\pm$ 1.\%)
and its contribution decreases a little along the major axis
(53 $\pm$ 1.\% at 0.3 r$_{e}^{corr}$)
and minor axis as well
(54 $\pm$ 11.\% at 0.3 r$_{e}^{corr}$).
The fraction of the under-solar metallicity component
is quite the same in the nucleus (31 $\pm$ 1.\%)
and along both axes (34 $\pm$ 2.\% at 0.3 r$_{e}^{corr}$).
The extremely metal poor component G3 has a very small contributions
around all observed region:
5 $\pm$ 1.\% at the nucleus,
8 $\pm$ 1.\% at 0.3 r$_{e}^{corr}$ of the major axis and
9 $\pm$ 1.\% at 0.3 r$_{e}^{corr}$ of the minor axis.
Therefore, the template contributions along the major axis
are analogous to these ones along the minor axis, i.e.
there is a homogeneous stellar population spatial distribution
along both directions.
The RHII component does not show any considerable contribution.
The synthesis results of NGC\,7796 are given in Tabs. 23-24 (all apertures)
and Figure 25 (nucleus, major axis observation).

%
\renewcommand{\arraystretch}{0.8}
\renewcommand{\tabcolsep}{2.0mm}
\begin{table*}[htbp]
\footnotesize
\caption{Synthesis results in terms of flux fractions for NGC\,1052
along the major axis.}
\begin{tabular} {l|c c c c c c c c}
\hline
\hline
\multicolumn{9}{c}{\it{Flux fraction at $\lambda$5870 \,\AA }} \\
\hline
R (arcsec) &  G1 &  G2 &  G3
&  Y1 &  Y2 &  Y3
&  RHII & $E(B-V)_i$ \\
           &(\%) &(\%) &(\%)
&(\%) &(\%) &(\%)
& (\%)  & (mag)      \\
\hline
0.00       & 61.8 $\pm$ 8.5  &  22.8 $\pm$ 9.9 &  4.2 $\pm$ 4.4
& 3.4 $\pm$ 3.8 & 3.4 $\pm$ 3.7 & 3.2 $\pm$ 3.8
& 1.4 $\pm$ 2.3 & 0.05\\
\hline
 1.10\,SE         & 79.7 $\pm$ 4.9 &  3.9 $\pm$ 4.0 &  8.8 $\pm$ 5.5
&  1.4 $\pm$ 2.5 &  1.5 $\pm$ 2.3 &  3.9 $\pm$ 3.8
&  0.8 $\pm$ 1.8 & 0.00 \\
 3.56\,SE         & 55.0 $\pm$ 6.1 & 8.6 $\pm$ 7.3 &  5.4 $\pm$ 5.4
& 3.7 $\pm$ 3.9 & 2.4 $\pm$ 3.1 & 24.6 $\pm$ 3.7
& 0.3 $\pm$ 1.1 & 0.00 \\
 6.84\,SE         & 69.4 $\pm$ 4.2 & 10.9 $\pm$ 6.9 &  4.4 $\pm$ 4.2
& 0.6 $\pm$ 1.6 & 10.6 $\pm$ 3.8 & 3.5 $\pm$ 4.1
& 0.6 $\pm$ 1.6 & 0.00 \\
 11.80\,SE        & 57.1 $\pm$ 5.1 & 11.4 $\pm$ 6.4 & 2.6 $\pm$ 3.5
& 3.7 $\pm$ 4.1 & 20.4 $\pm$ 5.1 & 4.2 $\pm$ 4.5
& 0.4 $\pm$ 1.4 & 0.00\\
 20.21\,SE        & 24.7 $\pm$ 4.6 & 35.3 $\pm$ 9.1 & 13.7 $\pm$ 6.7
& 2.5 $\pm$ 3.2 & 6.1 $\pm$ 4.7 & 16.9 $\pm$ 5.0
& 0.7 $\pm$ 1.7 & 0.00 \\
\hline
1.08\,NW          & 61.0 $\pm$ 3.3  &  19.7 $\pm$ 3.4 & 11.0 $\pm$ 5.5
& 1.0 $\pm$ 2.0 & 3.0 $\pm$ 3.6 & 3.3 $\pm$ 4.3
& 1.0 $\pm$ 2.0 & 0.04\\
3.54\,NW          & 38.6 $\pm$ 5.1  &  23.6 $\pm$ 5.8 & 18.7 $\pm$ 4.4
& 2.4 $\pm$ 2.9 & 5.4 $\pm$ 4.0 & 9.4 $\pm$ 5.3
& 2.0 $\pm$ 2.7 & 0.00\\
6.81\,NW          & 39.5 $\pm$ 6.7  &  26.9 $\pm$ 7.0 & 1.7  $\pm$ 2.9
& 7.6 $\pm$ 5.6 & 8.3 $\pm$ 5.7 & 6.4 $\pm$ 5.2
& 9.6 $\pm$ 3.3 & 0.00\\
11.78\,NW         & 32.9 $\pm$ 4.5  &  27.5 $\pm$ 4.1 & 10.4 $\pm$ 5.6
& 3.7 $\pm$ 3.3 & 8.3 $\pm$ 2.8 & 9.0 $\pm$ 6.4
& 8.1 $\pm$ 2.8 & 0.00\\
20.19\,NW         & 36.3 $\pm$ 9.3 &  29.5 $\pm$ 9.7 & 3.6  $\pm$ 4.5
& 5.9 $\pm$ 5.4 & 7.1 $\pm$ 5.6 & 6.0 $\pm$ 6.0
& 11.6 $\pm$ 4.5 & 0.00\\
\hline
\end{tabular}
\begin{list} {Tables Notes.}
\item  Last column: internal reddening, corresponding to the Galactic law.
\end{list}
\label{tabcontrib1052major}
\end{table*}
%

%
\renewcommand{\arraystretch}{0.8}
\renewcommand{\tabcolsep}{2.0mm}
\begin{table*}[htbp]
\footnotesize
\caption{Synthesis results in terms of flux fractions for NGC\,1052
along the minor axis.}
\begin{tabular} {l|c c c c c c c c}
\hline
\hline
\multicolumn{9}{c}{\it{Flux fraction at $\lambda$5870 \,\AA }} \\
\hline
R (arcsec) &  G1 &  G2 &  G3 &  Y1 &  Y2 &  Y3 &  RHII & $E(B-V)_i$ \\
           &(\%) &(\%) &(\%) &(\%) &(\%) &(\%) & (\%)  & (mag)      \\
\hline
0.00     & 60.9 $\pm$ 7.8 &  8.5 $\pm$ 9.2 &  5.1 $\pm$ 6.1
& 9.1 $\pm$ 9.0 & 11.1 $\pm$ 7.4 & 3.4 $\pm$ 4.6
& 2.0 $\pm$ 3.2 & 0.05 \\
\hline
2.17\,SW & 60.7 $\pm$ 3.2 &  1.7 $\pm$ 2.8 & 30.3 $\pm$ 5.4 
& 1.0 $\pm$ 2.0 &  2.5 $\pm$ 3.2 & 3.7 $\pm$ 3.5 
& 0.0 $\pm$ 0.0 & 0.02 \\
4.90\,SW & 69.0 $\pm$ 5.9 &  7.3 $\pm$ 5.7 &  5.5 $\pm$ 5.3
& 3.7 $\pm$ 4.0 &  9.0 $\pm$ 5.5 & 2.7 $\pm$ 3.4
& 2.8 $\pm$ 3.2 & 0.03 \\
9.41\,SW & 66.9 $\pm$ 6.1 & 13.7 $\pm$ 6.6 &  8.4 $\pm$ 6.3
& 2.2 $\pm$ 3.1 &  3.9 $\pm$ 4.1 & 3.3 $\pm$ 3.9
& 1.5 $\pm$ 2.5 & 0.02 \\
\hline
2.19\,NE & 70.3 $\pm$ 7.3 & 11.2 $\pm$ 7.0 &  9.7 $\pm$ 6.4
& 2.4 $\pm$ 3.2 &  2.3 $\pm$ 3.1 & 2.7 $\pm$ 3.5
& 1.4 $\pm$ 2.4 & 0.05 \\
4.92\,NE & 57.3 $\pm$ 8.0 & 19.2 $\pm$ 9.1 &  2.7 $\pm$ 3.9
& 4.4 $\pm$ 4.7 &  9.3 $\pm$ 6.7 & 2.1 $\pm$ 3.1
& 5.0 $\pm$ 4.0 & 0.01 \\
9.43\,NE & 56.5 $\pm$ 6.6 & 12.2 $\pm$ 7.4 &  4.1 $\pm$ 4.5
& 4.4 $\pm$ 4.5 &  14.5 $\pm$ 6.8 & 2.9 $\pm$ 3.9
& 5.3 $\pm$ 3.9 & 0.02 \\
\hline
\end{tabular}
\begin{list} {Tables Notes.}
\item  Last column: internal reddening, corresponding to the Galactic law.
\end{list}
\label{tabcontrib1052minor}
\end{table*}
%

%
\renewcommand{\arraystretch}{0.8}
\renewcommand{\tabcolsep}{2.0mm}
\begin{table*}[htbp]
\footnotesize
\caption{Synthesis results in terms of flux fractions for NGC\,7796
along the major axis.}
\begin{tabular} {l|c c c c c c c c}
\hline
\hline
\multicolumn{9}{c}{\it{Flux fraction at $\lambda5870$ \AA }} \\
\hline
R (arcsec) & G1 & G2 & G3 & Y1 & Y2 & Y3 & RHII & $E(B-V)_i$ \\
      & (\%) & (\%) & (\%) &  (\%) & (\%) & (\%) & (\%) & (mag) \\
\hline
0.00     & 54.2 $\pm$ 6.3 & 22.3 $\pm$ 6.9  &  5.9 $\pm$ 8.1
&  3.3 $\pm$ 3.9 & 4.3 $\pm$ 4.5 &  5.2 $\pm$ 5.4
& 4.6 $\pm$ 3.8 & 0.05 \\
\hline
1.09\,S  & 44.3 $\pm$ 9.5  & 29.7 $\pm$ 9.5  &  7.2 $\pm$ 6.4
&  3.9 $\pm$ 4.4 & 7.0 $\pm$ 5.9 &  2.7 $\pm$ 3.5
& 5.1 $\pm$ 3.9 & 0.04 \\
3.55\,S  & 35.7 $\pm$ 7.5  & 33.0 $\pm$ 8.7  &  5.1 $\pm$ 5.1
&  7.1 $\pm$ 4.9 & 4.4 $\pm$ 4.4 & 10.2 $\pm$ 5.2
& 4.4 $\pm$ 3.3 & 0.03 \\
7.10\,S  & 49.3 $\pm$ 6.8  & 22.9 $\pm$ 7.0  &  8.4 $\pm$ 6.2
&  3.3 $\pm$ 4.0 & 5.6 $\pm$ 4.9 &  7.5 $\pm$ 5.7
& 2.9 $\pm$ 3.2 & 0.01 \\
13.21\,S & 50.7 $\pm$ 9.0  & 20.8 $\pm$ 9.5  &  7.0 $\pm$ 6.2
&  7.3 $\pm$ 5.6 & 4.2 $\pm$ 4.2 &  6.5 $\pm$ 5.6
& 3.5 $\pm$ 3.4 & 0.01 \\
\hline
1.09\,N  & 51.7 $\pm$ 9.5  & 21.8 $\pm$ 10.1 &  6.0 $\pm$ 6.2
&  4.9 $\pm$ 4.9 & 7.6 $\pm$ 5.9 &  3.2 $\pm$ 4.1
& 4.8 $\pm$ 4.0 & 0.03 \\
3.55\,N  & 48.0 $\pm$ 10.6 & 15.5 $\pm$ 11.2 & 12.2 $\pm$ 8.5
&  6.3 $\pm$ 5.9 & 6.0 $\pm$ 5.2 &  7.2 $\pm$ 5.9
& 4.7 $\pm$ 3.9 & 0.02 \\
7.10\,N  & 43.0 $\pm$ 10.5 & 18.7 $\pm$ 10.0 &  8.1 $\pm$ 6.9
&  9.6 $\pm$ 6.4 & 6.6 $\pm$ 5.3 &  7.6 $\pm$ 6.2
& 6.3 $\pm$ 4.5 & 0.02 \\
13.21\,N & 44.6 $\pm$ 10.5 & 16.0 $\pm$ 9.8  &  9.4 $\pm$ 7.4
& 10.8 $\pm$ 6.4 & 6.7 $\pm$ 5.2 &  8.2 $\pm$ 4.8
& 4.3 $\pm$ 3.7 & 0.01 \\
\hline
\end{tabular}
\begin{list} {Tables Notes.}
\item  Last column: internal reddening, corresponding to the Galactic law.
\end{list}
\label{tabcontrib7796major}
\end{table*}
%

%
\renewcommand{\arraystretch}{0.8}
\renewcommand{\tabcolsep}{1.0mm}
\begin{table*}[htbp]
\footnotesize
\caption{Synthesis results in terms of flux fractions for NGC\,7796
along the minor axis}
\begin{tabular} {l|c c c c c c c c}
\hline
\hline
\multicolumn{9}{c}{\it{Flux fraction at $\lambda5870$ \AA }} \\
\hline
R (arcsec) & G1 & G2 & G3 & Y1 & Y2 & Y3 & RHII & $E(B-V)_i$ \\
           & (\%) & (\%) & (\%) &  (\%) & (\%) & (\%) & (\%) & (mag) \\
\hline 
0.00     & 55.0 $\pm$ 9.2   & 21.2 $\pm$ 9.8  & 5.4 $\pm$ 5.6
&  4.4 $\pm$ 4.7 &  6.0 $\pm$ 5.3 &  4.2 $\pm$ 4.5
& 3.7 $\pm$ 3.7 & 0.02 \\
\hline
1.09\,E  & 54.5 $\pm$ 9.0  & 25.2 $\pm$ 9.9 & 4.5 $\pm$ 5.2
&  3.4 $\pm$ 3.6 &  2.5 $\pm$ 3.2 &  3.4 $\pm$ 3.9
& 6.5 $\pm$ 3.4 & 0.01 \\
3.55\,E  & 38.1 $\pm$ 10.5 & 23.7 $\pm$ 11.5 & 6.9 $\pm$ 6.7
& 11.1 $\pm$ 7.2 &  5.9 $\pm$ 5.0 & 10.3 $\pm$ 6.3
& 3.9 $\pm$ 3.6 & 0.01 \\
6.93\,E  & 36.9 $\pm$ 8.0  & 24.0 $\pm$ 9.3 & 10.0 $\pm$ 7.6
& 10.6 $\pm$ 6.1 &  8.0 $\pm$ 5.2  &  7.7 $\pm$ 5.5
& 2.9 $\pm$ 3.1 & 0.02 \\
12.50\,E & 38.4 $\pm$ 6.0  & 21.3 $\pm$ 8.6 & 9.0 $\pm$ 7.7
&  7.00 $\pm$ 5.8 &  2.1 $\pm$ 3.1 & 20.5 $\pm$ 6.1
& 1.8 $\pm$ 2.6 & 0.01 \\
\hline
1.09\,W  & 52.2 $\pm$ 6.7  & 20.5 $\pm$ 7.3 & 5.8 $\pm$ 5.1
&  4.8 $\pm$ 4.8 & 10.9 $\pm$ 5.1 &  3.9 $\pm$ 4.2
& 1.8 $\pm$ 2.7 & 0.02 \\
3.55\,W  & 44.0 $\pm$ 8.9  & 12.7 $\pm$ 7.9 & 6.4 $\pm$ 6.2
&  7.4 $\pm$ 5.8 & 13.9 $\pm$ 5.9 & 11.6 $\pm$ 7.1
& 4.0 $\pm$ 3.9 & 0.01 \\
6.93\,W  & 54.0 $\pm$ 8.2  &  9.5 $\pm$ 7.3 & 9.0 $\pm$ 7.3
&  7.6 $\pm$ 5.4 & 10.8 $\pm$ 5.1 &  5.1 $\pm$ 5.1
& 4.0 $\pm$ 3.8 & 0.02 \\
12.50\,W & 32.5 $\pm$ 6.4  & 37.4 $\pm$ 8.3 & 8.7 $\pm$ 7.1
&  8.5 $\pm$ 5.9 &  2.6 $\pm$ 3.2 &  4.3 $\pm$ 4.6
& 5.9 $\pm$ 4.0 & 0.03 \\
\hline
\end{tabular}
\begin{list} {Tables Notes.}
\item  Last column: internal reddening, corresponding to the Galactic law.
\end{list}
\label{tabcontrib7796minor}
\end{table*}
%

%
\begin{figure}[htbp]
\resizebox{\hsize}{!}{\includegraphics{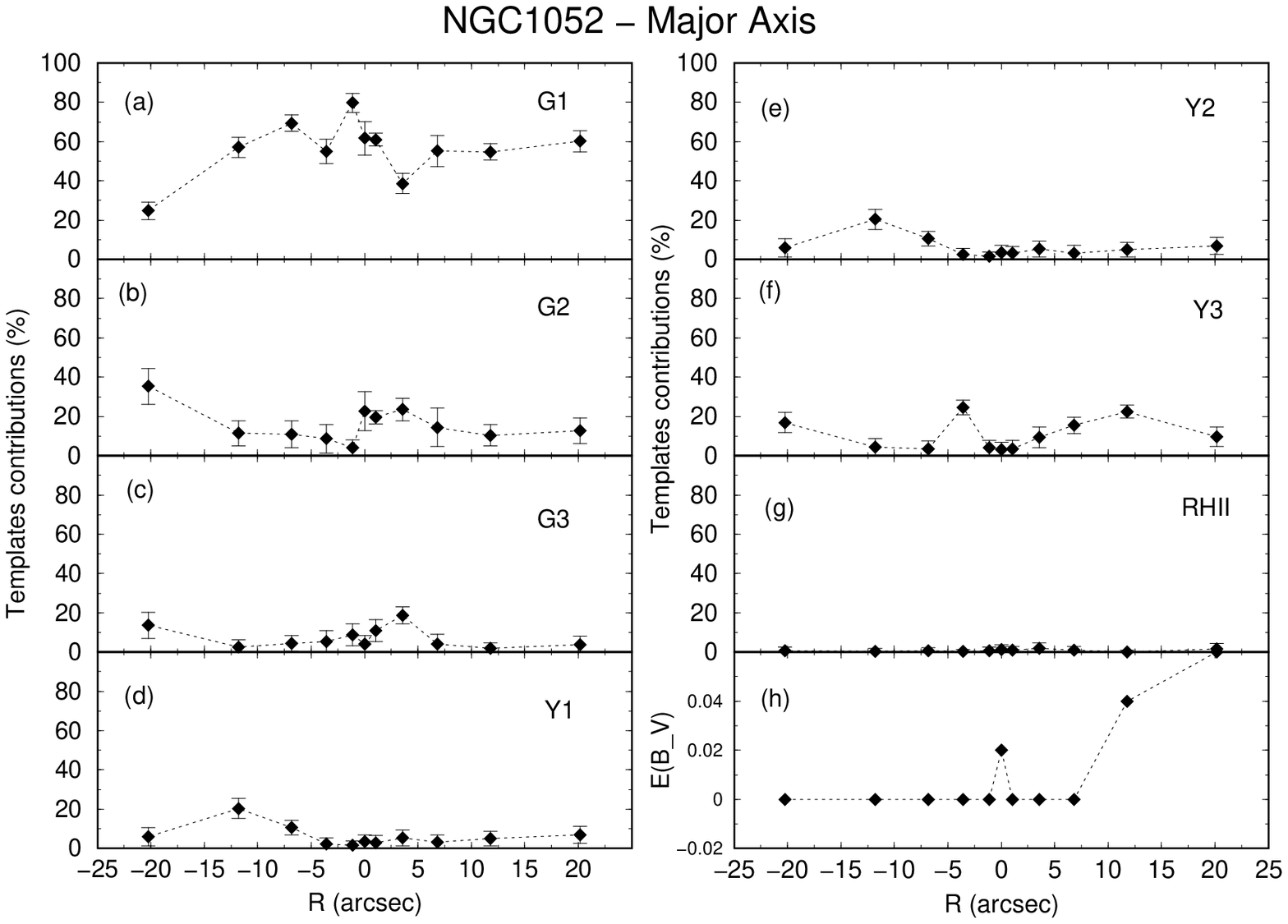}}
\resizebox{\hsize}{!}{\includegraphics{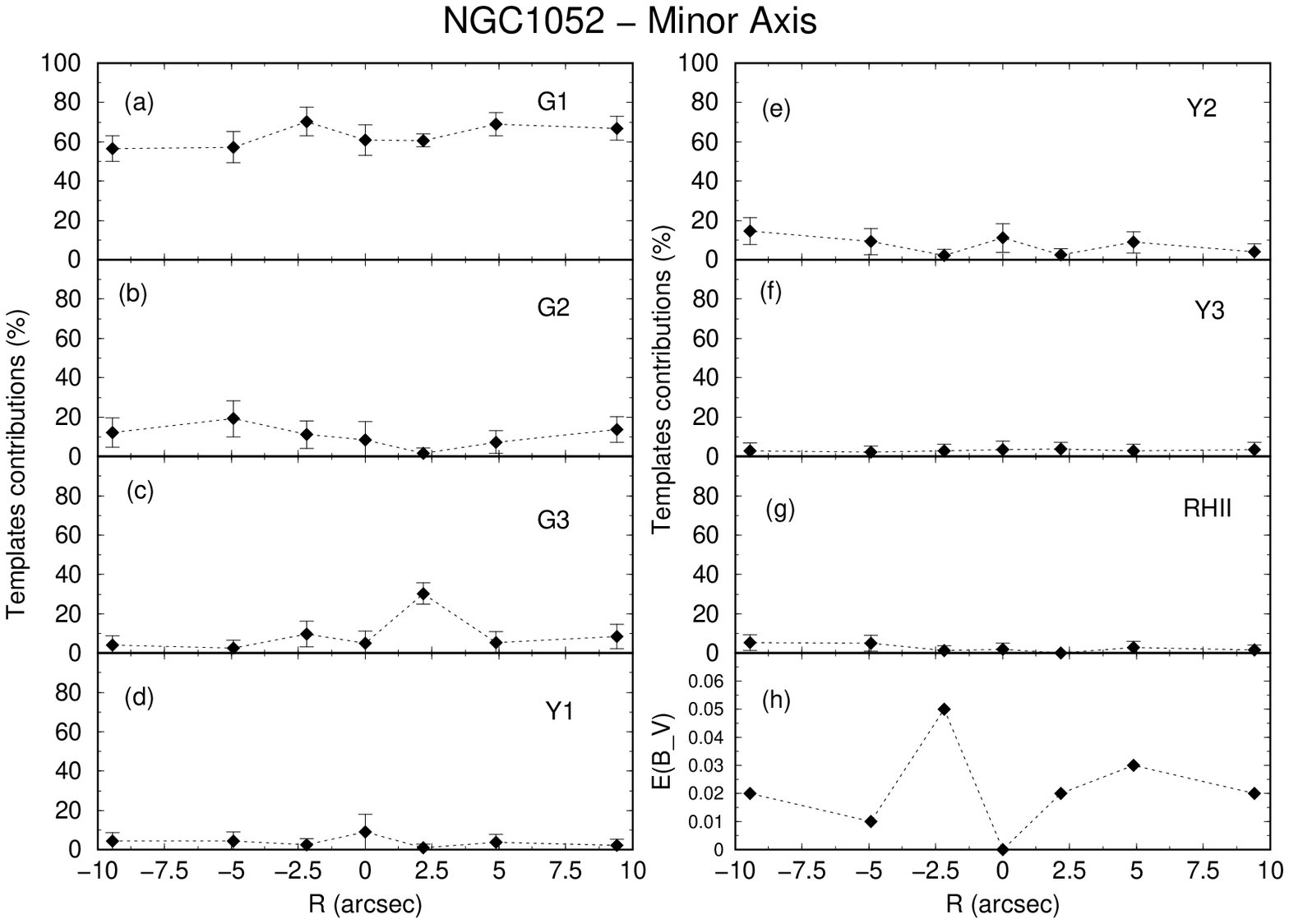}}
\caption{Synthesis results
of NGC\,1052
in flux fractions as a function of the distance to the center
along both photometric axes -
panels {\bf a)}, {\bf b)}, {\bf c)}, {\bf d)}, {\bf e)}, {\bf f)}
and {\bf g)} for the G1, G2, G3, Y1, Y2, Y3 and RHII components respectively;
panel {\bf h)} - spatial distribution of the internal reddening.
Note that the abscissa has different scale for each axis plot.}
\label{spatial_contribution1052}
\end{figure}
%

%
\begin{figure}[htbp]
\resizebox{\hsize}{!}{\includegraphics{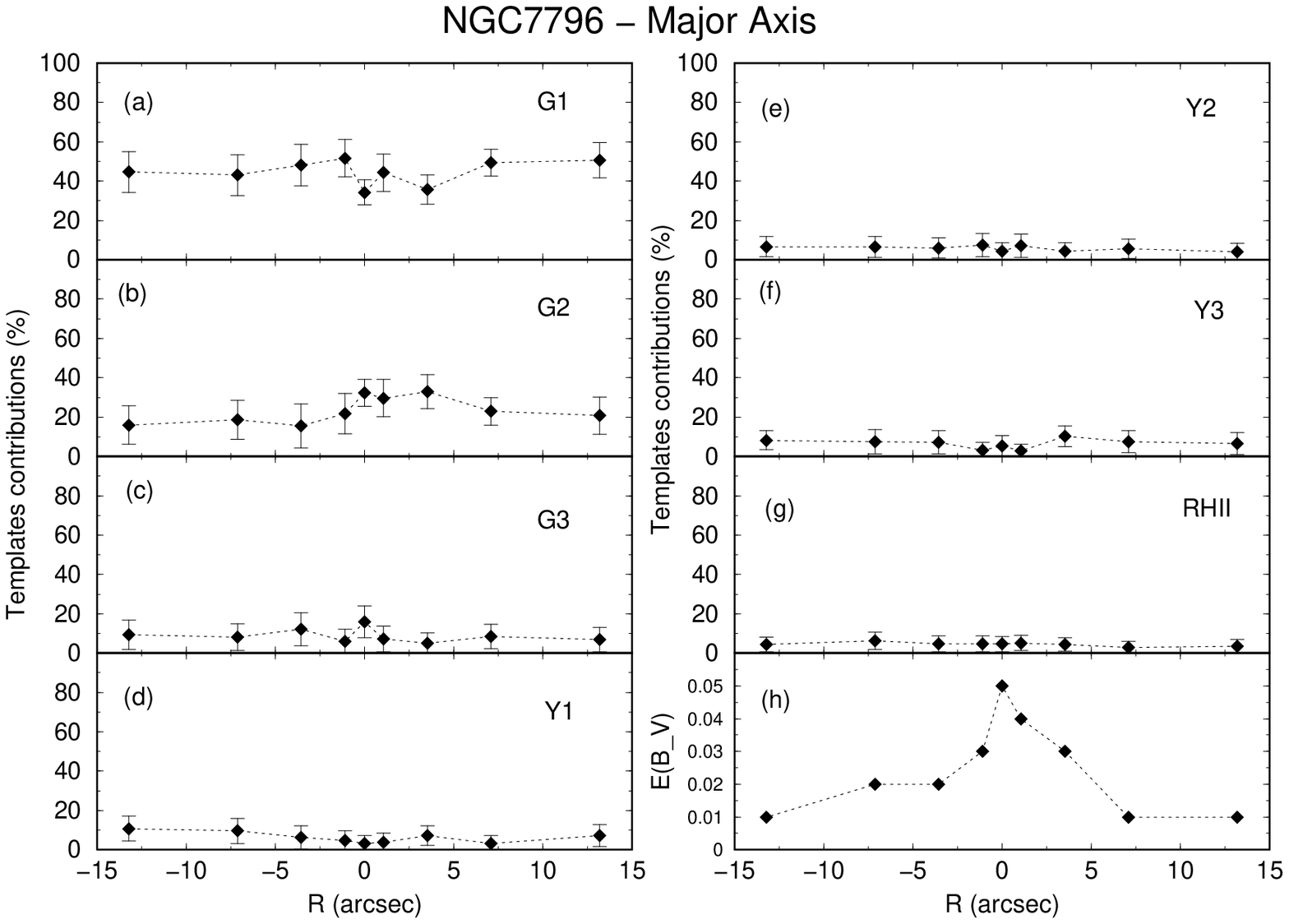}}
\resizebox{\hsize}{!}{\includegraphics{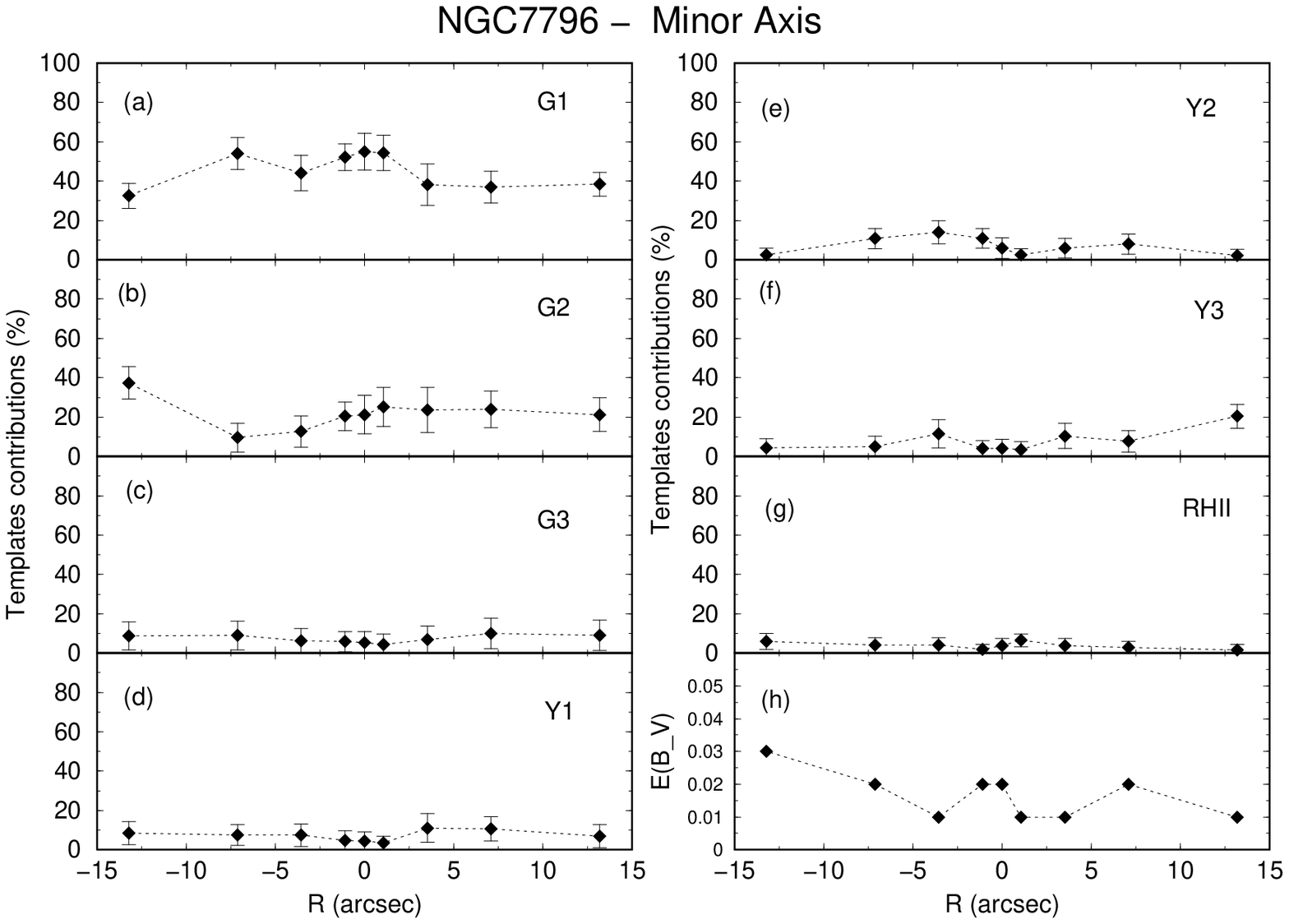}}
\caption{Synthesis results
of NGC\,7796
in flux fractions as a function of the distance to the center
along both photometric axes -
panels {\bf a)}, {\bf b)}, {\bf c)}, {\bf d)}, {\bf e)}, {\bf f)}
and {\bf g)} for each base component;
panel {\bf h)} - spatial distribution of the internal reddening
(like Fig. 24).}
\label{spatial_contribution7796}
\end{figure}
%

%
\begin{figure}[htbp]
\resizebox{\hsize}{!}{\includegraphics{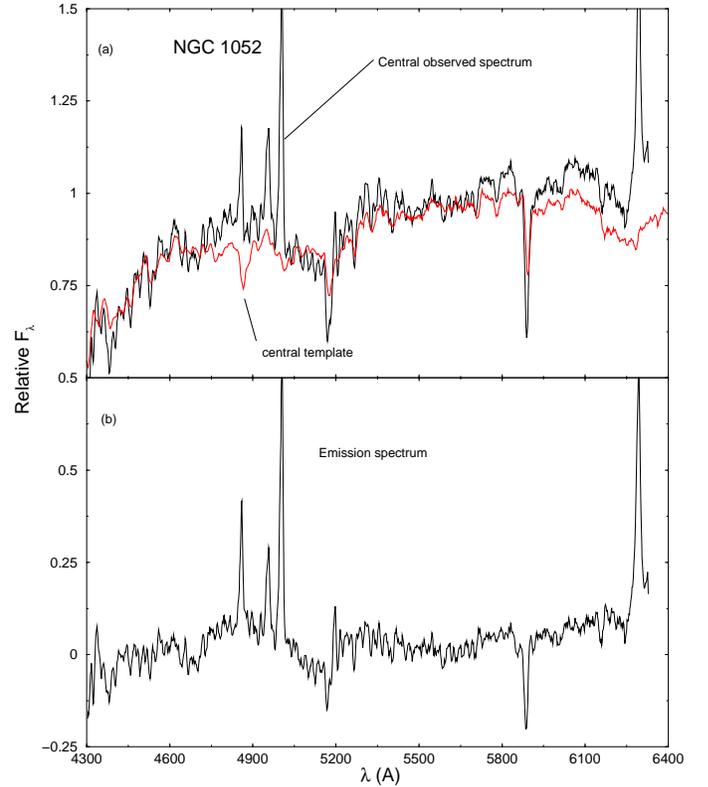}}
\caption{Stellar population synthesis of the central extraction of
NGC\,1052 (major axis extraction).
{\bf a)} Observed spectrum corrected for reddening and template spectrum.
{\bf b)} Residual spectrum as a pure emission spectrum.}
\label{result_center1052}
\end{figure}
%

%
\begin{figure}[htbp]
\resizebox{\hsize}{!}{\includegraphics{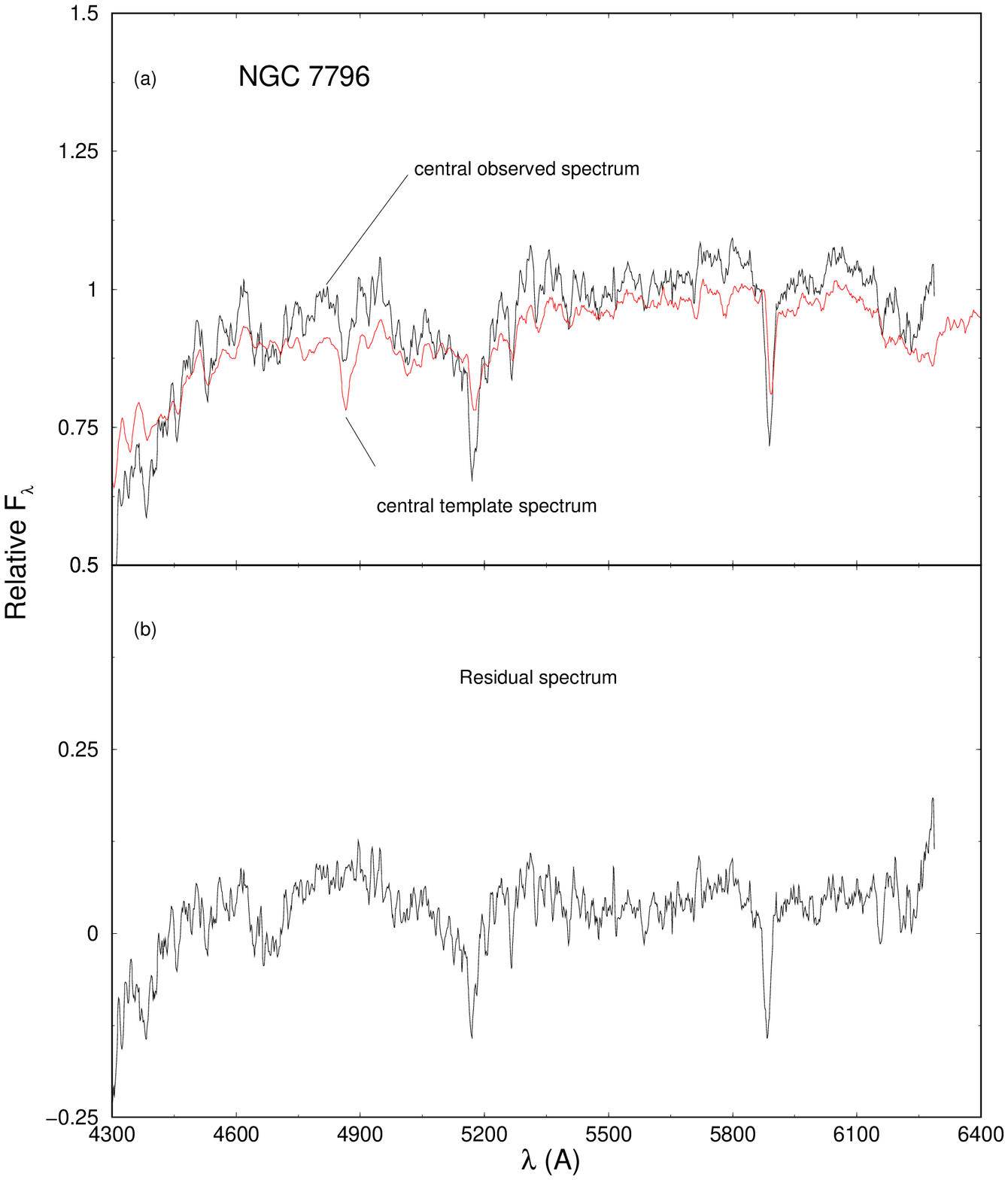}}
\caption{Stellar population synthesis of the central extraction of
NGC\,7796 (major axis extraction).
{\bf a)} Observed spectrum corrected for reddening and template spectrum.
{\bf b)} Residual spectrum.}
\label{result_center7796}
\end{figure}

The spectra representing the composite stellar population 
of NGC\,1052 and NGC\,7796 have been constructed
using the star cluster templates combined
according to the proportions given by each synthesis result.
We illustrate this procedure in Figures 26 and 27 respectively
for NGC\,1052 and NGC\,7796,
in which we show the synthesis for the central aperture 
using their major axis extractions only (top panel).
The resulting residual spectrum for each galaxy,
which is obtained after the subtraction of the population template,
is shown in the bottom panel of both figures.
For NGC\,1052, it is a pure emission spectrum with some absorption lines.
For NGC\,7796, it is a pure absorption spectrum.
On the synthetic spectra of both galaxy nuclei compared to the observed ones,
we can note there are residuals for the Mg b and Na D absorptions
that are greater than for the Fe lines
possibly indicating some overabundance of those elements relative to iron.

The goodness of the stellar population synthesis method can be
assessed by looking at the observed and synthetic spectra
(Figs. 26 and 27).

%
%
\section{Star formation history in NGC\,1052 and NGC\,7796}

The star formation history of some region of a galaxy can be characterized
by the luminosity-weighted mean age of the bulk of its stars and
the timescale of the correspondent main episode.
In order to find the luminosity-averaged age of a galaxy region
we have made the stellar population synthesis described in Section 6.
The formation timescale has been obtained through the
the luminosity-averaged $\alpha$/Fe abundance ratio (see Sect. 5).
The [$\alpha$/Fe] solar relative ratio is linearly proportional to
the logarithm of the timescale of a star forming episode
$\Delta t_{SFR}$ (Gyr),
which is the FWHM of a gaussian star formation rate
(Thomas et al. 2005):

$$ [\alpha/Fe] \approx \frac{1}{5} - \frac{1}{6} \log \Delta t_{SFR} $$

Our results for [$\alpha$/Fe] and stellar age for {\bf NGC\,1052}
are divided into the bulge (minor axis data) 
and along the stellar disc (major axis data) as following:
(i) $\alpha$/Fe is little oversolar and quite homogeneous
in the galaxy center ([$\alpha$/Fe] = +0.2 $\pm$ 0.2 dex)
or around the bulge ([$\alpha$/Fe] = +0.4 $\pm$ 0.1 dex)
but it varies along the disc
from the solar value up to [$\alpha$/Fe] = +0.5 dex
and does not show a monotonic radial dependency
(we point that the average dispersion of [$\alpha$/Fe]
is greater than the error of a single estimation of it, see Sect. 5);
(ii) the bulk of the stars is old, 10-15 Gyr, in the bulge
(dominates 82.\% of the V light)
and there is only along the disc
an outwards radial increasing of a intermediate-aged (or young)
stellar population together to a small decreasing of the old component;
(iii) the iron abundance seems to be constant over whole observed region
applying the relation of
Salaris, Chieffi \& Straniero (1993)
cited in Section 5;
(iv) and the global metallicity changes from the solar value up to
[Z/Z$_{\odot}$] = +0.35 dex with a possible decreasing outwards
along the disc only.
These results corroborate part of the conclusions of other works
(Raimamm et al. 2001,
Thomas et al. 2005);
read Section 1.
Therefore, the star formation history in the bulge of NGC\,1052
was dominated by an ancient episode (at $\approx$ 13 Gyr ago)
with a timescale of 0.1 or 1 Gyr at least.
The star formation timescale in the disc was heterogeoneous
based on an unknown galaxy formation process.

For {\bf NGC\,7796}, the results for [$\alpha$/Fe] and stellar age
are divided into the nucleus and along both photometric directions
outwards:
(i) $\alpha$/Fe is oversolar with a small dispersion in the nucleus
([$\alpha$/Fe] = +0.4 $\pm$ 0.1 dex)
and possibly it shows a monotonic radial dependency along both axes 
decreasing to the solar value outwards;
(ii) the stars are old in the nucleus, 10-15 Gyr
(dominate 83.\% of the V light),
and there is along both directions
a small outwards decreasing of this component with
a little homogeneous contribution of a intermediate-aged or young component;
(iii) the iron abundance seems to be constant over whole observed region
if we adopt the cited relation of
Salaris, Chieffi \& Straniero (1993), Section 5;
(iv) and the mean global metallicity is oversolar
with a possible outwards monotonic radial decreasing along both directions
(or it has a high dispersion of 0.35 dex).
The stellar population parameters of the nucleus agree with the results of 
Thomas et al. (2005);
read Section 1.
Therefore, the star formation in the center of NGC\,7796
happened at 10-15 Gyr ago whose bulk of stars has a very small age spread (0.1 Gyr).
The star formation should have continued outwards
in order to explain the strong radial decreasing of the $\alpha$/Fe ratio.

The mass of the stellar content of each galaxy has been estimated
applying the relation of
Thomas et al. (2005)
between the stellar mass and the central velocity dispersion valid to any environment.
The stellar masses of NGC\,1052 and NGC\,7796 are
1.4$\times$10$^{11}$\,$M_{\odot}$
and 2.9$\times$10$^{11}$\,$M_{\odot}$ respectively.

%
%
\section{Summary and conclusions} 

We have made a detailed stellar population analysis, adopting
long slit spectroscopic observations, of two different
early-type galaxies of low density regions of the local Universe:
NGC\,1052, a Liner of a group that shows a stellar rotating disc,
and NGC\,7796, a field spheroid with a counter rotating core.
Both galaxies have stellar masses around 10$^{11}$\,$M_{\odot}$.
The spatial distributions of
the mean luminosity-weighted stellar age, metallicity, and $\alpha$/Fe ratio
along both photometric axes of them
have been obtained in order to reconstruct the star formation history
in their kinematically distinct subsystems.
We have measured several Lick line-strength indices on the aperture spectra
which were extracted from the galaxy center up to almost one effective radius.
The Lick index radial gradients were also measured.
We have confirmed the stellar kinematics of these spheroidal systems as well.
The measured Lick indices were compared to the theoretical ones
of the simple stellar population models of
Thomas, Maraston and Bender (2003a)
directly on the index versus index planes.
We have applied a stellar population synthesis 
using a grid of star cluster templates from
Bica (1988).

Based on the Lick index comparisons with the SSP models,
both galaxies show a stellar overabundance of Magnesium relative to Iron,
which shows a strong dispersion along the disc of NGC\,1052
and it is almost constant in its bulge.
It is higher in the nucleus of NGC\,7796
and has an outwards radial decreasing along both axes of it.
The variation of [$\alpha$/Fe] goes from 0.0 up to +0.5 dex.
A plausible explanation for this points to a big spread of
the star formation timescale inside these galaxies on their formation process.
In the case of NGC\,7796,
the SN-Ia/SN-II ratio should has been crescent from the galaxy center outwards
on a kind of an inside-out galaxy formation process.

The results of the population synthesis indicate that
the nucleus of both ellipticals is dominate by old metal rich stars
($\sim$13 Gyr, $\sim$Z$_{\odot}$).
The populations are more homogeneous in the bulge of NGC\,1052 than along its disc,
where there is an outwards radial decreasing of the contribution of the older-richer components
together with a respective rising proportion of the younger-metal poor ones.
The results for NGC\,7796 are analogous:
older ages and higher metallicities in the nucleus
but with a similar radial behavior of the age and Z variations along both axes.

Differently from
Mehlert et al. (2003),
the observed regions of NGC\,1052 and NGC\,7796 have a strong dispersion
for the parameters of their stellar populations, specially for 
the $\alpha$/Fe ratio and age.
Moreover, the star characteristics are associated with their kinematics:
the stellar populations are older and $\alpha$-enhanced
in the not rotating bulge of NGC\,1052
and counter rotating core of NGC\,7796, while
they show a strong spread of $\alpha$/Fe and age along the rotating disc of NGC\,1052
and an outwards radial decreasing of them outside the core of NGC\,7796.
The global metallicity has an outwards radial decreasing, while
the iron abundance is nearly constant.
The age variation is possibly connected to the $\alpha$/Fe one.

The abundances of Sodium and Carbon of the stellar content
seem to have some spatial dispersion in both galaxies
(maybe an outwards radial decreasing).
However, probably there is some interstellar contribution to the Na D index.
A careful analysis based on SSP models is still necessary because
the models must consider the influence of the abundance variations of Na and C
on the Lick indices.

Therefore, the bulk of the stars in the bulge of NGC\,1052 and KDC of NGC\,7796
was formed in an ancient short episode (maybe at 13 Gyr ago with a timescale of 0.1-1 Gyr)
providing an efficient chemical enrichment by SN-II, while
in the NGC\,1052 disc and outer parts of NGC\,7796
the star formation occurred later with larger temporal scales,
having made the enrichment by SN-Ia important.
Specifically, for NGC\,7796 an inside-out formation is plausible, while
a merging episode with a drawn out star formation is more acceptable for NGC\,1052.

%
%
\begin{acknowledgements}

A. Milone is thankful
to \emph{FAPESP}, Funda\c c\~ao de Amparo \`a Pesquisa do Estado de S\~ao Paulo
(process numbers 2006/05029-3 and 2000/06695-0),
and \emph{Proap-Capes/INPE} as well.
The authors thank to the referees by the relevant suggestions and comments
that improved the manuscript.
The \emph{IRAF} (Image Reduction Analysis Facility) system
was adopted for the data reduction and analysis.
IRAF is distributed by the National Optical Astronomy Observatories,
which are operated by the Association of Universities for Research
in Astronomy, Inc., under cooperative agreement with the National
Science Foundation.

\end{acknowledgements}

%
%

\end{document}